\documentclass{aa}

\usepackage{graphicx,xspace}
\usepackage{natbib}
\usepackage{url}
\usepackage[varg]{txfonts}
\usepackage{threeparttable}
\usepackage{amsmath, amssymb}
\usepackage{graphicx,color}
\usepackage{microtype}
\usepackage{lscape}
\usepackage{threeparttable}
\usepackage{booktabs, makecell}

\def\cxo{\textit{Chandra}\xspace}
\def\ero{\textit{eROSITA}\xspace}
\def\ede{eROSITA\_DE}
\def\eins{\textit{Einstein}\xspace}
\def\exos{\textit{EXOSAT}\xspace}
\def\fxo{${f_{\rm X}}/f_{\rm opt}$}
\def\gai{\textit{Gaia}\xspace}
\def\heo{He{\sc I}\xspace}
\def\het{He{\sc II}\xspace}
\def\integ{\textit{INTEGRAL}\xspace}

\def\ros{\textit{ROSAT}\xspace}
\def\rxte{\textit{RXTE}\xspace}
\def\srgero{\textit{SRG}/eROSITA\xspace}
\def\srg{\textit{SRG}\xspace}

\def\swb{\textit{Swift}/BAT\xspace}

\def\xmmn{\textit{XMM-Newton}\xspace}

\newcommand\fergs{\ensuremath{\mathrm{erg}\,\mathrm{cm}^{-2}\,\mathrm{s}^{-1}}\xspace}
\newcommand\fcgs{\ensuremath{\mathrm{erg}\,\mathrm{cm}^{-2}\,\mathrm{s}^{-1}\,\mbox{\AA}^{-1}}\xspace}
\newcommand\kmps{\ensuremath{\mathrm{km}\,\mathrm{s}^{-1}}\xspace}
\newcommand\lx{\ensuremath{\mathrm{erg}\,\mathrm{s}^{-1}}\xspace}

\begin{document}

\title{Compact white-dwarf binaries in the combined \srgero/SDSS eFEDS survey}

\author{Schwope, A.\inst{1}
\and Kurpas, J.\inst{1,2}
\and Baecke, P.\inst{1}
\and Knauff, K.\inst{1}
\and Stütz, L.\inst{1}
\and Tub\'in-Arenas, D.\inst{1,2}
\and Standke, A. \inst{1,2}
\and Anderson, S.F. \inst{3}
\and Bauer, F. \inst{4,5,6,7}
\and Brandt, W.N. \inst{8,9,10}
\and Covey, K.\inst{11}
\and Demasi, S.\inst{3}
\and Dwelly, T.\inst{12}
\and Freund, S.\inst{12}
\and Friedrich, S. \inst{12}
\and Gänsicke, B.T. \inst{13}
\and Maitra, C.\inst{12}
\and Merloni, A. \inst{12}
\and Munoz-Giraldo, D.\inst{14}
\and Rodriguez, A.\inst{15}
\and Salvato, M.\inst{12}
\and Stassun, K. \inst{16}
\and Stelzer, B.\inst{14}
\and Strong, A.\inst{12}
\and Morrison, S.\inst{17}
}
\institute{Leibniz-Institut f\"ur Astrophysik Potsdam (AIP), An der Sternwarte 16, 14482 Potsdam, Germany\\
\email{aschwope@aip.de}
\and 
Potsdam University, Institute for Physics and Astronomy, Karl-Liebknecht-Straße 24/25, 14476 Potsdam, Germany
\and
Astronomy Department, Box 351580, University of Washington, Seattle, WA 98195, USA
\and
Instituto de Astrof{\'{\i}}sica, Facultad de F{\'{i}}sica, Pontificia Universidad Cat{\'{o}}lica de Chile, Campus San Joaquín, Av. Vicuña Mackenna 4860, Macul Santiago, Chile, 7820436
\and
Centro de Astroingenier{\'{\i}}a, Facultad de F{\'{i}}sica, Pontificia Universidad Cat{\'{o}}lica de Chile, Campus San Joaquín, Av. Vicuña Mackenna 4860, Macul Santiago, Chile, 7820436
\and
Millennium Institute of Astrophysics, Nuncio Monse{\~{n}}or S{\'{o}}tero Sanz 100, Of 104, Providencia, Santiago, Chile
\and
Space Science Institute, 4750 Walnut Street, Suite 205, Boulder, Colorado 80301
\and
Department of Astronomy \& Astrophysics, The Pennsylvania State University, University Park, PA 16802, USA
\and
Institute for Gravitation and the Cosmos, The Pennsylvania State University, University Park, PA 16802, USA
\and
 Department of Physics, The Pennsylvania State University, University Park, PA 16802, USA
\and
 Department of Physics and Astronomy, Western Washington University, 516 High Street, Bellingham, WA 98225, USA
\and
Max-Planck-Institut f\"ur extraterrestrische Physik, Gie{\ss}enbachstra{\ss}e, 85748 Garching, Germany
\and 
Department of Physics, University of Warwick, Coventry CV4 7AL, UK
\and
Institut für Astronomie \& Astrophysik, Eberhard-Karls-Universität Tübingen, Sand 1, 72076 Tübingen, Germany
\and
Department of Astronomy, California Institute of Technology, 1200 East California Boulevard, Pasadena, CA 91125, USA
\and
Department of Physics and Astronomy, Vanderbilt University, Nashville, TN 37235, USA
\and
Department of Astronomy, University of Illinois at Urbana-Champaign, Urbana, IL 61801, USA
}
\authorrunning{Schwope et al.}
\titlerunning{CWDBs in eFEDS}
\date{\today}


\abstract
{X-ray surveys combined with optical follow-up observations are used to generate complete flux-limited samples of the main X-ray emitting source classes. \ero on the Spectrum-Roentgen-Gamma (SRG) mission provides sufficient sensitivity to build significantly enhanced samples of rare X-ray emitting sources. }
{We strive to identify and classify compact white-dwarf binaries, cataclysmic variables (CVs) and related objects, that were detected in the sky area of eFEDS, the \ero Final Equatorial Depths Survey, and were observed in the plate program of SDSS-V.}
{Compact white-dwarf binaries are selected from spectra obtained in the early SDSS-V plate program. A dedicated set of SDSS plate observations were carried out in the eFEDS field, providing spectroscopic classifications for a significant fraction of the optically bright end ($r<22.5$) of the X-ray sample. The identification and subclassification rests on visual inspections of the SDSS spectra, spectral variability, color-magnitude and color-color diagrams involving optical and X-ray fluxes, optical variability and literature work.}
{Upon visual inspection of SDSS spectra and various auxiliary data products we have identified 26  accreting compact white-dwarf binaries (aCWDBs) in eFEDS, of which 24 are proven X-ray emitters. Among those 26 objects are 12 dwarf novae, three WZ Sge-like disk-accreting non-magnetic CVs with low accretion rates, five likely non-magnetic high accretion rate novalike CVs, two magnetic CVs of the polar subcategory, and three double degenerates (AM CVn objects). Period bouncing candidates and magnetic systems are rarer than expected in this sample, but it is too small for a thorough statistical analysis. Fourteen of the systems are new discoveries, of which five are fainter than the \gai magnitude limit. Thirteen aCWDBs have measured or estimated orbital periods, of which five were presented here. Through a Zeeman analysis we revise the magnetic field estimate of the polar system J0926$+0105$, which is likely a low-field polar at $B=16$\,MG. We quantify the success of X-ray versus optical/UV selection of compact white-dwarf binaries which will be relevant for the full SDSS-V survey. We also identify six white-dwarf main-sequence (WDMS) systems, among them one confirmed pre-CV at an orbital period of 17.6 hours and another pre-CV candidate.
}
{This work presents successful initial work in building large samples of all kinds of accreting and X-ray emitting compact white-dwarf binaries that will be continued over the full hemisphere in the years to come.}
\maketitle

\section{Introduction}

\ero on the \srg observatory was launched on July 13, 2019, from the Baikonur cosmodrome into space \citep{predehl+21, sunyaev+21}. In its main mission, which started in December 2019, it performs an all-sky X-ray survey between 0.2 and 10 keV with  unprecedented depth, spatial, and spectral resolution. To reach the depth that is needed to achieve its main cosmological goals the exposure was planned to be built up sequentially through a number of subsequent all-sky surveys with shorter exposure time. To get an idea about the source content and source density eventually achieved over the full sky, \srgero performed a survey over a limited sky area down to its expected all-sky depth after several years. This survey, called eFEDS (eROSITA Final Equatorial Depth Survey), was performed in the CalPV (Calibration and Performance Verification) phase of the mission, prior to the start of the regular survey observations. With 140 deg$^2$, eFEDS is small compared to the whole sky, but the largest contiguous X-ray survey. All its parameters, the special mode of observation, the construction of the source catalog and the identification of likely optical counterparts were described in a number of papers accompanying the Early Data Release of all the CalPV observations performed with \srgero \citep{brunner+22, salvato+22, boller+22, liu+22a, liu+22b}.

\srgero is {\it the} mission to generate large samples of all classes of X-ray emitting objects \citep[see][for a description of the expected impact on all source classes]{merloni+12}. While the survey will detect new objects as X-ray sources in large numbers, the nature of the new sources is seldomly revealed just from the X-ray data. Only exceptionally bright sources will be discovered with sufficient numbers of counts to classify based on the X-ray morphology, the spectrum and/or the light curve without reference to external data. For the overwhelming majority of weak sources, only follow-up observations will uncover their nature. To this end, comprehensive spectroscopic identifications of X-ray sources were performed in SDSS-IV \citep{menzel+16,comparat+20, abdurrouf+22}, they were continued in SDSS-V \citep{kollmeier+17} in its initial plate program \citep{almeida+23} and will further be continued with the robotic fibre positioning system. Here we report results obtained in SDSS-IV and -V with special plates where holes were drilled primarily at positions of candidate optical counterparts of eFEDS-discovered X-ray sources. 

Our focus lies on the class of (accreting) compact white dwarf binaries (aCWDBs), which has much a lower sky density of sources compared to e.g.~coronal emitters and the frequent AGN. These may be addressed shortly as cataclysmic binaries or cataclysmic variables (CBs, CVs) in a general sense, thus comprising also double degenerates (DDs) and other related objects like Symbiotic binaries (SySts). A few of these objects were discovered by chance due to their pronounced variability pattern in \ero CalPV and survey observations \citep[e.g.~][]{ok+23, schwope+22a, schwope+22b} but the current work is the first attempting systematic follow-up with the SDSS telescope and with the SDSS/BOSS spectrograph \citep{gunn+06,smee+13}. Previous surveys with emphasis on aCWDBs were undertaken with \eins, \exos, \ros, \integ, \rxte, \xmmn, and \swb  \citep{hertz+90, giommi+91,  beuermann_schwope94, schwope+02, revnivtsev+08, sazonov+06, motch+10, suleimanov+22}. The focus of such studies, besides just characterizing the source content of X-ray surveys at high and low Galactic latitudes was an assessment of the composition of the famous Galactic ridge X-ray emission \citep[GRXE,][]{worrall+82}. An assessment of CVs contributing to the GRXE as seen by \integ is given by \cite{bouchet+11}. Using the various surveys, space densities and luminosity functions of the relevant source classes were constrained \citep[e.g.,][]{pretorius_knigge12,pretorius+13, pretorius_mukai14,schwope18, suleimanov+22}, although such attempts suffer from small number statistics. The number of known CBs with well-determined distances was simply too small, typically of order 2 or 3 dozen. This is going to change in the years to come through comprehensive and systematic follow-up of \ero discovered sources with the SDSS and with 4MOST \citep{kollmeier+17,dejong+19}, to be combined with \gai-determined distances. The current work presents the very first step in this enterprise.

This manuscript presents initial identification work on Galactic X-ray emitters found in eROSITA surveys, and the chosen field is eFEDS.
Soon after the X-ray observations were completed, X-ray source lists were made and likely counterparts identified. 
The X-ray source list that was used to identify the counterparts for the subsequent SDSS observations was preliminary. Software and calibration updates that became available later were used to generate the final published source lists and counterpart catalogs and explain some differences between the early preliminary and the final catalogs. For a discussion of the X-ray properties in this paper we always use the final catalogs \citep{brunner+22, salvato+22}.

While the focus of this paper lies on aCWDBs, we wish to address briefly also CWDBs, i.e.~objects that are not or not yet accreting and might be possible progenitor objects of aCWDBs. These objects were not recognized as X-ray sources in the \ero observations but were targeted for spectroscopic observations in SDSS-V as CB candidates through, e.g., their ultraviolet excess.

We begin with a short description of the basic \ero X-ray and SDSS-V optical data, followed by a description of auxiliary data and data products that were used to classify the objects. We then describe in some detail the properties and the classification of the individual objects. We then give a concise overview of the main findings of this paper and conclude by giving an outlook into the full SDSS-V survey that has just began with the fiber positioning system (FPS). We refer to eFEDS as the survey and the resulting catalog with a limiting X-ray flux of about $1 \times 10^{-14}$\,\fergs (0.2-2.3 keV). We refer to eRASS1, eRASS2, eRASS3, eRASS4 as the four completed individual surveys performed with \srgero and their corresponding catalogs, and to eRASS:3 as the stack of the first three complete all-sky surveys with a limiting point-source flux of about $2 \times 10^{-14}$\,\fergs.

If not otherwise mentioned, we use distances determined by \cite{bailer-jones+21} which are based on \gai DR3 astrometric parameters and give X-ray fluxes and luminosities in the $0.2-2.3$\,keV energy band. Magnitudes from the legacy survey or SDSS are given in the AB system, those from \gai are given in the VEGAMAG system.

\section{Observations and analysis\label{s:obs}}
\subsection{X-ray observations with \srgero}
The X-ray observations in the eFEDS field with \srgero were performed during the CalPV phase of the mission in November 2019 in the so-called field scanning mode. The survey area is about 140 square degrees. All the details of the observations, the data processing, and the generation of a catalog are described in \cite{brunner+22}. 

\newpage
\begin{landscape}
\begin{table}[t]
\caption{CWDBs in the eFEDS area sorted with increasing right ascension. The column SrcID lists the source ID in the published X-ray source list from \cite{brunner+22}. The X-ray flux in the energy band $0.2-2.3$ keV is given in units of $10^{-14}$\,\fergs. The optical magnitude and color are from the legacy survey. A parallax is given, if the source is listed in \gai DR3. The distance $D$ is $r_{\rm geo}$ from \cite{bailer-jones+21} and the logarithm of the X-ray luminosity ($0.2-2.3$ keV) was calculated using this distance with geometry parameter $4\pi$. The 'Type' column gives the likely subtype with following meaning: AMC -- AM CVn, CV -- not further specified CV, DN -- dwarf nova (not further specified), SU -- SU UMa type dwarf nova, UG -- U Gem type dwarf nova, WZ -- WZ Sge type dwarf nova, P -- polar. The second to last column lists the orbital period in hours, letters in parenthesis refer to explanations in the foot of the table. In the last column references to original publications are given, the given numbers are expanded below the table. The lower part lists CVs that were not X-ray selected. For those X-ray upper limits from eFEDS are given.
\label{t:cvs}
}
\begin{tabular}{|l|l|r|r@{$\pm$}l|r|r|r|r@{$\pm$}l|r|r|l|l|r|c|}
\hline
  \multicolumn{1}{|c|}{SDSS J} &
  \multicolumn{1}{|c|}{eFEDS J} &
  \multicolumn{1}{r|}{ID} &
  \multicolumn{2}{c|}{X-ray flux} &
  \multicolumn{1}{c|}{$g$} &
  \multicolumn{1}{c|}{$g-r$} &
  \multicolumn{1}{r|}{Sep} & 
  \multicolumn{2}{c|}{$\pi$} &
  \multicolumn{1}{r|}{$D$} &
  \multicolumn{1}{r|}{$L_{\rm X}$} &  
  \multicolumn{1}{l|}{Type} &
  \multicolumn{1}{c|}{$P$}&
  \multicolumn{1}{c|}{Ref}
\\
    &   &  & \multicolumn{2}{c|}{} & (mag) & (mag) & (") & \multicolumn{2}{c|}{(mas)} & (pc) & erg/s & & (hr)& \\
\hline
  08:40:41.4$+$00:05:20.3 & 084041.3$+$000517 & 2343 & \multicolumn{2}{c|}{$5.31\pm0.74$}   & 20.80 & $-$0.01 & 3.2 & \multicolumn{2}{c|}{} & & & CV & & 2,3 \\
  08:43:03.5$-$01:48:58.5 & 084303.5$-$014859 & 1445 & \multicolumn{2}{c|}{$11.94 \pm 1.43$}& 18.61 & 0.81 & 1.0 & 0.49 & 0.14 & 2045 & 31.8& NL & 8.31 & 1\\
  08:44:00.1$+$02:39:19.3 & 084400.0$+$023917 & 4654 & \multicolumn{2}{c|}{$3.76 \pm 0.63$} & 18.33 & 0.38 & 2.3 & 0.73 & 0.17 & 1470 & 31.0&DN & 4.97 & 4,12\\
  08:44:13.7$-$01:28:07.8 & 084413.5$-$012806 & 7747 & \multicolumn{2}{c|}{$1.81 \pm 0.38$} & 20.12 & $-$0.02 & 2.4 & 0.29 & 0.32 & & & AMC & 0.53$^{a}$ &2,3\\
  08:46:41.0$+$02:18:23.1 & 084641.0$+$021826 & 7802 & \multicolumn{2}{c|}{$2.07 \pm 0.47$} & 18.61 & 0.70 & 3.2 & 0.11 & 0.14 & 3580 & 31.5 & DN & $>$6$^{b}$ &1\\
  08:47:08.3$+$01:19:39.5 & 084708.2$+$011942 & 13836 & \multicolumn{2}{c|}{$1.82 \pm 0.49$}& 21.91 & $-$0.04 & 3.5 &\multicolumn{2}{c|}{}& & & AMC & & 1\\
  08:47:35.4$+$01:45:33.9 & 084735.5$+$014535 & 7743 & \multicolumn{2}{c|}{$3.44 \pm 0.64$} & 22.02 & 0.05 & 2.2& \multicolumn{2}{c|}{} & & & NL: & & 1\\
  08:50:37.2$+$04:43:57.0 & 085037.2$+$044359 & 532 & \multicolumn{2}{c|}{$17.96 \pm 1.60$} & 18.69 & $-$5.33 & 2.4 & 0.98 & 0.33 & 1080 & 31.4 & P & 1.72& 5 \\
  08:51:07.4$+$03:08:34.4 & 085107.4$+$030835 & 1243 & \multicolumn{2}{c|}{$9.98 \pm 1.14$} & 18.77 & $-$0.08 & 1.6 & 1.72 & 0.12 & 580 & 30.6 & SU & 1.56 & 6\\
  08:53:00.6$+$02:04:24.8 & 085300.6$+$020421 & 6050 & \multicolumn{2}{c|}{$3.74 \pm 0.73$} & 14.97 & $-$0.26 & 3.3 & \multicolumn{2}{c|}{}& & & DN & & 1\\
  08:55:50.9$-$01:54:28.6 & 085550.8$-$015431 & 3469 & \multicolumn{2}{c|}{$3.49 \pm 0.58$} & 21.00 & 0.10 & 3.0 & 6.96 & 1.95 & 247 & 29.4 & DN & & 1\\
  09:02:46.5$-$01:42:01.8 & 090246.7$-$014203 & 10211 & \multicolumn{2}{c|}{$4.97 \pm 1.27$}& 20.85 & 0.17 & 3.4 & 0.46 & 0.73 & 1730 & 31.3& SU: & 2.04: & 1,12\\
  09:03:44.2$-$01:33:26.2 & 090344.3$-$013323 & 2080 & \multicolumn{2}{c|}{$6.45 \pm 0.82$} & 19.55 & $-$0.07 & 3.0 & 3.68 & 0.36 & 280 & 29.8 & AMC &$\sim 1$ & 12,13,1 \\
  09:12:48.2$-$00:07:21.3 & 091247.9$-$000717 & 12260 & \multicolumn{2}{c|}{$1.64 \pm 0.47$}& 21.14 & 0.71 & 5.4 & 0.55 & 1.40 & 990 & 30.3 & DNc & & 1\\
  09:14:10.7$+$01:37:33.0 & 091410.5$+$013733 & 2056 & \multicolumn{2}{c|}{$5.75 \pm 0.78$} & 18.58 & 0.73 & 2.2 & 0.98 & 0.12 & 1020 & 30.9 & UG & 6.043 & 12,14,15\\
  09:18:18.5$+$04:36:10.0 & 091818.4$+$043611 & 15325 & \multicolumn{2}{c|}{$1.35 \pm 0.38$}& 21.58 & 0.23 & 1.9 & \multicolumn{2}{c|}{} & & & DNc & & 1 \\
  09:26:14.3$+$01:05:57.3 & 092614.1$+$010558 & 175 & \multicolumn{2}{c|}{$27.88 \pm 1.63$} & 17.30 & 0.04 & 2.0 & 2.68 & 0.34 & 390 & 30.7 & P & 1.48& 5 \\
  09:26:20.4$+$03:45:42.2 & 092620.9$+$034531 & 14091 & \multicolumn{2}{c|}{$1.22 \pm 0.40$}& 19.82 & 0.21 & 13.2 & 1.21 & 0.32 & 930 & 30.1 & DN & & 9,10\\
  09:29:02.9$+$00:53:34.0 & 092902.8$+$005329 & 2919 & \multicolumn{2}{c|}{$2.96 \pm 0.59$} & 21.13 & 0.48 & 4.0 & \multicolumn{2}{c|}{}& 300:$^{c}$ & 29.5: & WZ: & 1.85: &  1\\
  09:29:21.7$+$04:01:35.2 & 092921.8$+$040136 & 610 & \multicolumn{2}{c|}{$4.15 \pm 0.80$}  & 19.98 & 1.17 & 2.5 & 0.58 & 0.23 & 1850 & 31.2 & CVc & & 1\\
  09:32:05.2$+$03:43:32.7 & 093205.1$+$034336 & 3911 & \multicolumn{2}{c|}{$3.88 \pm 0.65$} & 16.85 & $-$0.08 & 4.0 & 1.57 & 0.10 & 630 & 30.3 & DN & & 11,1\\
  09:32:38.2$+$01:09:02.5 & 093238.3$+$010901 & 223 & \multicolumn{2}{c|}{$22.87 \pm 1.42$} & 20.26 & 0.73 & 1.8 & 0.07 & 0.39 & 3150 & 32.4 & NL & & 4,1\\
  09:35:55.1$+$04:29:15.9 & 093555.0$+$042920 & 1261 & \multicolumn{2}{c|}{$6.69 \pm 0.76$} & 20.18 & 0.81 & 4.5 & 2.77 & 0.23 & 360 & 30.0 & CVc & 1.76 & 1\\
\hline
  08:45:55.1$+$03:39:29.3 & --- & --- & \multicolumn{2}{c|}{$< 2$}    & 20.59 & $-$0.01 & -- & 3.43 & 1.94 & 302 & 29.3 & WZ & 1.42 &3,12 \\
  09:04:03.5$+$03:55:01.3 & --- & --- & \multicolumn{2}{c|}{$<2.9$}  & 19.31 & $-$0.05 & -- & 3.52 & 0.35 &  295 & $<$ 29.5 & WZ & 1.43 & 1,7,8 \\
  09:20:09.5$+$00:42:44.9 & 092009.3$+$004240$^d$ & 27233 & \multicolumn{2}{c|}{$0.69 \pm 0.30$}  & 17.54 & 0.17    & 5.2 & 0.86 & 0.12 & 1170 & & NLc & 3.6 & 4 \\

\hline
\end{tabular}

$^{a)}$ The period was estimated from outburst parameters (likely recurrence time and outburst magnitude), using relations derived by \cite{levitan+15};
$^{b)}$ Period estimate based on SpT of the donor;
$^{c)}$ The parameters (distance, period, luminosity, sub-type) are based on the estimated spectral type of the donor and its brightness using the sequence tabulated by \cite{knigge+11};
$^{(d)}$ The X-ray source is not listed in the main, but is in the supplementary eFEDS X-ray catalog.

References: 
(1) this work,
(2) \cite{drake+14}, 
(3) \cite{breedt+14},
(4) \cite{szkody+03},
(5) \cite{rodriguez+23},
(6) \cite{nogami+96},
(7) \cite{szkody+04},
(8) \cite{woudt+12},
(9) \cite{welsh+05},
(10) \cite{wils+10},
(11) \cite{necker+22},
(12) \cite{inight+23b},
(13) \cite{carter+13}
(14) VSNET 19440
(15) \cite{thorstensen+17}
\end{table}
\newpage
\end{landscape}

The X-ray information we give here is derived from the final catalog comprising 27910 unique sources, 542 of them being significantly extended. It is available from the web-page maintained by the \ede \, collaboration as a fits file named {\tt eFEDS\_c001\_main\_V7.4.fits}\footnote{\url{https://erosita.mpe.mpg.de/edr/eROSITAObservations/Catalogues/}}. Additionally, also a supplementary table of X-ray sources was inspected called {\tt eFEDS\_c001\_supp\_V7.4.fits}, which lists sources with lower detection likelihood.

A word of caution on the utilized X-ray fluxes seems appropriate. Like any other X-ray observatory \ero cannot measure astrophysical fluxes directly. What \ero delivers through the detection process in a certain energy band is a count rate in this band. Only with the assumption of a spectral model can the rate be transformed into a flux, by folding a modeled photon distribution through the response of the X-ray optics and the detector. Pre-calculated energy conversion factors (ECFs in units of \fergs / rate) are used for this purpose. Most of the X-ray sources are AGN which typically follow a 
power-law distribution of the photons. The ECF used in Brunner's catalog also assumes a power-law spectrum for all the sources. aCWDBs often have thermal or multi-component spectra, but in general not power-law spectra. It would thus be more appropriate to model individual sources and determine their best-fit spectral parameters to determine their fluxes or, given the typically low number of photons per source, use a more appropriate ECF. However, the expected deviations are small. The ECF-ratio between the standard power law ECF used for catalog production and that of a 10 keV thermal spectrum absorbed by some cool interstellar matter (same column density of $4\times 10^{20}$\,cm$^{-2}$ assumed) is just 1.037. A less than 4\% uncertainty or bias is small and thus acceptable. We expect and show below (see Tab.~\ref{t:succ}) that the flux uncertainties by using a non-appropriate ECF are small compared to flux variability.

\subsection{SDSS follow-up}
Special plates were drilled for eFEDS follow-up, originally planned to be performed in the spring season of 2020 during SDSS-IV and continued in SDSS-V.
More details are given by \cite{almeida+23}, who describe the 18th data release of the SDSS collaboration. All 13269 spectra from eFEDS science targets were released in DR18, and include spectra obtained in SDSS-V and in earlier phases of the SDSS. All of the eFEDS spectra having a stellar pipeline classification 
and many additional spectra were visually screened to check the pipeline classification and improve the redshifts. The results are published as a value-added-catalog (VAC) via the DR18 VAC web-site: \url{https://www.sdss.org/ dr18/data access/value-added-catalogs/}. 

All SDSS spectra with low signal-to-noise were visually screened to check if the pipeline classification was correct. The spectra were initially sorted into three extragalactic (BLAZARS, AGN, GALAXY) and one galactic category (objects with or without emission lines at zero redshift). A few pipeline-classified extragalactic sources were re-classified as stars. In a second step, all galactic objects were further sub-classified as being either likely single stars (coronal emitters, if found as X-ray sources) or binaries. For this, color-color and color-magnitude diagrams were used in addition to the spectra and finding charts. This exercise revealed the 26 aCWDBs and 6 WDMS binaries which form the backbone of this paper. The spectra of the more than 800 coronal X-emitters are analyzed separately and will be published elsewhere. Without going into any detail, only the careful visual inspection of the spectra led to the identification of the sample presented here. The currently available pipeline classification is rather easily confused to either miss aCWDB objects or mistakenly assign other objects in the CV bin. The completeness of the spectroscopic identification program is discussed below in Sect.~3.3.

All objects that were eventually classified as aCWDBs are listed in Tab.~\ref{t:cvs}. Given are the coordinates of the SDSS-observed counterparts, the associated source ID in the eFEDS X-ray catalog, the X-ray flux in the energy band 0.2$-$2.3\,keV, the SDSS $g$-band magnitude and the $g-r$ color, and, if the object has a counterpart in \gai DR3, the parallax. The table also lists the likely distance, the implied X-ray luminosity, $L_{\rm X} = 4\pi D^2 F_{\rm X}$, the likely aCWDB sub-category, the measured or estimated orbital period and a reference to previous work, if the object was mentioned in the literature earlier. The given luminosities might be regarded indicative only in cases. They are based on the X-ray flux which is some kind of average over a orbital phase interval and would be different if a different phase interval would have been covered. They are based on the nominal distance from \cite{bailer-jones+21} and do not take distance uncertainties into account, and the assumed geometry factor to convert fluxes into luminosities. 

We also list non- or not-yet- accreting WDMS objects to complete the picture. These were selected from the pool of Galactic sources as objects whose spectra are in some sense symbiotic, i.e.~show K- or M-star features with a a blue excess. Sometimes the extra blue component shows broad Balmer absorption lines and thus can easily be recognized as a white dwarf (WDMS object consisting of a white dwarf and a main sequence star), sometimes there is a blue featureless excess. The WDMS objects are listed in Tab.~\ref{t:other}. Given are the coordinates of the SDSS object, the $g$-band magnitude and the $g-r$ optical color from the legacy survey, the measured parallax from \gai, H$\alpha$ emission line parameter, the spectral type of the object and a key to a reference, which is given below the table.

\subsection{Auxiliary data and data products: profiles of individual objects}
In this section we describe the data and derived data products that were used for further characterization and sub-classification of the individual sources. For each of the spectroscopically identified aCWDB or WDMS objects, graphical products were generated and arranged as a profile like the example shown in Fig.~\ref{f:prof}. Spectra and color-magnitude diagrams of all objects discussed in this paper are shown for completeness in the appendix.

\begin{itemize}

\item Finding chart\\
The chart shows a 1 arcmin $\times$ 1 arcmin $g$-band image from the PanSTARRs image server, centered on the position of the SDSS-target, with the \ero X-ray and \gai source positions indicated. The 1 $\sigma$ X-ray positional uncertainty is indicated with a circle.

\item Spectrum plot\\
The mean spectra were retrieved from the SDSS database. The thick black line indicates the measured flux, the grey line behind the 1 $\sigma$ flux error. For aCWDB objects, vertical color bars are highlighting the (expected) positions of the most important emission lines of hydrogen (light blue), neutral (pink) and ionized helium (green) typically encountered in such systems. Inserts were made to illustrate the profiles of the H$\alpha$, H$\beta$, and He{\sc ii}4686 emission lines. For WDMS objects, color bars illustrate just the positions of the H-Balmer lines 
\begin{figure*}[t]
\begin{minipage}[c]{0.39\hsize}
\resizebox{\hsize}{!}{\includegraphics{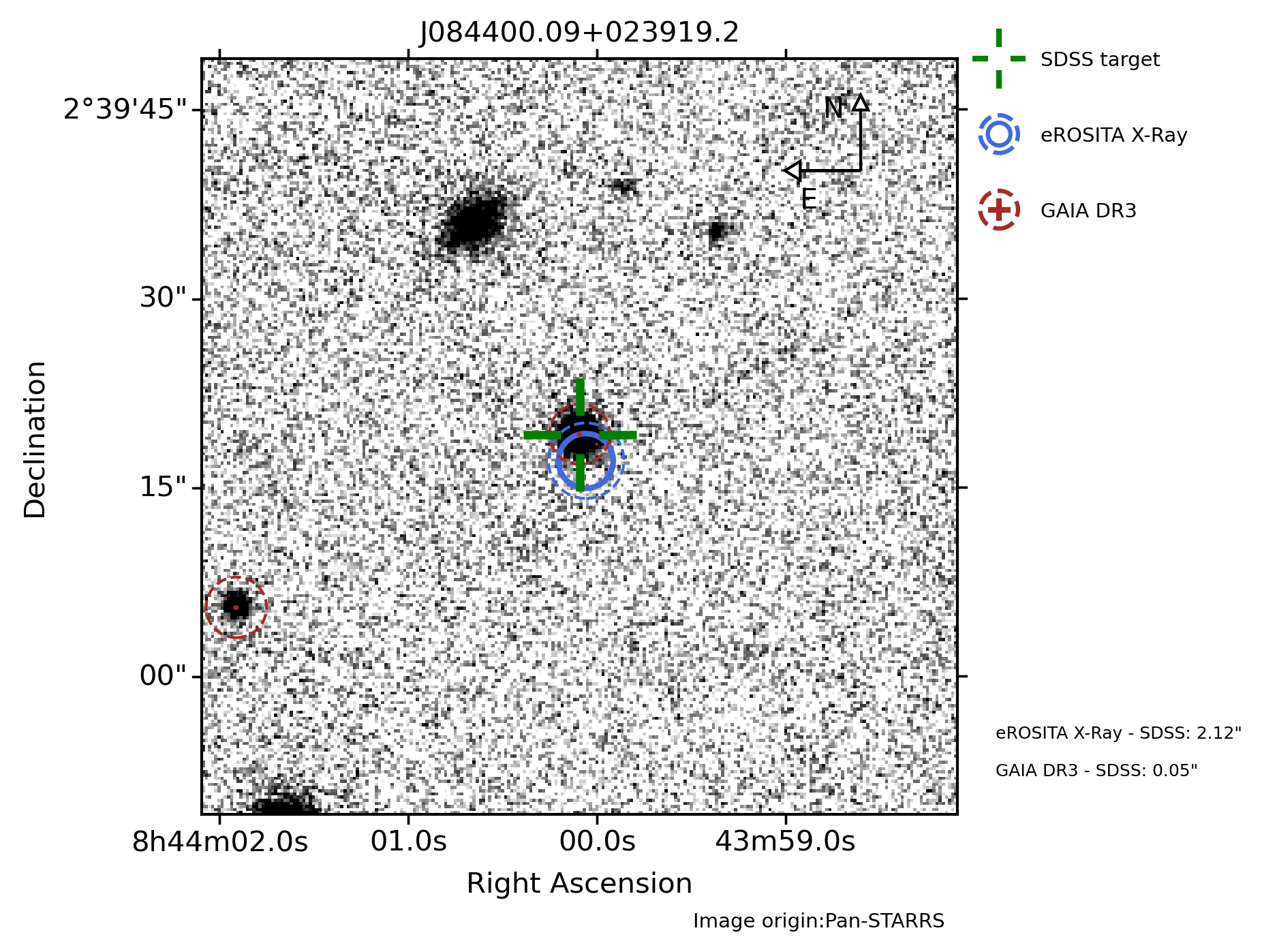}}
\end{minipage}
\hfill
\begin{minipage}[c]{0.60\hsize}
\resizebox{\hsize}{!}{\includegraphics{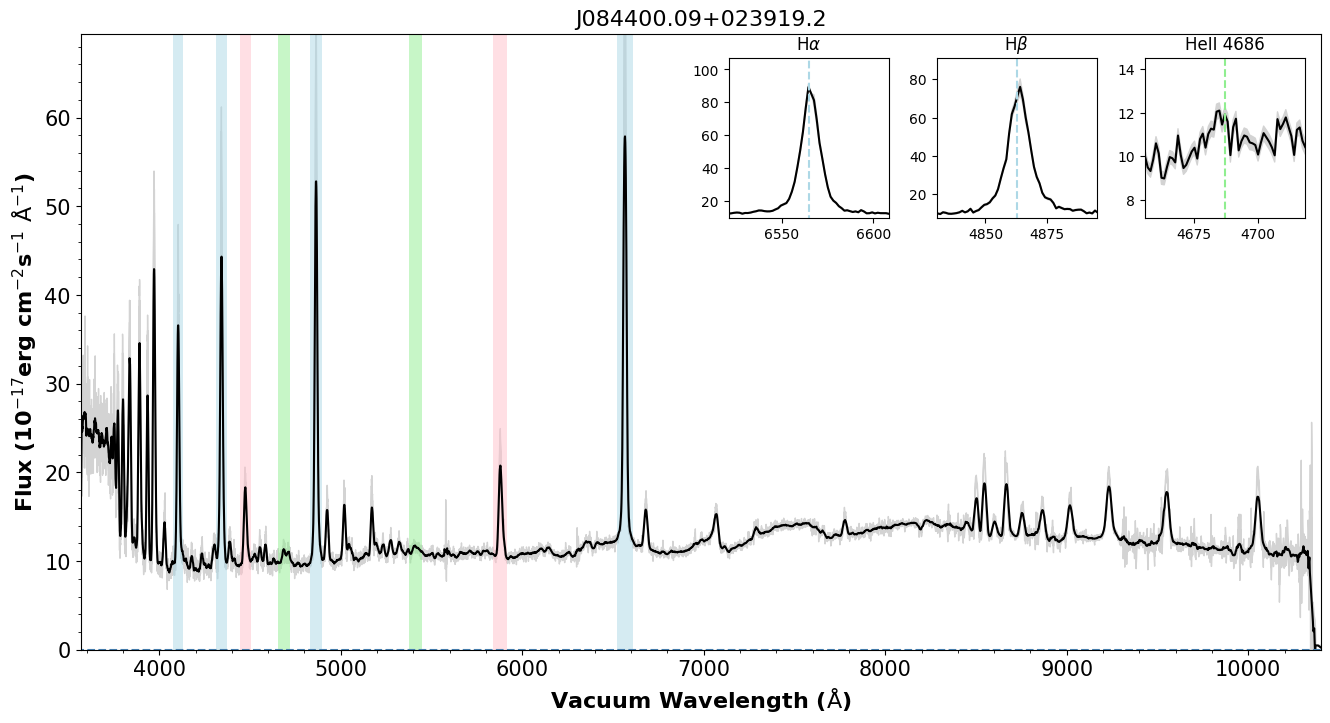}}
ß\end{minipage}
\vspace{1cm}

\begin{minipage}[c]{0.49\hsize}
\resizebox{\hsize}{!}{\includegraphics{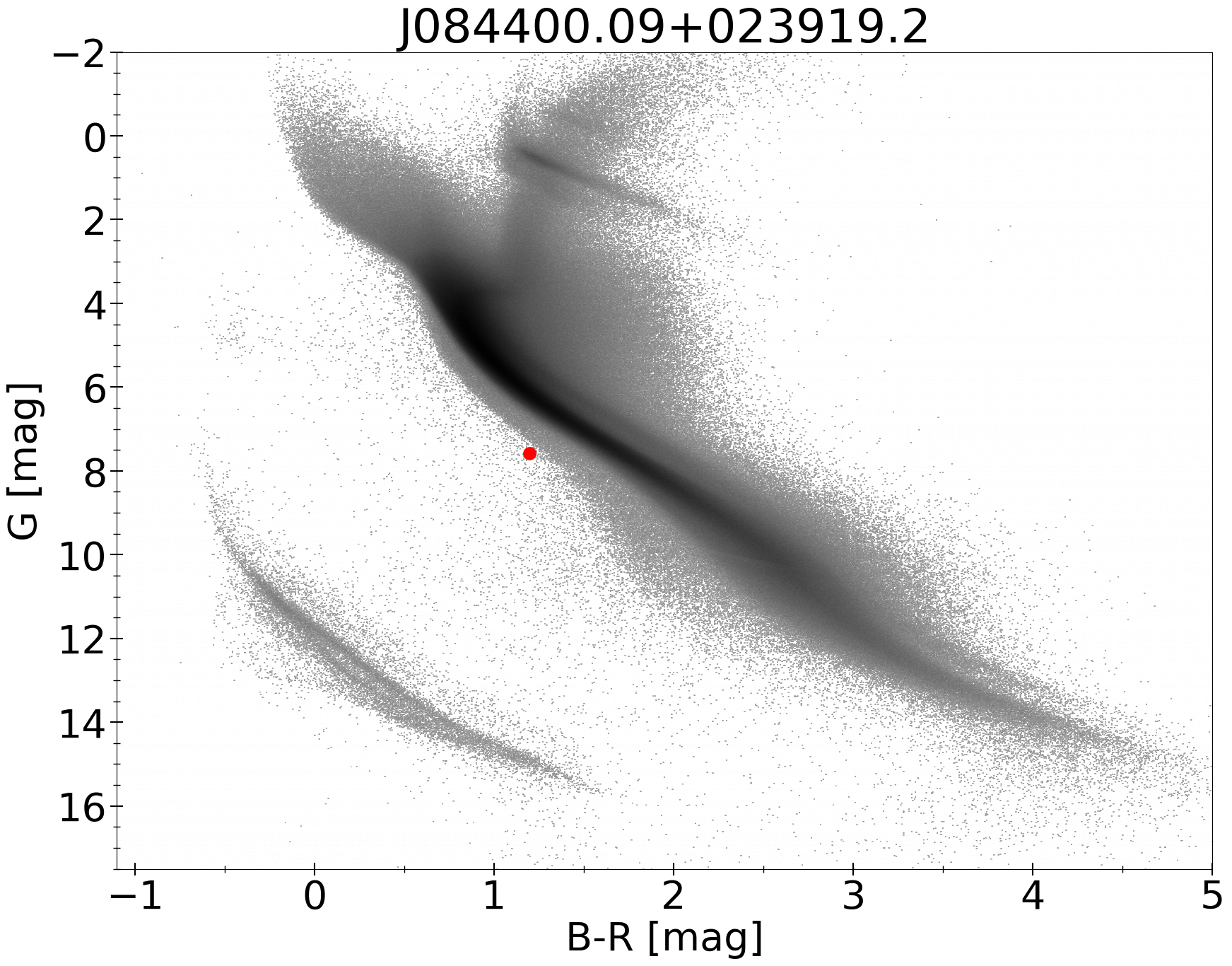}}
\end{minipage}
\hfill
\begin{minipage}[c]{0.49\hsize}
\resizebox{\hsize}{!}{\includegraphics{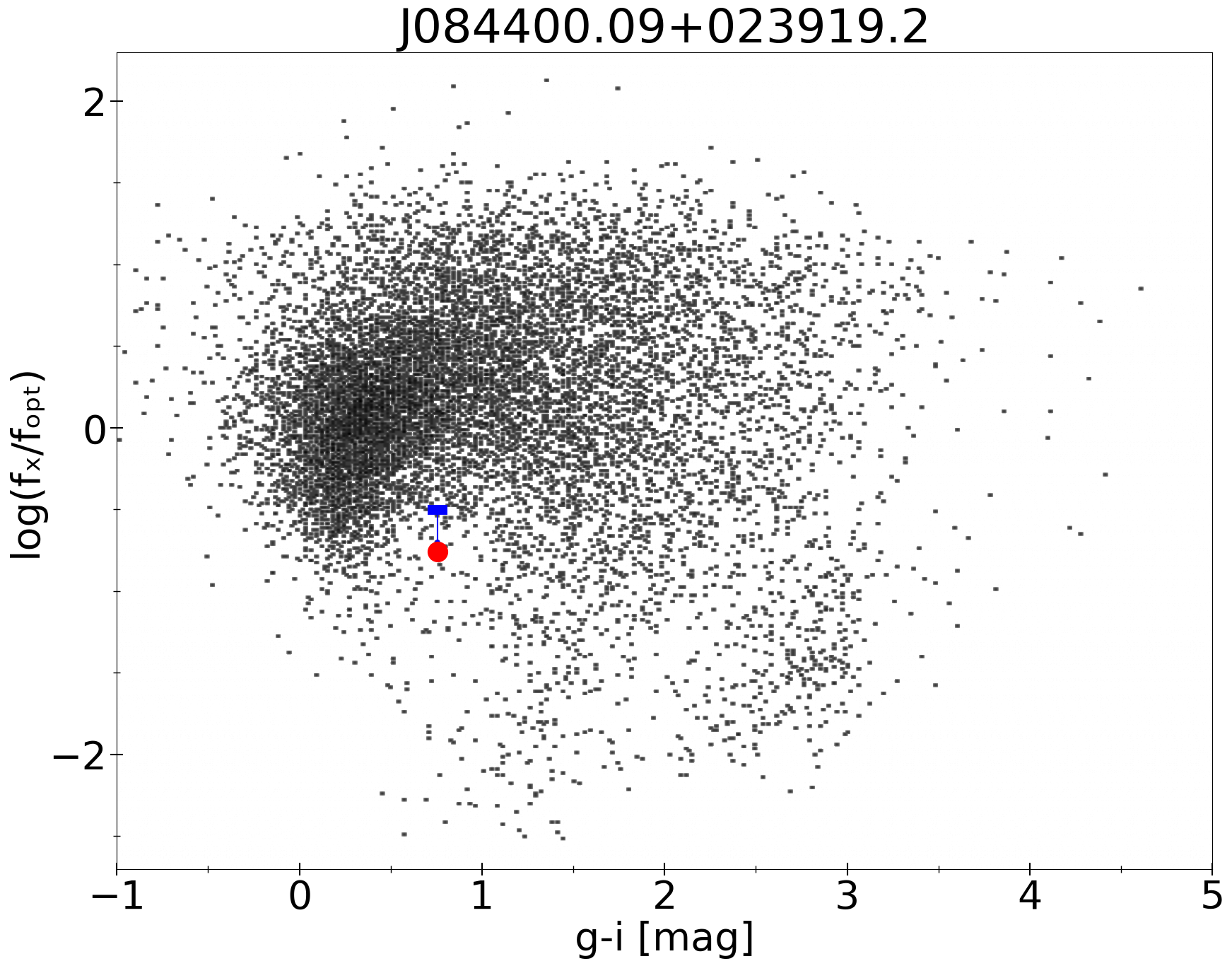}}
\end{minipage}
\vspace{1cm}

\begin{minipage}[c]{0.55\hsize}
\resizebox{\hsize}{!}{\includegraphics{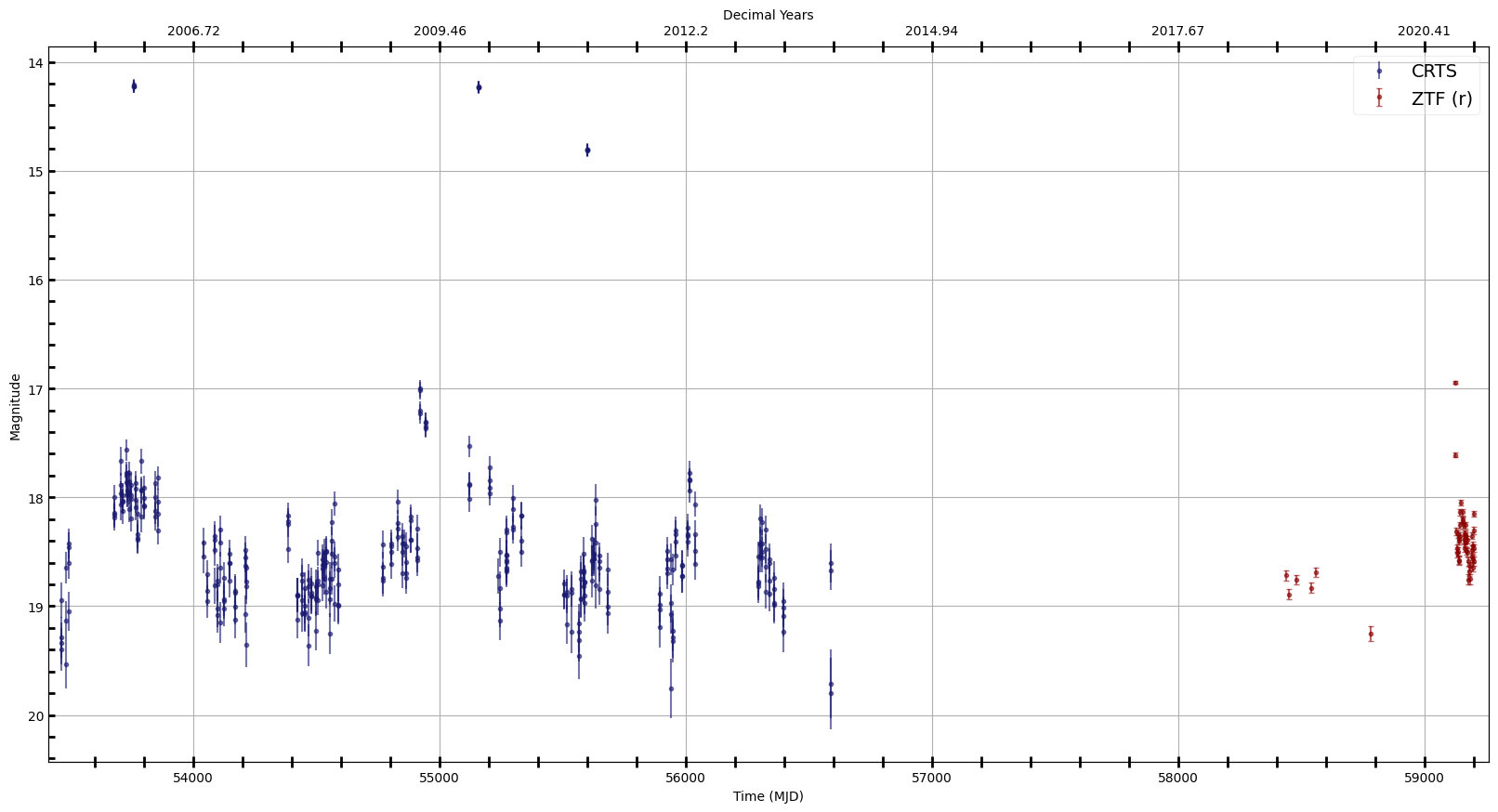}}
\end{minipage}
\hfill
\begin{minipage}[c]{0.40\hsize}
\resizebox{\hsize}{!}{\includegraphics{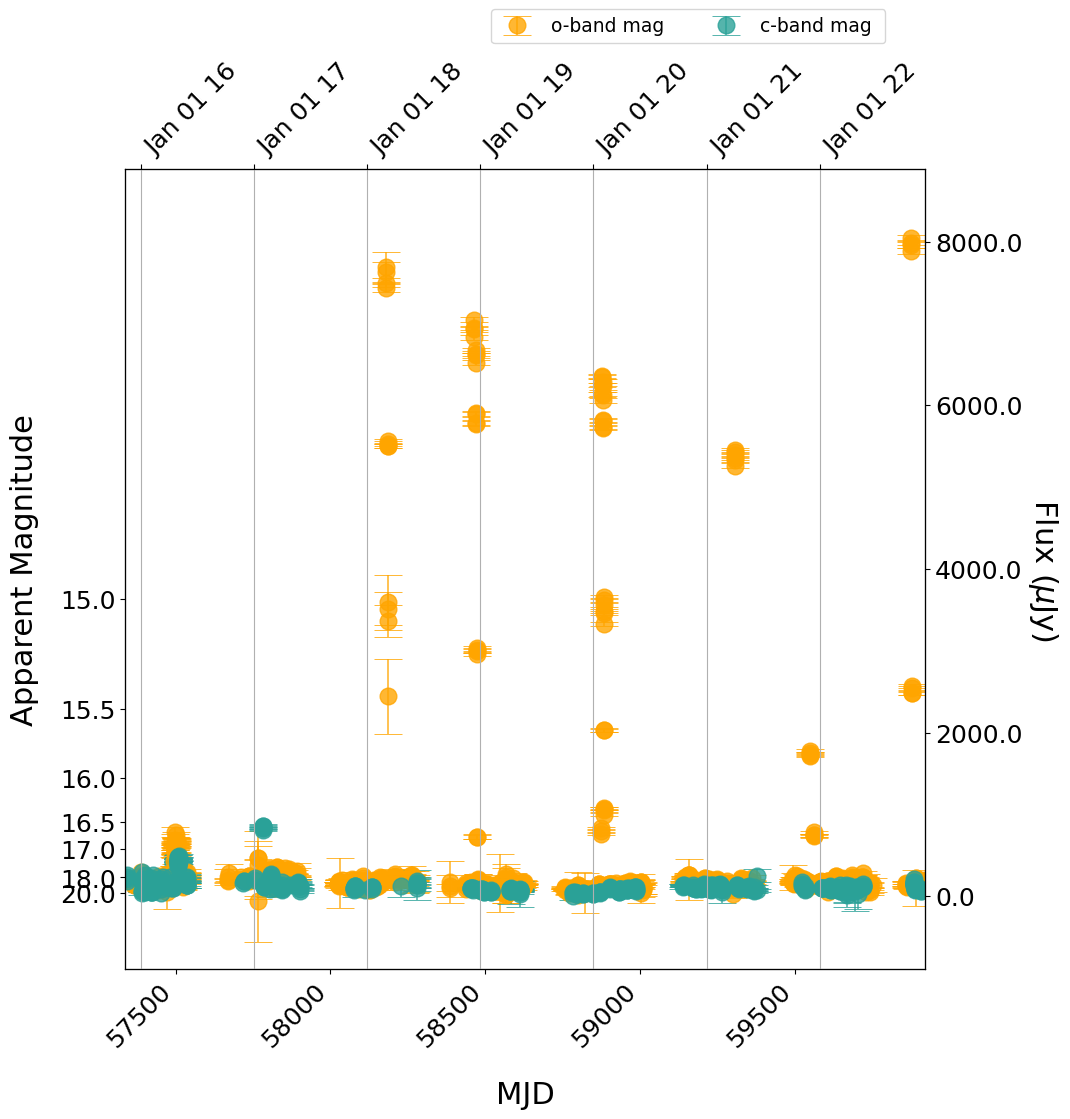}}
\end{minipage}
\caption{Profile of \ero SrcID 4654 with SDSS coordinates 08:44:00.1$-$02:39:19.3, a dwarf nova CV. Top row: finding chart and SDSS spectrum; Middle row: color-magnitude diagram (CMD, if available) and X-ray-to-optical color-color diagram (CCD); bottom row: (left panel) light curves from CRTS and ZTF ($r$-band) and from ATLAS (right panel)
\label{f:prof}}
\end{figure*}
 
($\alpha,\beta,\gamma,\delta$, in light blue), CaHK (pink), and the Na absorption doublet at 8200\,\AA\ (green).  Inserts are showing the H$\alpha$, CaHK, and the Na lines.

\item Color-magnitude diagram (CMD)\\
\gai counterparts were searched within a radius of 2 arcsec around the given SDSS coordinate. If an entry was found, the absolute $G$-band magnitude was determined using the distance $r_{\rm geo}$ determined by \cite{bailer-jones+21}. To better understand the location of the object in the CMD, a local sample of objects from \gai DR3 was retrieved with parallax/parallax\_error $>$ 100 and plotted as background.

\item X-ray optical color-color diagram (CCD)\\
The logarithm of the X-ray to optical flux ratio was computed as $\log{f_{\rm X}}/f_{\rm opt} = \log{f_{\rm X}} + \mbox{mag} / 2.5 + 5.49$. The X-ray flux in the detection band ($0.2-2.3$\,keV)  and the SDSS $g$-band flux were used. Both, the measured flux and/or an upper limit X-ray flux were derived and shown graphically. The optical color used was $g-i$. The values for the binaries studied here are put in perspective by plotting as background sources  all photometrically selected eFEDS counterparts from \cite{salvato+22}. Many background sources have $\log{f_{\rm X}}/f_{\rm opt}$ of order 0, hence similar fluxes at optical and X-ray wavelengths. These are typically AGN. The stellar branch at a flux ratio $\log{f_{\rm X}}/f_{\rm opt}$ below $-2$ is not well populated in eFEDS due to the relatively small survey area \citep[see, e.g.,][for a more comprehensive version of such a CCD including stars with low X-ray luminosity]{schwope+22b}. Only the cloud  at 
$g-i \simeq 2.8, \log{f_{\rm X}}/f_{\rm opt} \simeq -1.5$ is clearly dominated by coronal emitters at the bottom of the main sequence.

\begin{table*}
\caption{WDMS objects found in the eFEDS area. Listed are the magnitude, an optical color, the measured parallax from \gai DR3, H$\alpha$ emission line parameters, the spectral type or classification of the object and a reference. Keys to references are given below the table.
\label{t:other}
}
\begin{tabular}{|l|r|r|r@{$\pm$}l|rrr|l|c|}
\hline
  \multicolumn{1}{|c|}{SDSS J} &
  \multicolumn{1}{c|}{$g$} &
  \multicolumn{1}{c|}{$g-r$} &
  \multicolumn{2}{c|}{$\pi$} &
  \multicolumn{3}{|c|}{H$\alpha$} &
  \multicolumn{1}{l|}{Type}&
  \multicolumn{1}{|c|}{Ref}
  \\
    &  (mag)& (mag) & \multicolumn{2}{c|}{(mas)} & FW (\AA) & $F$ & EW(\AA) & & \\ 
\hline
08:32:22.6$+$03:31:02.2 &  19.88 & 0.62  & 1.886 & 0.205 & 4.4 & 39.7 & 6.2 & DA/MS & 1 \\
08:43:23.6$+$01:08:05.1 &  19.59 & -0.33 & 1.575 & 0.334 & \multicolumn{3}{|c|}{---}& DB/MS & 2,3\\
08:47:50.7$-$00:23:44.1 &  20.05 & 0.55  & 1.737 & 0.298 & 3.4 & 15.0 & 2.7 & DA/MS & 1\\
09:04:12.8$+$03:12:34.5 &  19.07 & 0.27  & 1.265 & 0.671 & 4.3 & 35.2 & 3.3 & preCVcand & 1\\
09:08:24.1$-$01:25:21.8 &  17.37 & -0.26 & 1.323 & 0.102 & 5.6 & 238  & 10.0 & preCV & 1,4 \\ 
09:36:57.1$+$02:53:42.5 &  19.84 & 0.28  & 1.245 & 0.312 & 5.5 & 14.9 &  3.3 & DA/MS& 1\\
\hline\end{tabular}

References: 
(1) this work,
(2) \cite{gentile_fusillo+19}
(3) \cite{culpan+22}
(4) \cite{geier+17}

\end{table*}

\item Optical light curves from CRTS, ZTF, and ATLAS\\
Photometric data were retrieved from the CRTS, ZTF, and ATLAS photometric surveys \citep[Catalina Real-Time Transient Survey, Zwicky Transient Facility, Asteroid Terrestrial-impact Last Alert System]{drake+09, masci+19, tonry+18}. By visually inspecting the light curves the nature of an object as a dwarf nova or an anti-dwarf nova, a VY Scl object, can easily be recognized if large-amplitude (several magnitudes) outbursts or long-lasting (several months) flux depressions are recorded.
\end{itemize}

In addition to this graphical profile, further analysis steps were undertaken per object for further sub-classification and parameter determination

\begin{itemize}
\item{Timing analysis}
We investigated periodicity in all objects by performing a Lomb-Scargle periodicity analysis on available r and g band light-curves from the Zwicky Transient Facility Data Release 19 (ZTF DR19). ZTF is a photometric survey that uses a wide 47 $\textrm{deg}^2$ field-of-view camera mounted on the Samuel Oschin 48-inch telescope at Palomar Observatory with $g$, $r$, and $i$ filters \citep{bellm+19, graham+19, dekany+20, masci+19}. Since entering Phase II, ZTF is at a 2-day cadence, and the median delivered image quality is 2.0" at FWHM.
 
We used the \texttt{gatspy} package to investigate whether sources showed periodicity in ZTF or not. To classify a source as periodic using ZTF data, we required a light-curve to show 1) quantitatively significant periodicity and 2) to pass a visual inspection. We define the significance of periodicity as the maximum Lomb-Scargle power subtracted by the median power, all divided by the median absolute deviation. We define ``significant periodicity" as light-curves where the significance is in the $86^\textrm{th}$ quantile. Furthermore, we excluded periodicities that fell on the sidereal day or harmonics of it. 

\begin{figure}
\resizebox{\hsize}{!}{\includegraphics{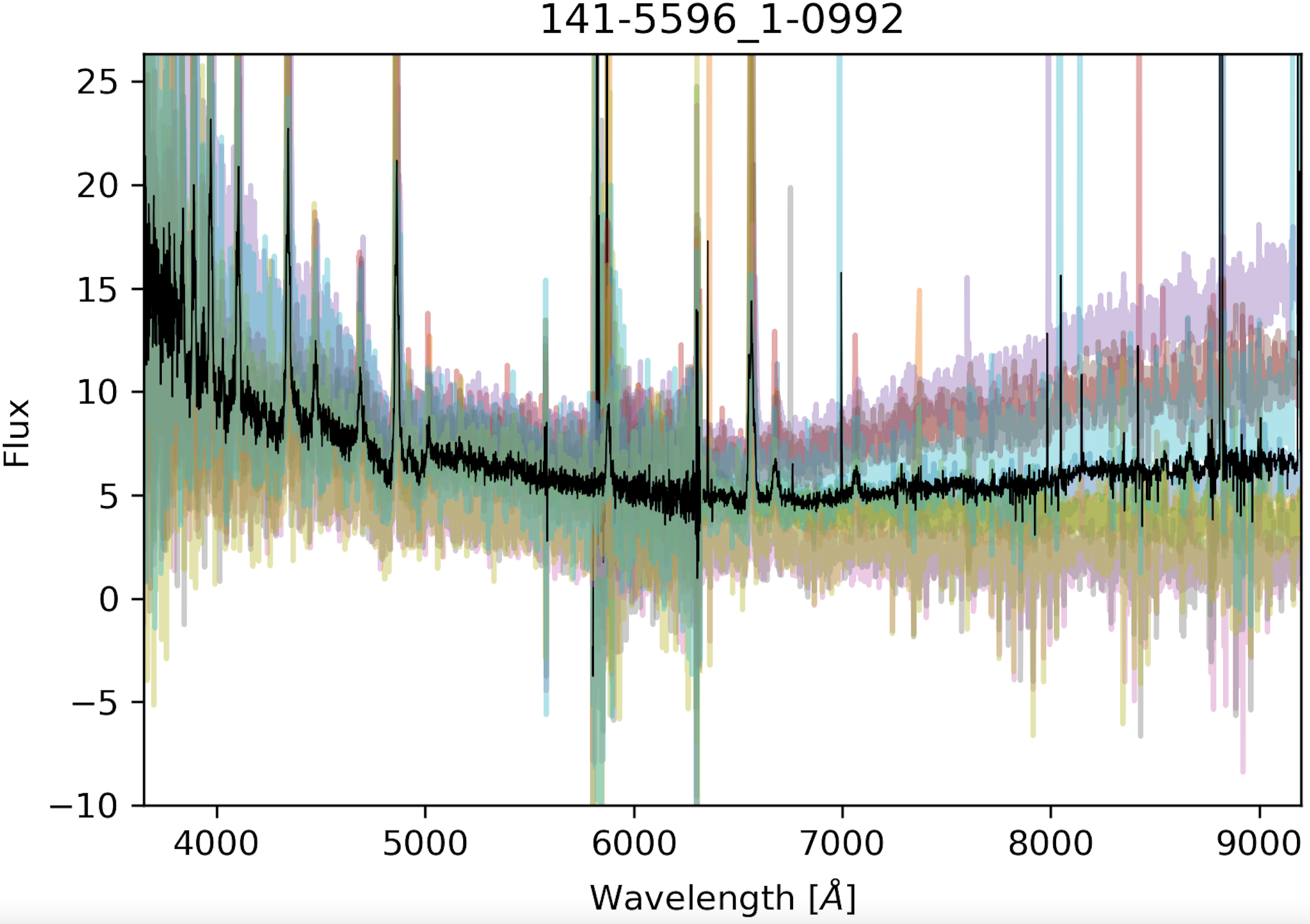}}
\resizebox{\hsize}{!}{\includegraphics{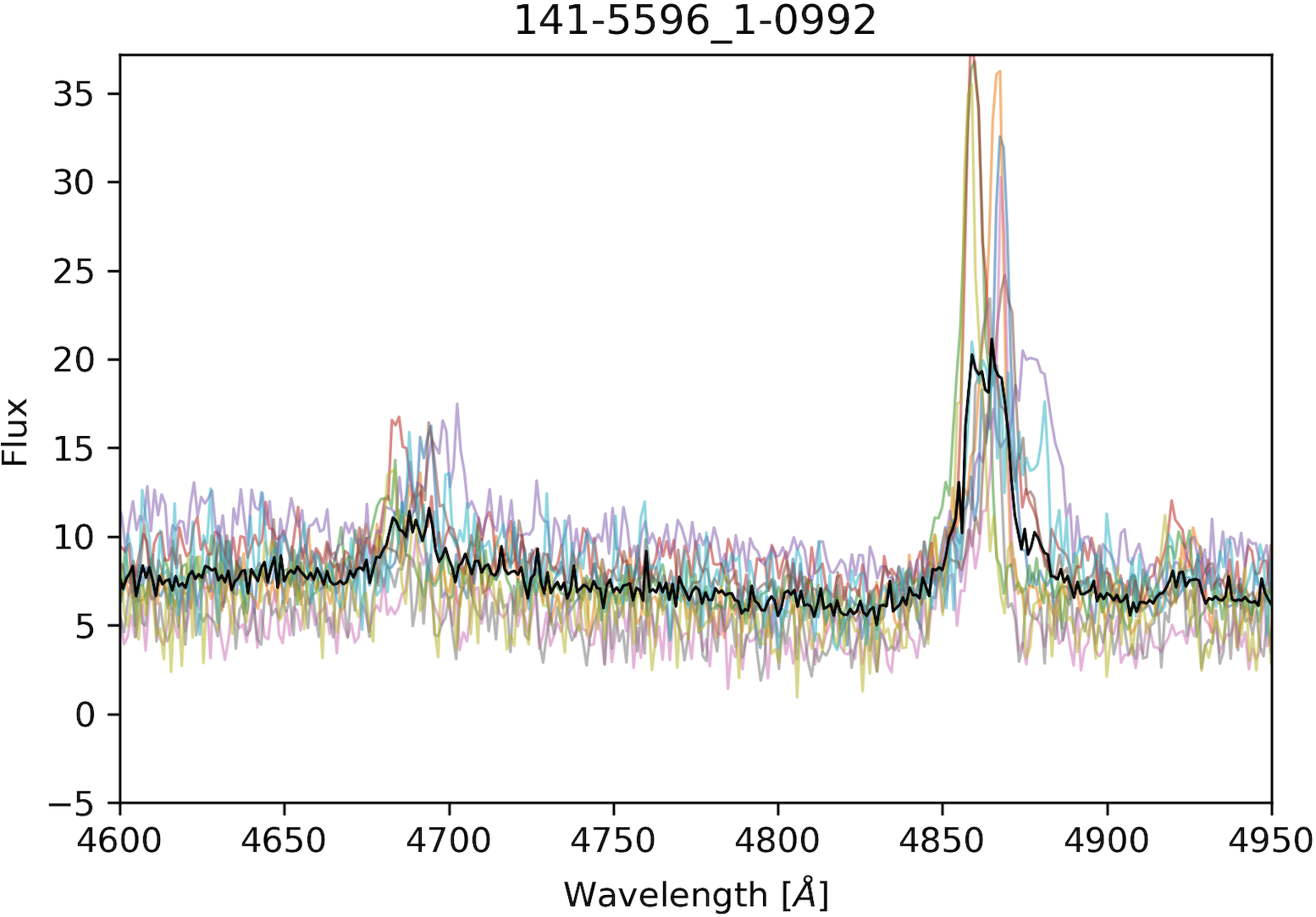}}
\caption{Presentation of the 10 sub-spectra of SDSS J092614.31$+$010557.3, a polar identified previously by \cite{rodriguez+23}. The upper panel shows the full spectral range, while the lower panel shows a zoom into the region with H$\beta$ and \het4686. Flux units are $10{-17}$\,\fcgs.
\label{f:subs}}
\end{figure}

\item Sub-spectral analysis \\
A mean SDSS spectrum is composed of a minimum of three, sometimes many more individual spectra with exposure times of 15 min each. The multiple 15\,min spectra for any target may have been accumulated over the space of several nights or weeks (and even from more than one plate). Although of lower signal-to-noise compared to the mean spectra and although sometimes sprinkled with cosmic rays these sub-spectra carry information about spectral line variability or photometric variability that happen on short time scales \citep[see, e.g.,][for a derivation of a binary orbital period based on just 7 SDSS sub-spectra]{schwope+09}. The individual spectra were initially visually inspected, if variability was found, a more thorough inspection was performed. The Na absorption line doublet 8183/8194 of all WDMS systems was fitted with double Gaussians to search for spectral variability indicating a close binary. Graphical products were generated for all objects for visual inspection but not systematically included in an object's graphical profile. 

As an example we shown in Fig.~\ref{f:subs} the 10 sub-spectra obtained for the polar CV J0926+0105 \citep{rodriguez+23}. The spectrum covering the full wavelength range in the upper panel displays strong variability towards short and long wavelengths. The infrared variability, red-ward of $\sim$6500\,\AA, is explained due to cyclotron beaming while variability in the near UV appears likely due to fore-shortening of an accretion-heated spot on the WD plus, perhaps, variable stream emission. The zoom into the blue spectral range shown in the lower panel covering the region around H$\beta$ and \het4686 shows pronounced line variability. The line profile, its radial velocity and its strength display pronounced changes typically found in polars.

\begin{figure}
\resizebox{\hsize}{!}{\includegraphics{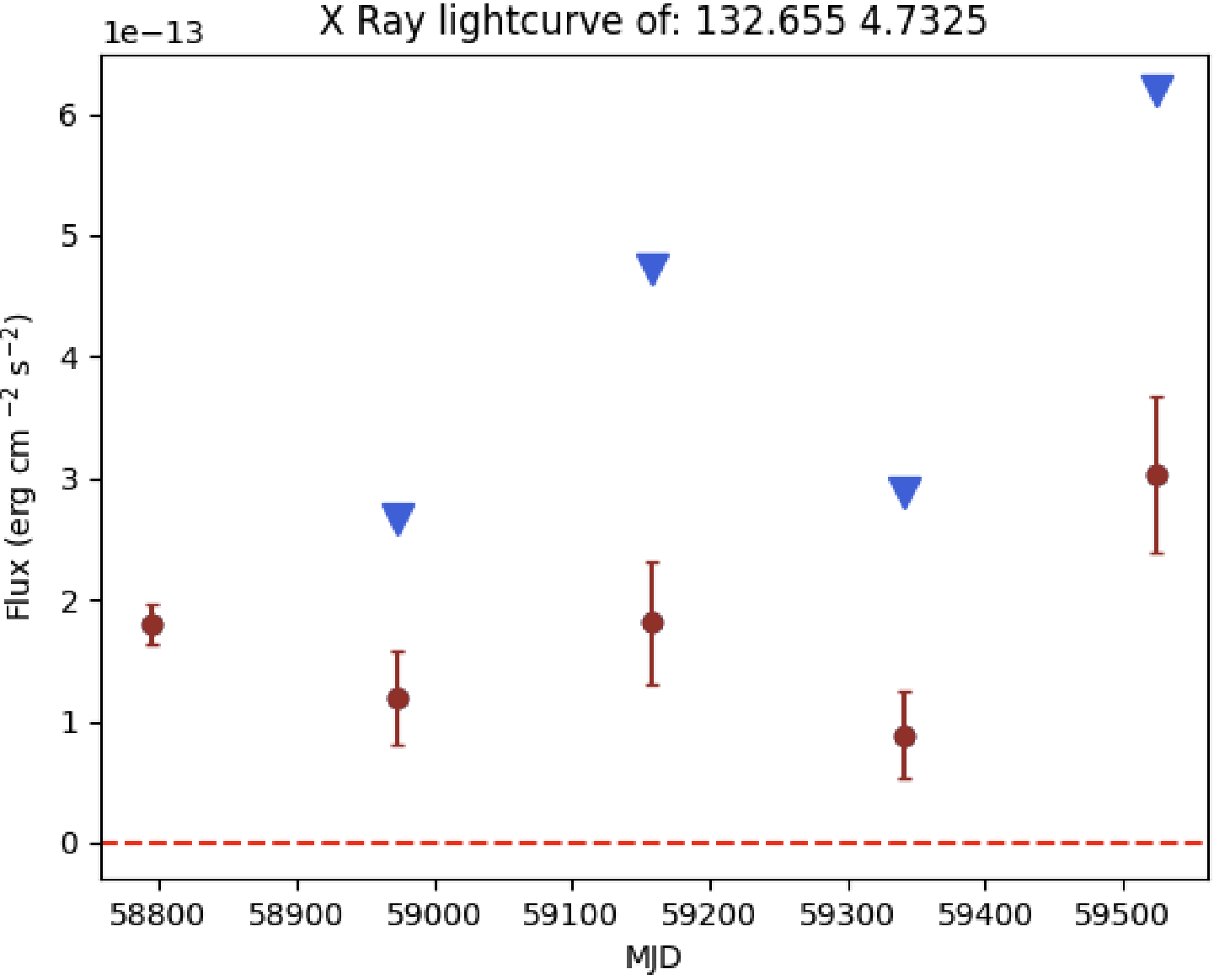}}
\caption{\ero X-ray light curve ($0.2-2.3$ keV) of the polar J0850$+$0443 combining eFEDS (first data point) with eRASS1 to eRASS4. Blue arrows indicate the upper limit fluxes at the given position \citep[see][for details of the \ero upper limit concept]{tubin+24}
\label{f:xlc}}
\end{figure}

\item X-ray variability \\
The X-ray detection in the eFEDS survey provided the starting point of our investigation. All our objects are (possible) close binaries and hence may undergo sometimes dramatic brightness changes through variability of their accretion rates. After the CalPV phase, \srgero completed four X-ray all-sky surveys called eRASS1 to eRASS4. The catalogs are available to the \ero collaboration. We thus generated low-cadence X-ray light curves which compare the eFEDS brightness with those of eRASS1 to  eRASS4. Such light curves with 5 data points were visually inspected. The type of variability could also give a hint to the underlying nature of the source. 
An example for the light curve of the polar SDSS J085037.20$+$044357.0 with detections in eFEDS and in the consecutive four eRASS scans is shown in Fig.~\ref{f:xlc}.

\item{Emission line parameters}
For all aCWDBs listed in Tab.~\ref{t:cvs} we measured parameters of the main emission lines, H$\alpha$, H$\beta$, \heo4471 and \het4686 for the CVs and \heo4471 and \het4686 for the AM CVn objects. The full width at half maximum (FWHM), the line flux $F$, and the equivalent width (EW) for the Balmer lines in the mean co-added were determined by Gaussian fits superposed on a local continuum. The latter was approximated by a 1st- or 2nd-order polynomial to adjacent wavelength ranges. For the weaker helium lines, a local continuum was determined and the line flux measured by integrating the continuum-subtracted spectrum over a suitably chosen wavelength interval. The results are listed in Tab.~\ref{t:linpar}, contrary to practice in stellar astrophysics, a positive value for the EW indicates an emission line.

Line parameters for the H$\alpha$ emission line were determined in the same way also for the WDMS objects and listed in Tab.~\ref{t:other}. In addition, we searched for radial velocity shifts of the Na absorption lines at 8183/8192 \AA\ among the sub-spectra. The Na-lines were fitted with a double Gaussian in absorption with fixed separation and assuming the same width of both lines.
\end{itemize}

\begin{table*}[]
\caption{Line parameters of new aCWDBs in the eFEDS area. For H$\alpha$ the FWHM of a single Gaussian fit to the line profile, the Flux $F$ and the equivalent width $EW$ are given, for the other lines (H$\beta$, He{\sc I}4471, He{\sc II}4686) only $F$ and EW are given. Fluxes are given in units of $10^{-17}$\,erg\,cm$^{-2}$\,s$^{-1}$, the FWHM and EW in \AA. Some objects were observed more than once in the SDSS project, older spectra (before the eFEDs survey) are labeled (S) in the first column. The three objects at the bottom of the table are the AM CVn objects that do not show Balmer emission lines. The typical uncertainties of the derived quantities are $5-10$\%.
\label{t:linpar}
}
\begin{tabular}{|l|rrr|rr|rr|rr|}
\hline
  \multicolumn{1}{|c|}{ID} &
  \multicolumn{3}{c|}{H$\alpha$} &
  \multicolumn{2}{c|}{H$\beta$} &
  \multicolumn{2}{c|}{He{\sc I}4471} &
  \multicolumn{2}{c|}{He{\sc II}4686} \\
  & FW & $F$ & EW & $F$ & EW & $F$ & EW & $F$ & EW \\
\hline
  J0840$+$0005 & 18.0 & 260 & 197 & 204 & 110.7 &  36 & 15.8 & --- & ---\\
  J0843$-$0148 & 22.6 & 635  & 28 & 391 &  19.7 &  69 &  4.2 & --- & ---\\
  J0844$+$0239 & 12.5 & 948 &  68 & 746 &  68.4 & 145 & 14.1 & --- & ---\\
  J0845$+$0339 & 32.5 & +82 & 105 &  70 &  61.0 &   9 &  3.6 & --- & ---\\
  J0846$+$0218 & 12.8 & 230 &  14 &  90 &   6.0 &  64 &  5.6 & 42 & 3.2\\
  J0847$+$0145 & 40.4 & 129 & 151 & 118 & 110.8 &  14 & 10.8 & --- & ---\\
  J0851$+$0308 & 34.5 & 953 & 100 & 953 &  65.8 & 164 &  9.3 & --- & ---\\
  J0853$+$0204 & 47.2 &  91 & 142 &  65 &  82.4 & --- & ---& --- & ---\\
  J0855$-$0154 & 43.2 &  78 &  80 & 133 & 100.6 & --- & ---& --- & ---\\
  J0902$-$0142 & 46.1 & 102 &  52 &  82 &  36.2 & --- & ---& --- & ---\\
  J0904$+$0355 & 44.6 & 176 &  48 & 109 &  22.0 & --- & ---& --- & --- \\
  J0912$-$0007 & 13.3 & 133 &  67 &  99 &  72.1 &  50 & 38.9 & --- & ---\\
  J0914$+$0137 & 19.4 & 567 &  25 & 446 &  20.0 & 161 &  8.0 & 55 & 2.6\\
  J0914$+$0137(S)& 15.7 & 620 & 26 & 460 &  19.0& 115 &  5.1 & --- & --- \\
  J0918$+$0436 & 37.8 &  85 & 196 &  43 & 108.6 & --- & ---& --- & ---\\
  J0920$+$0042 & 26.0 &2100 & 105 &1328 &  51.8 & 312 & 13.8 &226 & 8.6\\
  J0926$+$0105 & 17.5 & 206  & 51  &195 &  32.0 &  46 &  5.6 & 41 & 5.4\\
  J0926$+$0345 & 37.1 & 189  & 59 & 198 &  46.0 &  26 &  4.8& --- & ---\\
  J0926$+$0345(S) & 36.3 & 184 & 63 & 182 & 47.6& --- & ---& --- & ---\\
  J0929$+$0053 & 17.5 &  45 & 69 &  32 &   50.2& --- & ---& --- & ---\\
  J0929$+$0401 &  4.4 &  23 &  2 &   9 &    1.8& --- & ---& --- & ---\\
  J0932$+$0343 & 28.6 & 662 & 42 & 195 &    9.0& --- & ---& --- & ---\\
  J0932$+$0109 & 15.5 & 230 & 45 & 266 &   44.2 &  72 & 11.9 & 46 & 7.5\\
  J0935$+$0429 & 11.6 & 195 & 33 &  82 &   25.7 &   9 & 2.7 & --- & --- \\ 
\hline 
  J0844$-$0128 & --- & --- & ---& --- & --- & 28 &  8.9 &  3.4  & 1.1\\
  J0847$+$0119 & --- & --- & ---& --- & --- & 34 & 27.1 &  3.7  & 3.3\\
  J0903$-$0133 & --- & --- & ---& --- & --- & 97 & 12.9 & 16.7  & 2.5\\
\hline
\end{tabular}
\end{table*}

\section{Results\label{s:res}}
This section is divided in three parts. In the first two subsections we describe the main properties 
of the individual objects that led to the classification as documented in Tables \ref{t:cvs} and \ref{t:other}. When describing the main and unique properties of each source we take into account previous knowledge from e.g.~the CDS. We address the objects by using abbreviated coordinates  (JHHMMsDDMM) and firstly discuss aCWDBs followed by WDMS objects. The main aim of this exercise is to determine, if possible, the class and the sub-class for each system. The scheme we are following is similar to that of \cite{inight+23a}, who order the non-magnetic CVs according to increasing mass accretion rate and the magnetic systems according to increasing magnetic field strength. The non-magnetic low mass-transfer objects are the dwarf novae (DN, which show dwarf nova outbursts) and the nova-like (NL) high $\dot{M}$-systems. The magnetic systems are the intermediate polars (IPs), the polars, and the pre-polars. The latter are accreting from a stellar wind instead of Roche-lobe overflow (RLOF). All RLOF-magnetic CVs are also often referred to as NLs. If the classification is uncertain or remains ambiguous, we add the letter $c$ (candidate) to the tentatively assigned class. A somewhat separate class are the AM CVn systems, which are double-degenerate objects well below the CV orbital period minimum. In the following subsections we discuss the whole of the sample and give an outlook for the identification of more aCWDBs systems in the SDSS-V program.

\subsection{Notes on individual objects: aCWDBs}

\subsubsection{J0840$+$0005}
The object was mentioned as a CB candidate from the CRTS in \cite{drake+14}, a spectroscopic identification was provided by \cite{breedt+14}, who classify the object as DN. The classification rests on one recorded outburst in the CRTS with peak magnitude at $\sim$16 (outburst amplitude $>$4 mag). While their light curve plot looks convincing, it cannot be reproduced from public CRTS data\footnote{\url{http://nunuku.caltech.edu/cgi-bin/getcssconedb\_release\_img.cgi}}. One thus might regard the given identification as tentative. 

The spectrum displays a pronounced H Balmer series, has weaker HeI, and no HeII lines. The lines have symmetric profiles. The continuum appears to be flat and rather blue, there is a strong Balmer jump in emission. No signatures of the donor star are recognized. The sub-spectra do not show variability, neither in the line shapes nor in the continuum brightness.

The object is beyond the \gai limit, hence no CMD was generated. The CCD shows an extremely blue color (in accord with the spectral shape) and a high ${f_{\rm X}}/f_{\rm opt}$. ZTF and ATLAS show some apparently irregular variability with 0.5 mag amplitude but no outburst.

There is some weak evidence for X-ray variability. The eRASS1,3,4 fluxes are compatible with those of the eFEDS survey, only the eRASS2 flux was enhanced at $(1.5\pm 0.5) \times 10^{-13}$\,\fergs, a 2$\sigma$ result.

The object can certainly be classified as a non-magnetic CV. If the outburst reported by \cite{breedt+14} could be confirmed, it is a DN with long recurrence time. This would imply a short orbital period and a very low mass donor, in line with its non-detection in the SDSS spectrum. Finding an orbital period could be difficult, because the inclination seems to be rather low from the line profile shape.

\begin{figure}[]
\resizebox{\hsize}{!}{\includegraphics{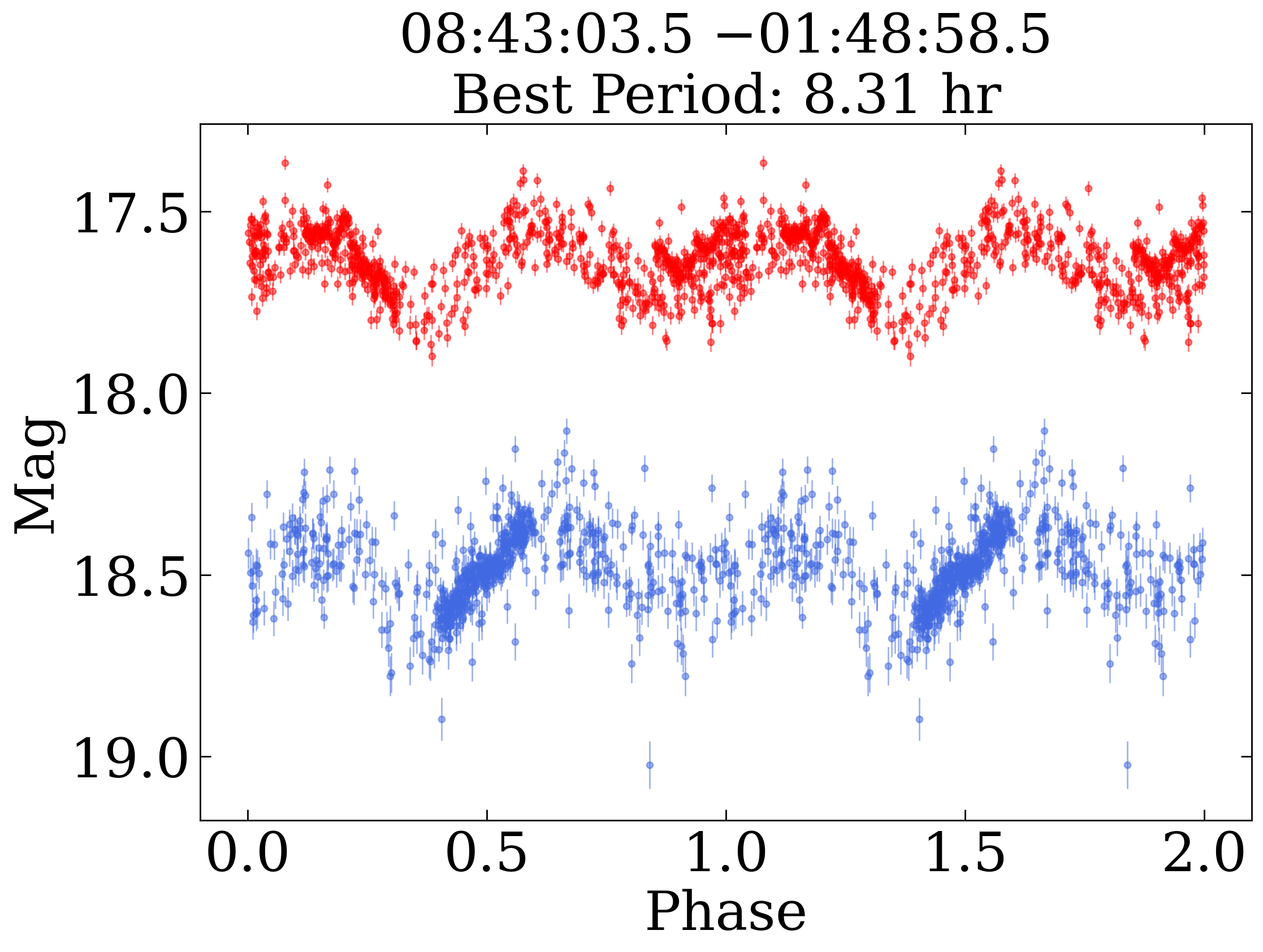}}
\caption{ZTF $g$ (blue symbols) and $r$ (red symbols) band light curves of J0843$-$0148 folded on the likely period of 8.31 hours.}
\label{f:0843ztf}
\end{figure}

\subsubsection{J0843$-$0148}
The object is new to the literature. It has a $G=17.9$ mag \gai counterpart at a distance $r_{\rm geo} = 2050^{+550}_{-350}$\,pc \citep{bailer-jones+21}. The SDSS spectrum has a smoothly variable continuum with a very broad spectral maximum at 6500\,\AA, Balmer emission lines, and Na5890 and Mg5175 absorption lines from the donor star. \heo\ lines are present but weak, \het\ lines are absent. The sub-spectra do not show variability. Using template spectra from \cite{kesseli+17}, the spectral type of the donor is determined as $\sim$K4. For an RL-filling donor the implied orbital period is longer than 6 hours \citep{knigge+11}. A period search on the ZTF data revealed a best period at 4.154 hours, at odds with the spectral type with an RL-filling donor. We thus choose as likely orbital period twice the value at 8.31 hours. The phase-folded light curve from the ZTF in $g$ and $r$ is shown in Fig.~\ref{f:0843ztf} and displays a double-humped structure with unequal minima which is interpreted as ellipsoidal modulation from the secondary. The full amplitude of the optical variation is about 0.3 mag.

At the nominal distance, the X-ray luminosity is $L_{\rm X} = 6\times10^{31}$\,\lx, the flux ratio \fxo\ is rather low. X-ray variability is not significant. The object is classified as a non-magnetic NL. 

\subsubsection{J0844$+$0239}
The object is a DN, originally found by \cite{szkody+03} in the second year data from the SDSS (52224-564-197). The new data obtained with the BOSS spectrograph (MJD-PLATE-FIBRE 58930-12527-972) confirm the spectral shape although at reduced flux level (about 65\% of the discovery spectrum). No sub-spectral variability was found. The \fxo-ratio is low for CV standards, a typical feature of non-magnetic CVs. Several DN outbursts are detected with outburst amplitudes $\Delta m > 4$ in CRTS and ATLAS. At the nominal distance of $1470^{+670}_{-350}$\,pc, the X-ray luminosity is $9.7\times 10^{30}$\,\lx. The X-ray brightness seems to be slightly enhanced during all following eRASS compared to eFEDS but remains compatible at the $1-2\sigma$ level.

\subsubsection{J0844$-$0128}
The object was selected as a CB candidate by \cite{drake+14} and identified as an AM CVn type object by \citep{breedt+14} through GMOS spectroscopy. The initial selection was based on several outbursts with amplitude 2 mag in CRTS data. ZTF and ATLAS have meanwhile recorded 4 mag outbursts on a quiescence level of about $r \simeq 21.3$. The BOSS spectrum confirms the AM CVn identification of \cite{breedt+14}. It has a very blue continuum, no hydrogen lines, but pronounced \heo\ and \het\ emission lines. No significant sub-spectral variability was found and no significant X-ray variability. 

While the object was known to belong to the AM CVn class previously, the current work reports its first X-ray detection. For the nominal distance of $2150^{+1100}_{-700}$\,pc, one derives $L_{\rm X} \simeq 1 \times 10^{31}$\,\lx. It thus seems to belong to the more X-ray luminous fraction of the known X-ray emitting AM CVns, which mostly have a factor $2-3$ lower luminosity. Following \cite{levitan+15}, the outburst amplitude of about 4 mag and the outburst recurrence time of about 225 days from ZTF suggest an orbital period of about $32$ min.

\subsubsection{J0845$+$0339}
V498 Hya was clasified as an SU UMa type DN, discovered by \cite{kato+09}, that was recently studied by \cite{vogt+21} to characterize the superoutburst cycle length. \cite{breedt+14} derive an orbital period of 85.18 min that is implied from the measured superhump period. It was re-observed in SDSS-V and while belonging to the general CV carton \cite[see][for the various ways in which CVs were selected for the plate program in SDSS-V]{inight+23b}, it was not X-ray selected. The SDSS/BOSS spectrum is similar to the GMOS spectrum by \cite{breedt+14}, but has slightly lower overall flux and weaker hydrogen emission lines. 

Its distance is rather uncertain; \cite{bailer-jones+21} give a range between 550 pc and 4100 pc. At the nominal value of 1635 pc, the object has at $G_{\rm abs} \simeq 10$, right in the middle between the WD sequence and the ZAMS. The object was also discussed in the recent paper by \cite{inight+23b} who re-classify it as a WZ Sge-type CV by using the earlier distance estimate from \cite{bailer-jones+18} or the inverted parallax at 302 pc. Both distance estimates put the object on the WD sequence and thus suggest a WD-dominated or WZ Sge-type CV. This is adopted here.

The object is not in the officially released eFEDS X-ray catalog. The upper-limit tool \citep{tubin+24} reveals an upper limit flux of $2 \times 10^{-14}$\,\fergs at the given position. Taking the upper limit at face value, one gets $\log L_{\rm X} (\mbox{erg s}^{-1}) = 29.3$ at the revised short distance 
($\log L_{\rm X} (\mbox{erg s}^{-1}) = 30.8$ at the old long distance).

\subsubsection{J0846$+$0218}
A reasonably bright, $g\simeq 18$, yet un-noticed DN. The SDSS spectrum has great similarity to that of 0843-0148, it has a smoothly variable continuum with a very broad spectral maximum at $\sim$6000\,\AA, Balmer emission lines, and Na5890 and Mg5175 absorption lines from the donor star. \heo\ lines are present but weak, and we might see some weak \het4686 emission. The sub-spectra do not show variability. Using templates from \cite{kesseli+17}, the spectral type of the donor is determined as K4 implying an orbital period longer than 6 hours \citep{knigge+11}.

The distance to the object is large, $D=3580$\,pc, but rather uncertain, within $2780 - 5280$\,pc. At the nominal distance the absolute magnitude is $G_{\rm abs}=5.3$, such that the object lies along the main sequence, and the X-ray luminosity is $L_{\rm X} = 3\times10^{31}$\,\lx. The flux ratio $\log$\fxo $\simeq -1$ is rather low compared to other objects studied here, which is due to its high optical brightness. There are short-term optical brightness changes with an amplitude of about 0.3 mag. X-ray variability is suggestively apparent but was not found to be significant.

\subsubsection{J0847$+$0119}
The SDSS spectrum shows a very blue continuum with \heo and \het emission lines superposed, but no apparent Balmer emission lines. The object can safely be identified as an AM CVn binary. The object is faint, has no parallax, and hence no X-ray luminosity. Given the optical faintness, $g=21.9$, the flux ratio $\log$ \fxo $=0.5$ is rather high compared to other aCWDBs studied here. The source was discovered only in eFEDS, not in the individual eRASS', and not in the stack of the first three eRASS (see Tab.~\ref{t:succ}), indicating X-ray variability by $\sim$30\% or larger. The ZTF data show variability with an amplitude of almost one magnitude, promising to search for orbital-period variations. No significant sub-spectral variability is observed.

\subsubsection{J0847$+$0145}
An optically very faint and very blue object with a very high flux ratio $\log$ \fxo $\simeq 0.8$. Superposed on the blue continuum are strong emission lines of the H-Balmer series and those of neutral helium. ZTF data display variability by about 1 mag. The object was discovered in eFEDS, not in the individual eRASS' but in eRASS:3, where it was found at 60\% of its flux in eFEDS. Since no outburst was recorded, the object is tentatively classified as NL, although from the spectral shape (optically thin gas) a DN classification seems possible.

\begin{figure}[]
\resizebox{\hsize}{!}{\includegraphics{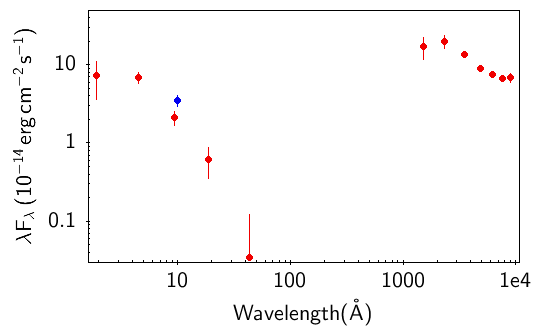}}
\caption{Spectral energy distribution of the DN J0855$-$0154 obtained from the 4XMM catalog. The blue symbol is from eFEDS}
\label{f:sed0855}
\end{figure}

\subsubsection{J0850$+$0443}
The object was selected as the counterpart of the X-ray source but the fibre remained unplugged, hence no SDSS spectrum was taken. The identification as a polar was made by \cite{rodriguez+23}, who selected the object as a likely CV from a correlation of eFEDS X-ray sources with ZTF variability. They present a comprehensive follow-up study of the new polar. The implied X-ray luminosity of $3.4 \times 10^{31} (D/1080\mathrm{pc})^2$\,\lx is typical for this class of source.

The X-ray flux is apparently variable at about 2 $\sigma$ between 1 (eRASS3) and 3 $\times 10^{-14}$\,\fergs (eRASS4) (see Fig.~\ref{f:xlc}). An X-ray spectrum was extracted from eFEDS and fitted initially with a thermal plasma model modified by some amount of cold interstellar matter using XSPEC. Such a fit left systematic residuals at the lowest energies. An additional blackbody component yielded a fit without systematic but only statistical residuals. Given the low total number of photons in the spectrum, the temperatures of the two emission models could not be determined and were thus fixed at typical values, $kT_{\rm BB}= 30$\,eV, $kT_{\rm plasma} = 10$\,keV. The implied cold interstellar column density was $N_{\rm H} = (2 \pm 1)\times 10^{21}$\,cm$^{-2}$ and the bolometric fluxes in the two emission components $F_{\rm BB} \simeq 3 \times 10^{-11}$\,\fergs ($(0.12 - 10)\times 10^{-11}$\,\fergs at 90\% confidence) and $F_{\rm plasma} \simeq 1 \times 10^{-12}$\,\fergs. The latter flux has an uncertainty of about 20\%. Hence, formally the source might display a soft excess, $F_{\rm BB} / F_{\rm plasma} > 1$ but the large uncertainties make us believe that the soft excess is mild or even insignificant. Nevertheless, this object is the first X-ray detected polar with a soft component after a long series of X-ray detected polars without soft emission at all \citep{vogel+08,ramsay+09,webb+18,schwope+20,schwope+22a,ok+23}.

\subsubsection{J0851$+$0308}
A well-studied SU UMa type DN, CT Hya, originally discovered by \cite{nogami+96}. A first SDSS spectrum obtained in November 2001 was presented by \cite{szkody+03}. The new BOSS spectrum was obtained when the source was twice as bright, but its spectral features (blue continuum, Balmer lines, \heo lines) remain unchanged. CRTS, ZTF and ATLAS recorded many DN outbursts with amplitudes up to 4\,mag. The X-ray luminosity is $4.0\times10^{30}(d/580\mathrm{pc})^2$\,\lx. 

\subsubsection{J0853$+$0204}
A very faint DN, $g=21.4$ in quiescence, with one ZTF-recorded outburst with an amplitude of $\sim$5.5 mag. Due to its optical faintness, the flux ratio $\log$\fxo $\simeq -1.8$ is unusually low. The source was detected only in eFEDS, not in the individual surveys, nor in eRASS:3. The spectrum shows the typical DN behavior, a blue continuum, H-Balmer and \heo emission lines. A spectrum of the source was taken in March 2008 in SEGUE (54529-2888-252), but the extraction of the spectrum was faulty (flux always below zero), so that the object was mis-classified as an A0 star despite weak H$\alpha$ and H$\beta$ in emission.

\subsubsection{J0855$-$0154}
The object has a typical dwarf nova spectrum in quiescence with a blue continuum, double-peaked H-Balmer emission lines from the disk and some weak He lines, including \het4686. It is located on the WD sequence in the CMD, but WD features cannot be recognized in the spectrum, likely due to its faintness. The outburst amplitude from ATLAS is $\Delta m > 3$ mag. It was discovered in eFEDS and is detected in all 4 surveys individually. It was also covered by \xmmn observations and is listed in 4XMM  \citep{webb+20,traulsen+20}. The SED from the near-IR to the X-ray spectral range is shown in Fig.~\ref{f:sed0855} with data from \xmmn, GALEX, and the SDSS. It was not detected with the OM on the \xmmn observatory. The system was about a factor 2 fainter in X-rays when observed with \xmmn compared to eFEDS. A total of $133\pm16$ photons were collected which is insufficient to obtain a robust spectral fit. We do not find sub-spectral variability within the SDSS.

\subsubsection{J0902$-$0142}
Another typical faint DN with blue continuum and double-peaked H-Balmer emission lines from the disk. Only the \heo5875 line can be recognized clearly, while all other \heo and \het lines remain undetected. Its position in the CMD is between the WD sequence and the ZAMS. Outburst magnitudes as high as 4 mag are observed in ATLAS, ZTF, and CRTS. It was discovered only in eFEDS, but remains undetected in all of the eRASS visits and in the stack of the first three, eRASS:3, indicating a dimming by at least a factor 2. We could not find sub-spectral variability among the SDSS sub-spectra.

\begin{figure}[]
\resizebox{\hsize}{!}{\includegraphics{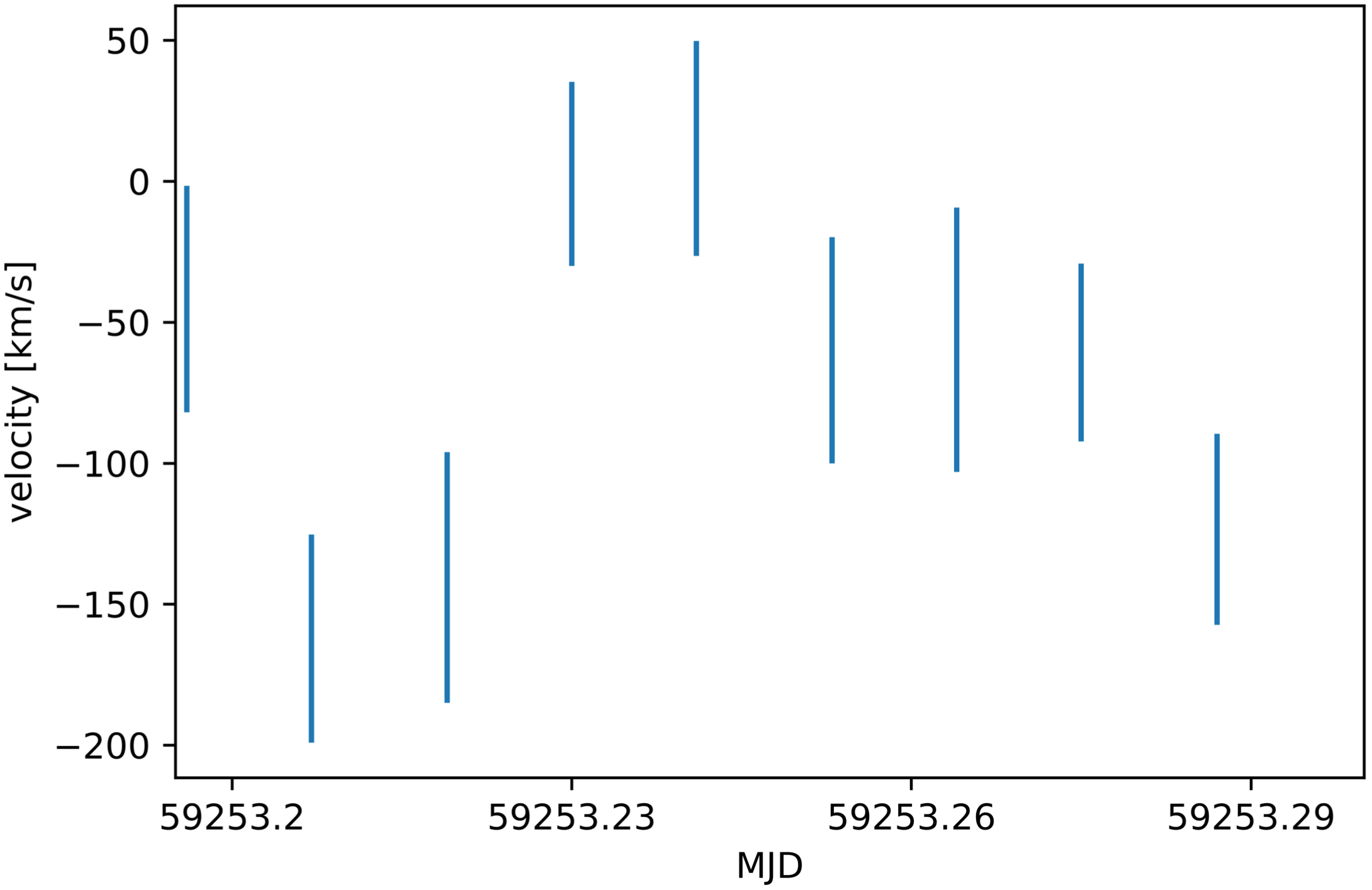}}
\caption{Radial velocities obtained for He{\sc I}5875 in the sub-spectra of the AM CVn object J0903$-$0133.}
\label{f:rvam1}
\end{figure}

\subsubsection{J0903$-$0133}
A very blue spectrum with emission lines of neutral and ionized helium, clearly an AM CVn object. It was listed as an AM CVn candidate in the extended list of \cite{carter+13} and confirmed being a DD by \cite{inight+23b}. It lies on the WD sequence and is one of the closest of its class at $D = 279$\,pc. Optical variability is observed, ranging between 19.1 and 20.1 mag. Whether the object is out-bursting or not is difficult to determine. The few optical data points at brighter magnitudes are not unique indicators of outbursts and could be outliers as well. The X-ray luminosity is quite low, $\log L_{\rm X} (\mbox{erg s}^{-1}) = 29.8$. 

The most prominent line is that of He{\sc I}5875 which was used to search for RV shifts among the sub-spectra. A template for a cross-correlation analysis was generated from the mean of the sub-spectra after continuum subtraction. All spectra were put on the same logarithmically equidistant wavelength grid. Peaks in the CCF were determined via Gaussian fits and the resulting RV curve is shown in Fig.~\ref{f:rvam1}. Some weak RV variability seems to be present perhaps with a timescale of about 1 hour but evidence remains weak. X-ray variability is apparent, but not significant.

\subsubsection{J0904$+$0355}
A white-dwarf dominated non-magnetic CV with typical double-peaked Balmer emission lines originating from an accretion disk. No obvious outbursts are found, although optical variability with an amplitude of about 1 mag is observed. The object was discovered by \cite{szkody+04} and later studied with high-speed photometry by \cite{woudt+12}, who determined a precise eclipse ephemeris. As such, this is a DN candidate. It is nearby, with a distance of $295^{+42}_{-35}$\,pc.

The object was not listed in the eFEDS X-ray catalog, but the upper limit server \citep{tubin+24} applied to the eFEDS imaging data suggests an upper limit flux of $f_{\rm X}\mbox{ (0.2-2.3)\,keV} = 2.9 \times 10^{-14}$\,\fergs. 
Taking the limit at face value this gives a luminosity below $10^{30}$\,\lx, hence it is a low-luminosity object at optical and X-ray wavelengths and thus classified as a WZ Sge-type dwarf nova.

\subsubsection{J0912$-$0007}
The SDSS spectrum and color of this faint object are mildly red. Symmetric, single peaked emission lines of the H-Balmer series and \heo are observed. \het seems to be weakly present. The object is located halfway between the ZAMS and the WD sequences in the CMD.  Optical variability of about 1 mag is observed, no outbursts were recorded. There is an X-ray detection in eFEDS, but not in any eRASS, nor in eRASS:3. No sub-spectral variability is seen. From its general spectral appearance, the object could be a low-inclination, weakly accreting polar, but the constancy of the emission lines in the sub-spectra argues against this. From its position in the CMD, compared to \cite{abril+20}, it is a DN candidate or a low-luminosity NL.

\subsubsection{J0914$+$0137}
The object is an optically bright, $G_{\rm abs} \simeq 7.6$, DN with a rather early-type donor. Its nature as a DN was reported in VSNET 19440, the period was determined spectroscopically by \cite{thorstensen+17}, and \cite{inight+23b} classify it as U Gem-type DN.
The Balmer lines are symmetric and single-peaked, likely due to a low inclination. \heo lines and weak \het lines can be recognized as well. Outbursts were recorded with CRTS, ZTF and ATLAS, with maximum outburst amplitude of about 3.5 mag. Apart from the outbursts there is additional continuum variability with an amplitude of about 0.5 mag. The object was detected in eFEDS and in eRASS:3 at a 20\% higher flux, but not in any of the eRASS' individually. 

\subsubsection{J0918$+$0436}
A faint object beyond the \gai limit that has limited time-domain coverage from the ZTF. The spectrum appears red and seems to show traces of a red M-type secondary star. The scarce data from ZTF imply variability by 1 mag. The forced-photometry light curve from ATLAS indicates constant brightness. The emission lines (prominent Balmer series and \heo5875) are broad and have flat-topped symmetric profiles implying a disk origin. The object is tentatively classified as a DN candidate.

\begin{figure}
\resizebox{\hsize}{!}{\includegraphics[angle=-90]{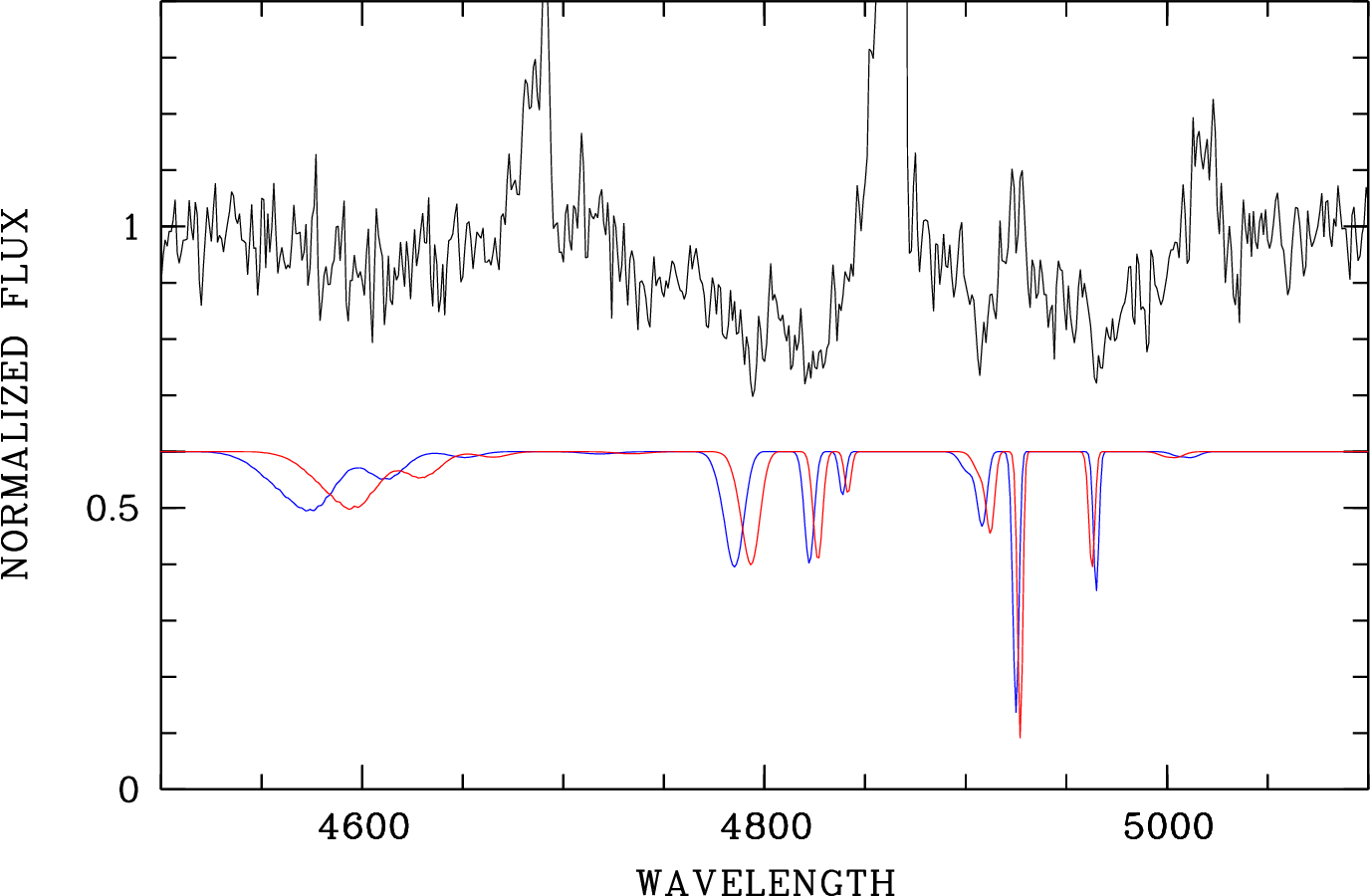}}
\caption{Normalized spectrum of the polar J0926$+$0105 around H$\beta$. The colored lines below the spectrum show Gaussian-weighted dipole matrix elements of all the H-Balmer lines in fields of $B=15.5$\,MG (red) and 16.5\,MG (blue), respectively. The FWHM of the Gaussian was 0.5 MG and the spectra smoothed to a FWHM resolution of 2\,\AA. Wavelengths are given in \AA.}
\label{f:zeem}
\end{figure}

\subsubsection{J0920$+$0042}
V524 Hya was described in \cite{szkody+03} as a high-inclination system showing eclipses, whose sub-class could not firmly be determined. The derived orbital period is 3.6 hr. ZTF, CRTS, and ATLAS data reveal strong variability, consistent with the interpretation that these data were obtained during eclipses. No outbursts were recorded. The emission lines of the H-Balmer series, of \heo and \het (4686 similarly bright as 4471) are skewed and single-peaked. No significant RV variations are seen in the sub-spectra. 
The object is tentatively classified as non-magnetic NL.

\subsubsection{J0926$+$0105}
The object was classified as a polar by \cite{rodriguez+23}, who selected it for follow-up work after cross-matching the eFEDS X-ray catalog with photometry from the ZTF. Time-resolved photometry and spectroscopy revealed the main binary parameters. 

The X-ray luminosity is at the low end for polars and the flux ratio $\log$ \fxo\ is also about half a dex lower compared to other polars (Schwope et al., in preparation). Because polars may show strong orbital variability and strong long-term variability due to accretion-rate changes, the non-simultaneity of X-ray and optical catalog data can be misleading. The mean X-ray flux shows a decrease from the eFEDS epoch ($27.9 \times 10^{-14}$\,\fergs) to eRASS4 ($6.3 \times 10^{-14}$\,\fergs). 

There is pronounced sub-spectral variability in the continuum and the emission lines. The object has a red cyclotron continuum, reminiscent to other low-field polars as presented and discussed by \cite{schwope+97}. In such low-field accretion plasmas, the individual high harmonics which fall into the optical window are smeared to a quasi-continuum as observed also in 0926$+$0105. If the magnetic field were as high as 36--42\,MG as suggested by \cite{rodriguez+23}, one would not see a cyclotron quasi-continuum peaking at around 9000\,\AA\ which is steeply falling off toward shorter wavelengths. We find further support for a low magnetic field from a closer inspection of the mean SDSS spectrum, which, in particular in the H$\beta$ region, shows some depressions which can be interpreted as Zeeman split H$\beta$ absorption lines. 

We tested this hypothesis by firstly normalizing the observed spectrum to a smoothly varying continuum and then by comparing it with Gaussian weighted absorption coefficients of all the H$\beta$ transitions. The results are shown in Fig.~\ref{f:zeem}. Transitions at longer than the laboratory wavelength are $\sigma^+$ transitions, those between 4700 and 4800\,\AA\ are $\pi$-transitions, and those at shorter wavelengths belong to $\sigma^-$. We show two models, one centered at a magnetic field strength of $B=15.5$\,MG (red line in Fig.~\ref{f:zeem}), the other with $B=16.5$\,MG. While the former seems to better match with the $\pi$-components, the latter shows a better match with the $\sigma^+$-components, in particular that at 4907\,\AA; the line at 4966\,\AA\ is a stationary line with no dependence of the line position as a function of the magnetic field. Both model curves were computed for a Gaussian FWHM of $\Delta B = 0.5$\,MG, and folded with another Gaussian with FWHM of 2\,\AA\ to match the observed spectral resolution. 

Time-resolved photometry obtained by \cite{rodriguez+23} shows the object to be variable by about 1 mag in the $r-$band. Such variability amplitudes are also seen mostly in the public ZTF data. However, there is a marked brightness change in the ATLAS data, an excursion from a mean level $o=19$ to $o=16$ for a short time in the beginning of 2017. This was followed by an excursion to $o\sim17.5$ about a year later lasting several weeks. We believe these episodes mark times of an extraordinarily high mass transfer rate between the two objects. 

\begin{figure}
\resizebox{\hsize}{!}{\includegraphics{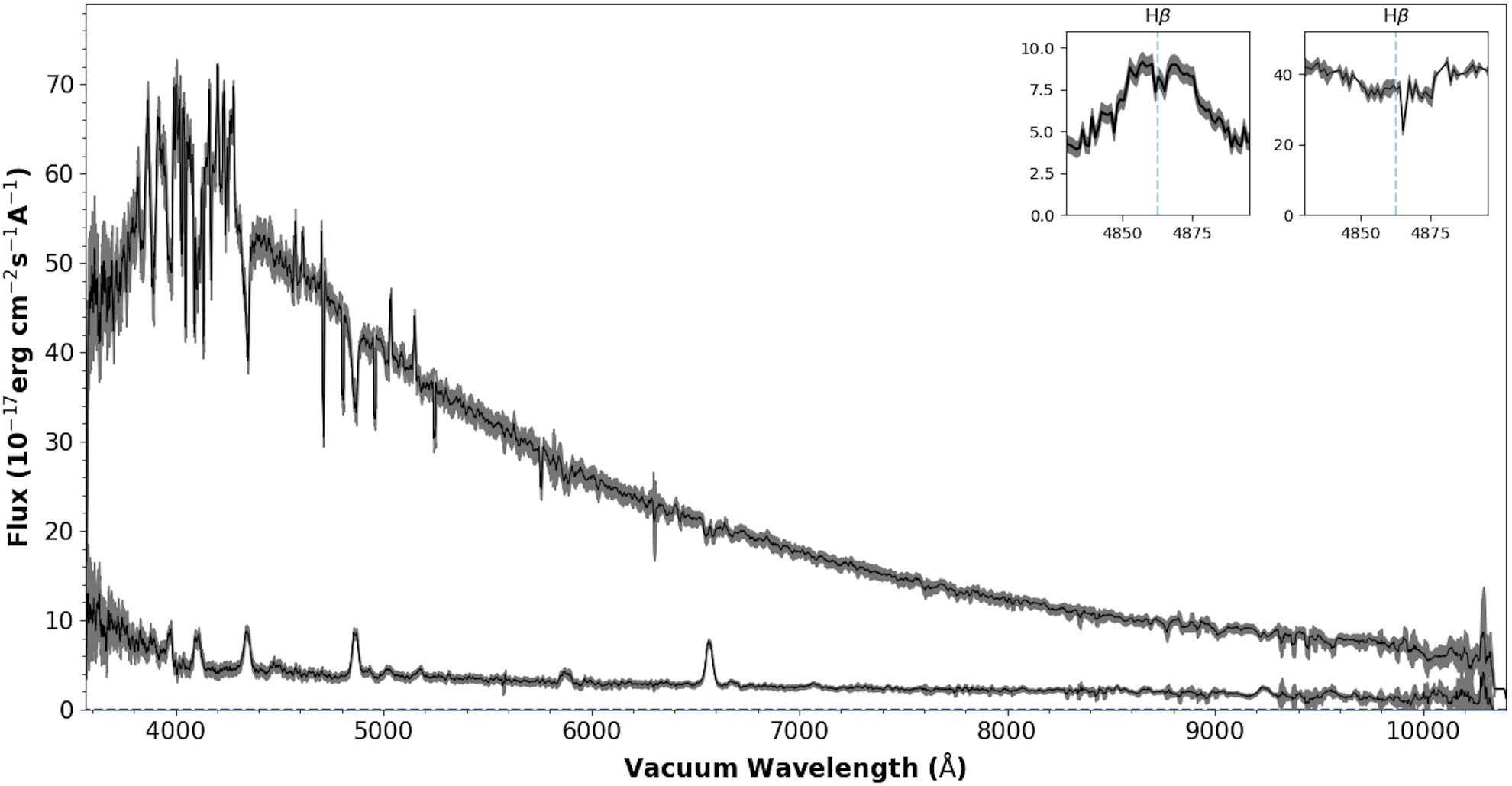}}
\caption{Spectra of the DN J0926$+$0345 obtained in March and April 2021. Five 15 min spectra were obtained in quiescence on MJD 59292, four additional in outburst on MJD 59316.
}
\label{f:dno0926}
\end{figure}

\subsubsection{J0926$+$0345} 
The object was identified as a DN or DN candidate by \cite{welsh+05} and \cite{wils+10} based on variability patterns, an UV excess and SDSS colors, but remained without spectroscopic identification. All photometric surveys considered here, CRTS, ZTF, and ATLAS show DN outbursts with amplitudes of about 3 mag. Spectra in SDSS-V were obtained at two epochs, 2021-Mar-19 and 2021-Apr-12, i.e.~separated by 24 days. In March, the object was in quiescence at $g\simeq 17.5$, while in April it was observed during  outburst at $g \simeq 15.1$. As usual, the outburst spectrum shows broad absorption lines with emission cores from the optically thick disk, while the spectrum taken in quiescence just shows emission lines from the thin disk (see Fig.~\ref{f:dno0926}). The emission lines are double-peaked indicating a high orbital inclination. The line parameters listed in Tab.~\ref{t:linpar} refer to the quiescence spectrum.

\subsubsection{J0929$+$0053} 
A faint, mildly blue object beyond the \gai limit. The spectrum obtained in SDSS-V is the second of this object, the first was obtained in January 2001 but remained unnoticed (plate-mjd-fib=475-51965-0217), probably due to its low signal-to-noise ratio. The spectrum shows an M-star donor and Balmer emission lines, and perhaps some weak \heo4471. The spectrum is noisy but the strength of the TiO band at 7600\,\AA\ and the shape of the continuum allow us to constrain the spectral type to $\sim$M5 with an uncertainty of one subclass. The implied V magnitude of a suitably scaled template star from \cite{kesseli+17} is $V=21.1$. According to the tables of \cite{knigge+11} of the CV donor sequence, the object has a short orbital period below the period gap ($P_{\rm orb} = 1.6-2.1$\,h). The distance is uncertain but of the order of 300\,pc, which implies a very low X-ray luminosity of $3.2\times 10^{29}$\,\lx.

The object was in the field of view of an \xmmn observation pointing to a nearby AGN, at an off-axis angle of about 15 arcmin at  the same flux level as during the eFEDS observations. Due to the large off-axis angle and the corresponding large vignetting factor, only $68 \pm 11$ photons were collected. We made no attempt to fit the data with a spectral model, as the parameters would be largely unconstrained.

The ZTF light curve shows variability between 20.5 and 21.5, the ATLAS forced photometry light curve remained flat, no CRTS data are available. The object was detected in eFEDS and all eRASS without significant X-ray variability. Given the late spectral type of the donor, and its low luminosity, the object is tentatively classified as a DN candidate, a possible WZ Sge-type object. The lines appear symmetric; it could be a low-inclination system.

\subsubsection{J0929$+$0401} 
A very peculiar object, that is difficult to classify. The SDSS pipeline classified the object as GALAXY, but the spectrum is clearly stellar. It is an M-type star with a large UV excess. It has stationary narrow Balmer emission lines. In the CMD it is located slightly above the ZAMS, while the $\log$\fxo$=-0.3$ is higher than for coronal emitter at this spectral type but below that of typical accreting systems. The X-ray luminosity points to an accreting system. The object displays marginal optical variability with an amplitude of about 0.1 -- 0.2 mag. It was detected in eFEDS and even brighter in eRASS:3 by 70\% but not in the eRASS' individually. We classify it as a CV candidate, although this is admittedly not fully convincing given the lack of pronounced typical emission lines.

\subsubsection{J0932$+$0343} 
J0932$+$0343 was reported as being a CV by \cite{necker+22} while searching for optical counterparts to IceCube neutrino alerts. It is a DN showing broad Balmer absorption lines from the WD and/or the disk and TiO absorption bands from the late-type donor. The object has frequent outbursts with outburst amplitudes as high as $\Delta m\simeq 4$. There is additional $\sim$0.5 mag level variability in the quiescence state. X-ray detections were made in eFEDS and all eRASS' although the signal in eRASS2 is compatible with zero. A total of 10 sub-spectra are available which were obtained during five nights. They show pronounced radial velocity variability but due to the spread of the sub-spectra over several nights a search for periodicity is not feasible. 

\subsubsection{J0932$+$0109} 
A first report of the object as a CV was given by \cite{szkody+03}. They mention a possible magnetic subclass due to the pronounced \het emission, stronger than  usual for non-magnetic CVs. \cite{southworth+06} took several VLT/FORS spectra revealing very symmetric H$\alpha$ lines, but a classification and period determination was not possible. We detected an X-ray signal in eFEDS and all eRASS with weak variability at the 2$\sigma$ level. The spectrum does not show any stellar features, neither from the WD nor from the donor. There is sub-spectral variability in line strength, but not in line position, which could indicate a low-inclination system. In the CMD it is located in the regime of NLs, while the optical light-curve exhibits variability by about 0.5 mag. No final classification is possible, it could be an IP but confirmation would need high-speed photometry and a dedicated X-ray observation to identify the spin period of the WD.

\subsubsection{J0935$+$0429}
On first inspection of its SDSS spectrum the object could be classified as a super-active coronal emitter, a flare star. The spectrum is dominated by an M-type star with intense H-Balmer and CaHK lines superposed. But it also has a remarkable blue component in its spectrum. In the CMD it is located slightly below the ZAMS. Its X-ray to optical flux ratio is about one dex and its X-ray luminosity is several dex above that of a coronal emitter. It was detected in eFEDS and in all individual eRASS and appears marginally variable at the 1$\sigma$ level. The emission lines show marked intensity variations between sub-spectra. 

The ZTF data show clear variability with a best-fit period of 1.76 hours confirming the close binary nature. 

\subsection{Notes on individual objects: Detached compact binaries}

\subsubsection{J0832$+$0331}
A WD/MS system with a DA white dwarf and an active M star as companion. The object resides just below the ZAMS in the CMD and displays some insignificant optical variability with low amplitude around a mean of $r= 19.2$ and occasional enhancements to $r=18.75$. The Na absorption lines measured in the four sub-spectra do not show radial velocity shifts. The object was selected as an UV-excess, i.e. candidate WDMS object, in the mwm\_cb\_gaiagalex carton.

\subsubsection{J0843$+$0108}
A WD/MS system with a DB white dwarf and an M star as companion. The object resides just above the WD sequence in a \gai-based CMD and was listed as WD candidate in the catalogs of \cite{gentile_fusillo+19,gentile_fusillo+21}. It is also listed in the catalog of known hot subdwarf stars of \cite{culpan+22}. Given the absolute magnitude and the spectral shape determined here (see bnelo in the Appendix), this is obviously a misclassification, the object is a WD/MS system. The companion is rather inactive. There is some ZTF variability with an amplitude of 0.2 mag. Nine sub-spectra were obtained grouped in 4 and 5 obtained in two successive nights. A $\chi^2$ test revealed marginal evidence for radial velocity variability with $\chi2_\nu = 1.23$. The object was selected by the mwm\_cb\_gaiagalex carton as an UV-excess, i.e.~candidate WDMS object.

\subsubsection{J0847$+$0023}
A WD/MS system with a DA white dwarf and an active M star as companion. It was selected by the mwm\_cb\_gaiagalex carton as an UV-excess object, i.e.~a candidate WDMS object. The object resides just below the ZAMS in a \gai CMD and displays some insignificant optical variability with low amplitudes around a mean of $r= 19.4$, with occasional outliers to $r=18.8$ or 18.9. The Na absorption lines measured in the seven available  sub-spectra that were obtained subsequently do not show radial velocity shifts.

\subsubsection{J0904$+$0312}
A WD/MS system with a DA white dwarf and an active M star as companion. The object was selected 
as an UV-excess, i.e. candidate WDMS object, in the the mwm\_cb\_gaiagalex carton. It resides just below the main sequence in a \gai based CMD and displays optical variability with an amplitude 0.3  to 0.4 mag, i.e. is a candidate for an interacting WD/MS pair. The five sub-spectra do not indicate RV variability on the timescale of one hour but the absolute mean value of about 100 \kmps implies a close binary nature. The object is thus regarded a Post Common Envelope Binary (PCEB) or pre CV candidate.

\begin{figure}[]
\resizebox{\hsize}{!}{\includegraphics{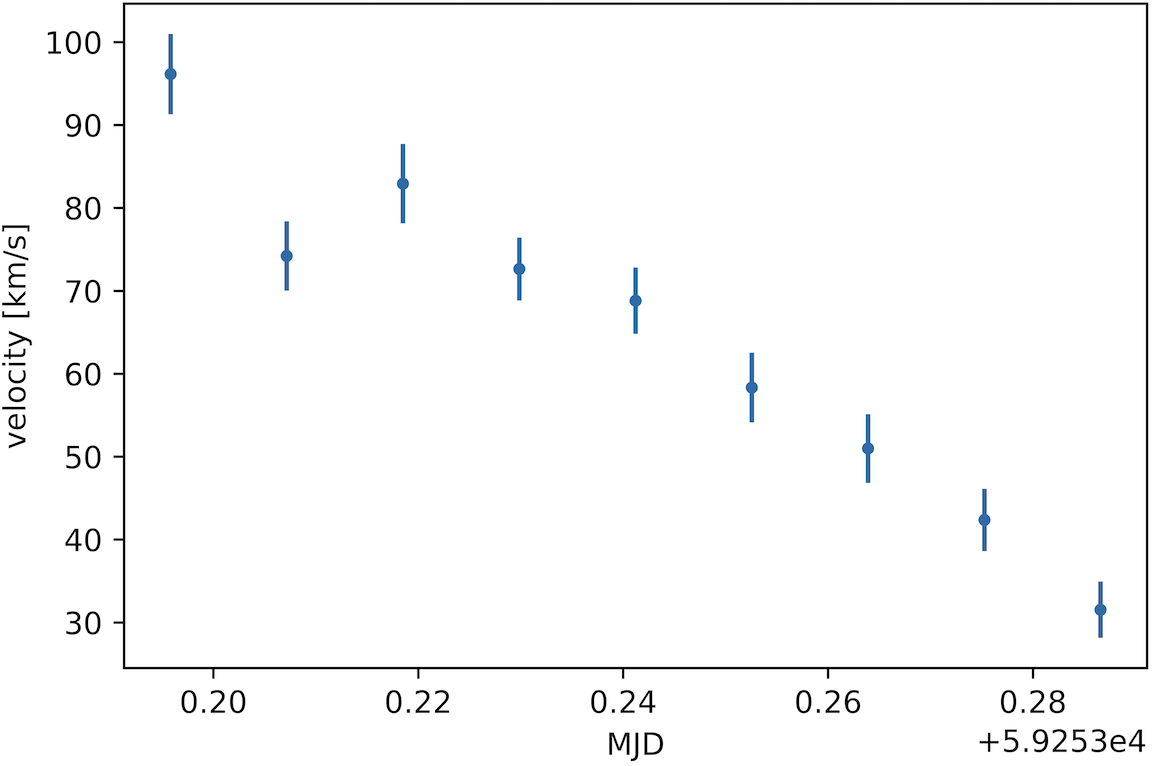}}
\caption{H$\alpha$ radial velocity obtained for the 9 sub-spectra of the NLc J0908$-$0125.}
\label{f:rvha0908}
\end{figure}

\begin{figure}[]
\resizebox{\hsize}{!}{\includegraphics{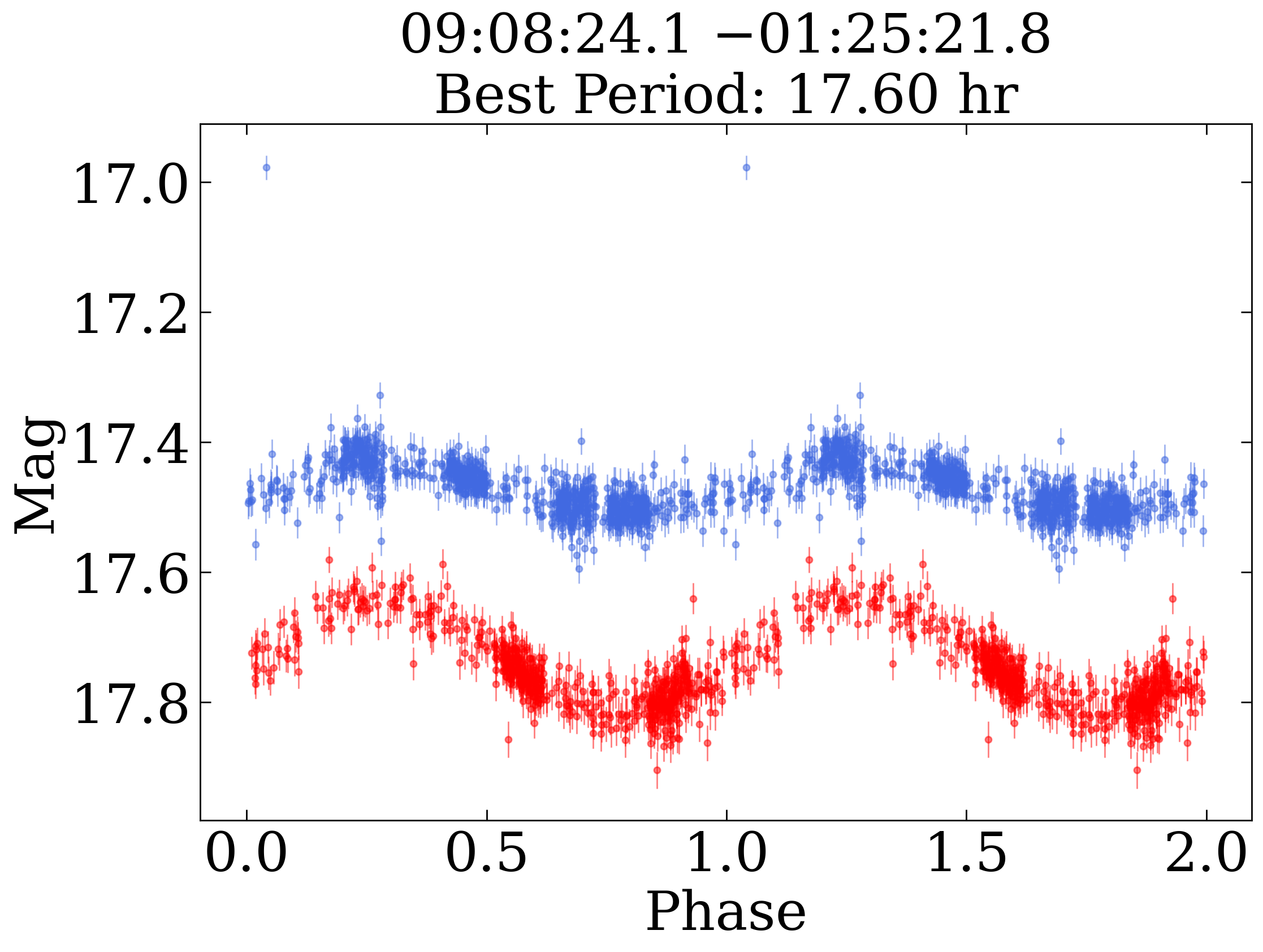}}
\caption{ZTF $g$ (blue) and $r$ (red) band light curves of J0908$-$0125 folded on the likely period of 17.6 hours.}
\label{f:0908ztf}
\end{figure}

\subsubsection{J0908$-$0125}
The object was selected as an UV-excess, i.e., candidate WDMS object, in the mwm\_cb\_gaiagalex carton and has no X-ray detection. It was previously recognized as a hot subdwarf \citep{perez-fernandez+16} but re-classified by \cite{geier20} as a WD. In the CMD it is located halfway between the WD and the main sequence. The SDSS spectrum reveals a very blue continuum with broad Balmer absorption lines (for lines above H$\alpha$) and narrow H Balmer and helium emission lines. The spectrum also reveals features of an M-type companion star. Nine consecutive subspectra were obtained and we searched for radial velocity variations of the H$\alpha$ emission line in each of those using a single Gaussian fit to the line. The result is shown in Fig.~\ref{f:rvha0908}. It shows mainly a continuous trend from 95\,\kmps to 30\,\kmps over a time interval of about 2 hours. This shows clearly that the object is a compact binary but the coverage was not sufficient to constrain the period spectroscopically.

The period search among the ZTF data revealed a best fit period of 17.6 hours. The phase-folded light curve is shown in Fig.~\ref{f:0908ztf}. It has a smoothly varying single-humped shape with an amplitude of about 0.1 mag in the $r$-band. At this period the M-star companion is not Roche-lobe filling, explaining the lack of X-ray emission. The width of the Balmer lines is much narrower than in the accreting systems (see Table~\ref{t:linpar}), with the exception of J0929$+$0401. We classify the object as a non-magnetic pre-CV (or PCEB) with a hot white dwarf irradiating the companion, thus giving rise to the photometric variability through ellipsoidal modulations.

\subsubsection{J0936$+$0253}
A WD/MS system with a DA white dwarf and an active M star as companion. The object resides just below the ZAMS in the CMD and displays negligible optical variability around $r= 19.5$. No sub-spectral RV variability is seen.
The object was selected by the mwm\_cb\_gaiagalex carton as an UV-excess, i.e. candidate WDMS object.

\subsection{Identification overview}
Upon visual inspection of SDSS spectra and various auxiliary data products we have identified 26  accreting compact white-dwarf binaries in eFEDS, of which 24 are proven X-ray emitters. CV subclasses were determined for most of those 26, but for about one third the subclass remains ambiguous. 
Six out of the 26 have no measured \gai-parallax.
Their detection and identification was possible only through the depth of the eFEDS X-ray observations and the availability of archival deeper optical imaging data from e.g.~the legacy survey. Tab.~\ref{t:cvs} lists one apparent exception, J0853+0204, at magnitude $g=14.97$. However, this magnitude was reached during a DN outburst, whereas the quiescence magnitude is $r=21.4$, hence also beyond the \gai limit.

The publicly released eFEDS X-ray catalog lists 27,910 X-ray sources of which 27,369 are considered point-like, i.e.~with  an X-ray extent {\tt EXT} $ < 6"$ \citep{brunner+22}. A positional match between the 27,369 point-like X-ray sources ({\tt RA\_corr, DEC\_corr}) and the fibre coordinates ({\tt plug\_ra, plug\_dec}) in {\tt spAll-v6\_0\_4-eFEDS.fits} within 60 arcsec reveals 9346 pairs. Hence, the identification fraction achieved through the plate program of eFEDS X-ray point sources is 34\%. This means that the total numbers of aCWDB objects in eFEDS is certainly larger than the sample discussed here. Formally this could be up to a factor 3, but the completeness fraction is not flat over magnitude. Furthermore, the SDSS-V completeness is also lower for samples that already had SDSS-IV and earlier spectroscopy (they had lower priority when competing for SDSS-V fibers).

All but two of the 26 objects in Table \ref{t:cvs} have a measured X-ray flux although one (J0920$+$0042) was not listed in the main eFEDS X-ray catalog but only in the supplement. The two remaining objects belong to the subclass of WZ Sge-type objects with low X-ray luminosities. 
Their distance of about 300 pc indicates the radius up to which members of this class can successfully be discovered through their X-ray emission, dependent of course on their intrinsic mass transfer rates and accretion luminosities which are not yet well determined. Objects that were studied so far indicate X-ray luminosities in the range of $\log L_{\rm X} (\mbox{erg s}^{-1}) \simeq 29 \dots 30.4$, but the sample is small, consisting of just six objects \citep{stelzer+17, schwope+21, munoz-giraldo+23}.

\begin{table}
\caption{CV subclasses of the 26 aCWDBs listed in Tab.~\ref{t:cvs} \label{t:subcl}}
\begin{tabular}{lr}
Subclass & members \\
\hline
DN or DNc (incl 2 SU UMa) & 11 \\
WZ or WZc & 3\\
NL or NLc (likely non-magnetic) & 4 \\
Polars & 2 \\
AMC & 3\\
CV or CVc & 3\\
\hline
\end{tabular}
\end{table}

The breakdown in CV subclasses of the 26 objects listed in Tab.~\ref{t:cvs} is given in Tab.~\ref{t:subcl}. The sample of aCWDBs presented here is not large, and thus caution should be exercised to not over-interpret the statistics based on it. Nevertheless a few features of the sample seem worth mentioning. 
\begin{enumerate} 
\item In this flux-limited sample one finds less than 10\% of secure magnetic objects. This contrasts with the more than 1/3 magnetic CVs found by \cite{pala+20} in their volume-limited sample. The determination of this ratio is relevant to further constrain CV evolution and angular momentum loss, and to test current scenarios for the generation and evolution of magnetic fields \citep{belloni+20, schreiber+21}. We might have missed the fraction of polars that were in a low state during eFEDS observations and thus escaped detection. The fraction of low-state MCVs is difficult to quantify but might be as large as 50\% as estimated from the accretion duty cycle in the prototype AM Herculis \citep{hessman+00}. Even when the sample presented here is corrected by a factor of two, the fractions of MCVs in our sample and that of \cite{pala+20} seem to be discrepant. We conclude that the current samples, both volume- and flux-limited, which are used to constrain the fraction of magnetic CVs are too small to derive robust constraints. 

\item
We identified three WZ Sge-type CVs, low-luminosity objects close to the white dwarf-sequence in the CMD, J0845$+$0339, J0929$+$0053 and J0904$+$0355. These are of interest because they might be period bouncing systems (or period bouncers, PBs), i.e.~systems that have passed the orbital period minimum and thus developed a degenerate donor star. They were predicted to make up a great portion of the cataclysmic variable population, between 40\% and 84\% \citep{goliasch+nelson15, belloni+20}, but few are known only \citep[see][]{munoz-giraldo+24} for a census. Every new detection thus helps to reduce statistical uncertainties and provide input to theoretical models.
One of those, J0929+0053, shows M-star features in its spectrum and is therefore not a bouncing candidate; J0845$+$0339 and J0904+0355 are the two remaining. Assuming an X-ray luminosity of $\log L_{\rm X}  \mbox{(erg s}^{-1}) = 29.5$ the maximum distance to detect PBs is $\sim$500\,pc which gives a survey volume of about $0.0034 \times 5 \times 10^8$\,pc, where the factor 0.0034 corrects for the fractional area of eFEDS. If one assumes that the two objects are PBs, that these objects have a scale height of 450 pc (the mean Galactic latitude of eFEDS is about 30\degr), and that the spectroscopic identification fraction is say 0.6, then one may estimate a mid-plane space density of $\rho_0 \lesssim 1.4 \times 10^{-6}$\,pc$^{-3}$. This is formally a factor 7 higher but still consistent with a recent estimate by \cite{inight+23b}, who analyzed 110 CVs that were observed in the SDSS-V plate program and derived $0.2 \times 10^{-6}$\,pc$^{-3}$, but more than a factor 10 lower than predicted recently \citep{belloni+20}.

\item There are three DD systems (AM CVn objects) found through their X-ray emission, two of them with \gai information. One (J0903$-$0133) is at about 270\,pc with a luminosity below $10^{30}$\,\lx, and hence among the closest and least luminous X-ray detected AM CVns, while another, J0844$-$0128, is among the most distant and most X-ray luminous (Tab.~\ref{t:cvs}) among those. This is a promising start to establish a comprehensive sample of accreting DDs from identification work of the eROSITA all-sky survey and eventually determine their unknown space density with high precision.

\item The identified aCWDB objects have Galactic latitudes between $b^{II} = 23\degr$ and 38\degr. Those with reliable \gai parallaxes ({\tt parallax\_over\_error $>$ 3}) lie at distances above the plane between 130\,pc and 820\,pc, and hence might extend to several scale heights above their likely parent distribution. The height above the plane will depend on the evolutionary state of the aCWDB. Long-period NLs are discussed to have scale heights of just 120 pc, while the old WZ-like objects might have scale heights of 450\,pc. Thirteen of the 26 aCWDB objects have measured or estimated orbital periods, five were added in this paper. The lack of orbital periods for half of the objects prevents a placement in an evolutionary context. Determining orbital periods of new aCWDBs is the next major task for eFEDS-CVs and the many other which will be unveiled through their X-ray emission and spectroscopic identification. A tremendous and important task ahead of us is the determination of orbital periods of complete samples, both nearby and distant, to constrain the evolutionary states of the objects and to measure the scale height of (sub-)classes so that this no longer need to be assumed, as was done previously \citep[e.g.,][]{pala+20,schwope18}.

\end{enumerate} 

We also classified six WDMS systems through their symbiotic appearance in the SDSS spectra. We searched for spectral variability, in particular for radial velocity variations of the H$\alpha$ emission or the Na absorption lines and found one object with pronounced radial velocity shifts, J0908$-$0125. Further inspection of its photometric variability in the ZTF revealed a clear periodicity of 17.6 hours. The object is thus classified as a PCEB or pre-CV. The found period is likely caused by the ellipsoidal shape of the donor and thus indicates the orbital period, which is on the long-period half of the sample presented by \cite{nebot+11}. A second such object, J0904$+$0312, was identified as a pre-CV candidate based on its high mean radial velocity and pronounced photometric variability, although sub-spectral variability could not yet been found.

In the final paragraphs we briefly discuss the current targeting strategies and prospects of finding new aCWDBs in the upcoming SDSS-V all-sky. New aCWDBs might be expected to be found as counterparts of \ero X-ray sources, through their UV excess, or, as in the past, just serendipitously \citep[see][for a comprehensive overview of aCWDBs observed in SDSS-I to -IV]{inight+23a}. Besides the genuine \ero identification programs 
two further programs are implemented to identify aCWDB candidates from a measured UV excess compared to main sequence stars \citep[see][for a description of the various cartons of objects]{almeida+23}. Here we ask, which of our 26 objects would have been selected for follow-up by any of those follow-up programs with the fiber positioning system (FPS). Section~\ref{s:fup} gives all the details. The FPS identification work will be based on eRASS:3 which is shallower than eFEDS by a bout a factor 2. If the sources listed in Table~\ref{t:cvs} were constant in flux, one could expect that 19 objects are detected in eRASS:3, i.e.~above a flux of $2 \times 10^{-14}$\,\fergs. 
Column (4) in Tab.~\ref{t:succ} shows that a constant flux is the exception, not the rule. Likely for this reason the eRASS:3 catalog detects just 16 out of the expected 19 aCWDBs. 

Of those 16, 15 are selected by one of the \ero-centric follow-up programs with the correct counterpart as determined in this paper. The 16th object is missed, because the target association procedures have picked the wrong counterpart. A UV-excess criterion selects 10 of the 16 X-ray detected sources for spectroscopy, and an additional 3, that are not detected in eRASS:3. This illustrates the complementarity of the two approaches. The overall impression we gain from this exercise is that SDSS-V will run a highly successful and almost complete aCWDB identification program, that the X-ray centric cartons select most of the counterparts correctly and that additional aCWDBs might be expected from UV-selection. UV-based selection may add aCWDBs if they are brighter than the \gai limit and have full astrometric information (position, proper motion, parallax). It is somewhat unfortunate in this regard is that the GALEX catalog omits most of the Galactic plane due to source confusion. In the plane, one is left with X-ray selected aCWDB candidates. We expect a significant fraction of X-ray detected aCWDBs being fainter than the \gai limit. These can be successfully targeted in sky areas surveyed by the Legacy Survey or other similarly deep surveys.

The 16 aCWDBs in eFEDS, corrected for an assumed identification fraction of 60\%, imply a surface density of 0.2 deg$^{-2}$ at a Galactic latitude of 30\degr. If one assumes this value to be representative for the extragalactic sky (10,000 deg$^2$), one may expect of order 2000 X-ray selected CVs identified in SDSS-V at high latitudes. The surface density will increase if one approaches the galactic plane. It is dependent a) on the luminosity function of the CVs and hence their mid-plane space density, which is different for each sub-class, b) on their scale height, which again is different for each sub-class and depends on their evolutionary age, c) on the relative fractions of the various sub-classes, and d) on the exposure map. Only the latter of these is known with sufficient accuracy, the others need to and will be determined as a result of the comprehensive identification programs that have just started. The mentioned several 1000 X-ray selected CVs are consistent with a former prediction based on the integration of a luminosity function \citep{schwope12}. However, the basic parameters that enter such forecasts were and are still unknown to a factor of 2 or more.

\begin{acknowledgements}
This work was supported by the German DLR under contract 50 OR 2203 and the Deutsche Forschungsgemeinschaft under grant numbers Schw536/37-1 and Schw536/38-1. \\

This work acknowledges support from ANID through the Millennium Science Initiative Program - ICN12\_009 (FEB), CATA-BASAL - FB210003 (FEB), and FONDECYT Regular - 1200495 (FEB). \\

This project has received funding from the European Research Council (ERC) under the European Union’s Horizon 2020 research and innovation programme (Grant agreement No. 101020057).\\

This work is based on data from \ero, the soft Cinstrument aboard SRG, a joint Russian-German science mission supported by the Russian Space Agency (Roskosmos), in the interests of the Russian Academy of Sciences represented by its Space Research Institute (IKI), and the Deutsches Zentrum für Luft- und Raumfahrt (DLR). The SRG spacecraft was built by Lavochkin Association (NPOL) and its subcontractors, and is operated by NPOL with support from the Max Planck Institute for Extraterrestrial Physics (MPE). \\

The development and construction of the eROSITA X-ray instrument was led by MPE, with contributions from the Dr. Karl Remeis Observatory Bamberg \& ECAP (FAU Erlangen-Nuernberg), the University of Hamburg Observatory, the Leibniz Institute for Astrophysics Potsdam (AIP), and the Institute for Astronomy and Astrophysics of the University of Tübingen, with the support of DLR and the Max Planck Society. The Argelander Institute for Astronomy of the University of Bonn and the Ludwig Maximilians Universität Munich also participated in the science preparation for eROSITA. \\

The eROSITA data shown here were processed using the eSASS/NRTA software system developed by the German eROSITA consortium. \\

Funding for the Sloan Digital Sky Survey V has been provided by the Alfred P. Sloan Foundation, the Heising-Simons Foundation, the National Science Foundation, and the Participating Institutions. SDSS acknowledges support and resources from the Center for High-Performance Computing at the University of Utah. SDSS telescopes are located at Apache Point Observatory, funded by the Astrophysical Research Consortium and operated by New Mexico State University, and at Las Campanas Observatory, operated by the Carnegie Institution for Science. The SDSS web site is \url{www.sdss.org}.\\

SDSS is managed by the Astrophysical Research Consortium for the Participating Institutions of the SDSS Collaboration, including Caltech, The Carnegie Institution for Science, Chilean National Time Allocation Committee (CNTAC) ratified researchers, The Flatiron Institute, the Gotham Participation Group, Harvard University, Heidelberg University, The Johns Hopkins University, L’Ecole polytechnique f\'{e}d\'{e}rale de Lausanne (EPFL), Leibniz-Institut f\"ur Astrophysik Potsdam (AIP), Max-Planck-Institut f\"ur Astronomie (MPIA Heidelberg), Max-Planck-Institut f\"ur Extraterrestrische Physik (MPE), Nanjing University, National Astronomical Observatories of China (NAOC), New Mexico State University, The Ohio State University, Pennsylvania State University, Smithsonian Astrophysical Observatory, Space Telescope Science Institute (STScI), the Stellar Astrophysics Participation Group, Universidad Nacional Aut\'{o}noma de M\'{e}xico, University of Arizona, University of Colorado Boulder, University of Illinois at Urbana-Champaign, University of Toronto, University of Utah, University of Virginia, Yale University, and Yunnan University.

Funding for the Sloan Digital Sky Survey IV has been provided by the Alfred P. Sloan Foundation, the U.S. Department of Energy Office of Science, and the Participating Institutions. \\

SDSS-IV acknowledges support and resources from the Center for High Performance Computing  at the 
University of Utah. The SDSS website is \url{www.sdss4.org}.\\

SDSS-IV is managed by the Astrophysical Research Consortium for the Participating Institutions of the SDSS Collaboration including the Brazilian Participation Group, the Carnegie Institution for Science, Carnegie Mellon University, Center for Astrophysics | Harvard \& Smithsonian, the Chilean Participation Group, the French Participation Group, Instituto de Astrof\'isica de Canarias, The Johns Hopkins University, Kavli Institute for the Physics and Mathematics of the Universe (IPMU) / University of Tokyo, the Korean Participation Group, Lawrence Berkeley National Laboratory, Leibniz Institut f\"ur Astrophysik Potsdam (AIP),  Max-Planck-Institut f\"ur Astronomie (MPIA Heidelberg), Max-Planck-Institut f\"ur Astrophysik (MPA Garching), Max-Planck-Institut f\"ur Extraterrestrische Physik (MPE), National Astronomical Observatories of China, New Mexico State University,  New York University, University of Notre Dame, Observat\'ario Nacional / MCTI, The Ohio State University, Pennsylvania State University, Shanghai Astronomical Observatory, United Kingdom Participation Group, Universidad Nacional Aut\'onoma de M\'exico, University of Arizona, University of Colorado Boulder, University of Oxford, University of Portsmouth, University of Utah, University of Virginia, University of Washington, University of Wisconsin, Vanderbilt University, and Yale University.\\

The Pan-STARRS1 Surveys (PS1) have been made possible through contributions of the Institute for Astronomy, the University of Hawaii, the Pan-STARRS Project Office, the Max-Planck Society and its participating institutes, the Max Planck Institute for Astronomy, Heidelberg and the Max Planck Institute for Extraterrestrial Physics, Garching, The Johns Hopkins University, Durham University, the University of Edinburgh, Queen's University Belfast, the Harvard-Smithsonian Center for Astrophysics, the Las Cumbres Observatory Global Telescope Network Incorporated, the National Central University of Taiwan, the Space Telescope Science Institute, the National Aeronautics and Space Administration under Grant No. NNX08AR22G issued through the Planetary Science Division of the NASA Science Mission Directorate, the National Science Foundation under Grant No. AST-1238877, the University of Maryland, and Eotvos Lorand University (ELTE). \\

This work is based on observations obtained with the Samuel Oschin Telescope 48-inch and the 60-inch Telescope at the Palomar Observatory as part of the Zwicky Transient Facility project. Major funding has been provided by the U.S National Science Foundation under Grant No. AST-1440341 and by the ZTF partner institutions: the California Institute of Technology, the Oskar Klein Centre, the Weizmann Institute of Science, the University of Maryland, the University of Washington, Deutsches Elektronen-Synchrotron, the University of Wisconsin-Milwaukee, and the TANGO Program of the University System of Taiwan.\\

This work has made use of data from the Asteroid Terrestrial-impact Last Alert System (ATLAS) project. The Asteroid Terrestrial-impact Last Alert System (ATLAS) project is primarily funded to search for near earth asteroids through NASA grants NN12AR55G, 80NSSC18K0284, and 80NSSC18K1575; byproducts of the NEO search include images and catalogs from the survey area. This work was partially funded by Kepler/K2 grant J1944/80NSSC19K0112 and HST GO-15889, and STFC grants ST/T000198/1 and ST/S006109/1. The ATLAS science products have been made possible through the contributions of the University of Hawaii Institute for Astronomy, the Queen’s University Belfast, the Space Telescope Science Institute, the South African Astronomical Observatory, and The Millennium Institute of Astrophysics (MAS), Chile. \\


\end{acknowledgements}

\bibliographystyle{aa}
\bibliography{efedscvs}


\begin{appendix}
\section{Follow-up of CWDBs in SDSS-V}
\label{s:fup}
Optical counterparts for eRASS:3 X-ray sources were selected using different input imaging data and with different priors by the \ero-collaboration. 
They were designed to either find compact binaries (mwm\_erosita\_compact), to find coronal emitters (mwm\_erosita\_stars) or to find the most likely counterpart to an X-ray source without particular focus on a source class (bhm\_spiders\_agn\_gaiadr3, bhm\_spiders\_agn\_lsdr10). Their exact targeting strategies will be described elsewhere, only very cursory information can given here. erosita\_compact tries to select aCWDBs using \gai DR3 and a RF-trained prior. Training was performed on a sample of 642 cataclysmic binaries with complete \gai (including parallax, proper motion, and a variability proxy) and X-ray information. X-ray information was collected from various input catalogs (\ros, \xmmn, \cxo) and fluxes transformed to the common \ero energy band. The erosita\_stars carton tries to select stars (coronal emitters) as X-ray counterparts using \gai DR3 \citep{schneider+22}. Both, the spiders\_agn\_gaiadr3 and spiders\_agn\_lsdr10 cartons are general purpose X-ray counterpart finders. They attempt to find the best optical counterpart not assuming a specific source class a priori. However, the training sample used is dominated by AGN and to a lesser extent by stars. The former carton is based on \gai DR3, and the latter on DR10 of the legacy survey (legacysurvey.org), particularly designed for objects beyond the \gai limit. Both SPIDERS cartons use NWAY as the tool to identify likely optical counterparts from the different input catalogs. The strategy is described in \cite{salvato+22}. 

The mwm\_cb\_galex\_mag and mwm\_cb\_galex\_vol galex\_mag cartons implemented in SDSS-V select likely CWDBs on the basis of a UV excess from matching \gai-GALEX sources in either the FUV- (mag) or the NUV-channel (vol).

The results of the various targeting routes are summarized in Tab.~\ref{t:succ}. The SDSS coordinates of the aCWDBs were used as input coordinates and a radius of 1 arcmin was chosen to search for matching sources in the mentioned target catalogs. Columns (2) and (3) give X-ray fluxes in the 0.2 -- 2.3 keV band from eFEDS and eRASS:3. Column (4) gives the measured X-ray flux ratio between eFEDS and eRASS:3. 
One of the objects that was originally not X-ray selected, J0904+0355 (WZ Sge type), is discovered with high confidence in eRASS:3. Of those 16 X-ray detected aCWDBs in eRASS:3, 14 are in \gai DR3, one of those has just photometric but no parallax information. The meaning of the numbers in columns (5) to (13) is described in the caption of the Table.

Matching stars (column 7) were tested here just for completeness, no matching source was expected because stars were selected (among other criteria) for their small X-ray to optical flux ratio. This explains the small number of only two matching sources. The stars carton would have selected one aCWDB correctly (J0851$+$0308) and would have chosen a nearby brighter star as a possible counterpart for J0932$+$0343 but with a small matching probability. This kind of contamination may of course occur more often among the selected targets. We do not derive a contamination fraction here given the small sample, but assume it being small.

The mwm\_erosita\_compact carton selects 8 counterparts correctly. The relatively small number of correct counterparts (50\%) is due to the chosen cut in the likelihood of association, $p\_{\rm any} > 0.5$. The eight chosen targets all have $p\_{\rm any} > 0.9$. Such a likelihood cut is not applied to the bhm\_spiders\_agn cartons, which both identify 16 possible counterparts of which each two were chosen incorrectly. Between bhm\_spiders\_agn\_gdr3 and bhm\_spiders\_agn\_lsdr10, one incorrectly chosen object is the same and another has changed. The mwm\_cb\_galex\_mag carton (column 10) selects eight objects, of which four are also X-ray discovered in eRASS:3, a further three were not in eRASS:3 but in eFEDS only and one is incorrectly assigned to an eRASS:3 and eFEDS X-ray source. The mwm\_cb\_galex\_vol carton (column 11) selects 12 objects in total, 9 of them are in eRASS:3 and in \gai DR3, a further two in eFEDS and \gai DR3 and one object in the vicinity of J0932$+$0109 at a distance of 22 arcsec, i.e.~it would have not picked the eFEDS discovered aCWDB. The same likely wrong association was made in galex\_mag. In total, of the 26, three have a chance to be recovered in SDSS-V (FPS) despite their missing X-ray flux in eRASS:3 due to their UV-properties. These three, J0844$-$0128, J0926$+$0345, and J0920$+$0042, are not new discoveries from this paper but were known in the literature beforehand. Three further aCWDBs from eFEDS have no X-ray emission in eRASS:3 but are \gai DR3 sources (J0846$+$0218, J0902$-$0142, J0912$-$0007), hence could have been picked by some other algorithm as interesting sources (e.g.~through a UV excess or variability) but were not. The reasons are unknown. There is only one source, J0929$+$0053, detected in eFEDS and eRASS:3, at \gai $G=20.88$ and therefore not considered by CB, STA, FUV, and NUV, that would not have been picked for spectroscopic follow-up by either carton, either for being too faint or for choosing the wrong counterpart (bhm\_spiders\_agn). For reasons that are difficult if not impossible to understand retrospectively it was chosen for follow-up from the initial preliminary X-ray source lists. 

In sum, of the 26 eFEDS-aCWDBs that were identified and discussed in this paper, 16 are re-covered as X-ray sources in eRASS:3. Of those, 15 are potential targets for spectroscopic follow-up in the SDSS-V fibre program through one of the X-ray centric cartons. UV-centric cartons select counterparts for 10 of those 16 X-ray discovered, but identify a further three that are not X-ray discovered in eRASS:3. 

\begin{table*}
\caption{Anticipated identifications on eFEDS aCWDBs in the SDSS-V with the FPS. X-ray sources are drawn from the eRASS:3 catalog. Columns have the following meaning: (2)-- X-ray flux in eFEDS (0.2-2.3 keV); (3) -- X-ray flux in eRASS:3 (0.2-2.3 keV); (4) X-ray flux ratio eRASS:3 / eFEDS; (5) aCWDB object is in \gai DR3 (1,0) = (Y,N); (6) to (9) X-ray centric cartons that selected the correct (=1) or the incorrect (=0) aCWDB object as counterpart to an eRASS:3 X-ray source; empty cells mean that no object was selected within 60 arcsec around the SDSS position of the given object; (10) and (11) UV centric cartons that selected the correct (=1) or the incorrect (=0) aCWDB object as counterpart to an eRASS:3 X-ray source; empty cell means that no object was selected within 60 arcsec around the SDSS position of the given object; Columns (12) and (13) give the number of X-ray or UV-based cartons that proposed to target the aCWDB object in SDSS-V; Column (14) indicates if an eRASS:3 detected aCWDB would be targeted by any of the discussed cartons. The ordering of the systems is the same as in Tab.~\ref{t:cvs}, the three objects at the bottom were not X-ray selected in eFEDS and were thus not foreseen for eFEDS follow-up in SDSS-V.
\label{t:succ}
}
\begin{tabular}{|l|r|r|r|c|c|c|c|c|c|c|c|c|c|}
\hline
  \multicolumn{1}{|c|}{Object} &
  \multicolumn{1}{c|}{$f_{\rm Xe}$} &
  \multicolumn{1}{c|}{$f_{\rm X3}$}&
  \multicolumn{1}{c|}{frat} &
  \multicolumn{1}{c|}{GDR3} &
  \multicolumn{1}{c|}{CB} &
  \multicolumn{1}{c|}{STA} &
  \multicolumn{1}{c|}{nw3} &
  \multicolumn{1}{c|}{nw10} &
  \multicolumn{1}{c|}{FUV} &
  \multicolumn{1}{c|}{NUV} &
  \multicolumn{1}{c|}{X-sel} &
  \multicolumn{1}{c|}{UV-sel} &
  \multicolumn{1}{c|}{eR3-XUV}
  \\
  \multicolumn{1}{|c|}{(1)} &
  \multicolumn{1}{c|}{(2)} &
  \multicolumn{1}{c|}{(3)} &
  \multicolumn{1}{c|}{(4)} &
  \multicolumn{1}{c|}{(5)} &
  \multicolumn{1}{c|}{(6)} &
  \multicolumn{1}{c|}{(7)} &
  \multicolumn{1}{c|}{(8)} &
  \multicolumn{1}{c|}{(9)} &
  \multicolumn{1}{c|}{(10)} &
  \multicolumn{1}{c|}{(11)} &
  \multicolumn{1}{c|}{(12)} &
  \multicolumn{1}{c|}{(13)} &
  \multicolumn{1}{c|}{(14)}
  \\
\hline
  J0840$+$0005 &  5.3 &  7.2 & 1.4 & 1 &   &  & 1 & 1 &   &   & 2 &   & 1\\
  J0843$-$0148 & 11.9 & 12.1 & 1.0 & 1 &   &  & 1 & 1 & 1 &   & 2 & 1 & 1\\
  J0844$+$0239 &  3.8 & 10.1 & 2.7 & 1 & 1 &  & 1 & 1 &   & 1 & 3 & 1 & 1\\
  J0844$-$0128 &  1.8 &      &     & 1 &   &  &   &   & 1 &   &   & 1 & \\
  J0846$+$0218 &  2.1 &      &     & 1 &   &  &   &   &   &   &   &   &\\
  J0847$+$0119 &  1.8 &      &     &   &   &  &   &   &   &   &   &   &\\
  J0847$+$0145 &  3.4 &  2.0 & 0.6 &   &   &  & 0 & 1 &   &   & 1 &   & 1\\
  J0850$+$0443 & 18.0 & 13.0 & 0.7 & 1 & 1 &  & 1 & 1 &   & 1 & 3 & 1 & 1\\
  J0851$+$0308 & 10.0 & 15.0 & 1.5 & 1 & 1 & 1& 1 & 1 &   & 1 & 4 & 1 & 1\\
  J0853$+$0204 &  3.7 &      &     &   &   &  &   &   &   &   &   &   &\\
  J0855$-$0154 &  3.5 &  4.8 & 1.4 & 1 &   &  & 1 & 1 &   & 1 & 2 & 1 & 1\\
  J0902$-$0142 &  5.0 &      &     & 1 &   &  &   &   &   &   &   &   &\\
  J0903$-$0133 &  6.4 &  5.0 & 0.8 & 1 & 1 &  & 1 & 1 &   & 1 & 3 & 1 & 1\\
  J0912$-$0007 &  1.6 &      &     & 1 &   &  &   &   &   &   &   &   &\\
  J0914$+$0137 &  5.8 &  7.0 & 1.2 & 1 & 1 &  & 1 & 1 &   &   & 3 &   & 1\\
  J0918$+$0436 &  1.4 &      &     &   &   &  &   &   &   &   &   &   &\\
  J0926$+$0105 & 27.9 & 17.9 & 0.6 & 1 & 1 &  & 1 & 1 & 1 & 1 & 3 & 2 & 1\\
  J0926$+$0345 &  1.2 &      &     & 1 &   &  &   &   & 1 & 1 &   & 2 &\\
  J0929$+$0053 &  3.0 &  5.0 & 1.7 &   &   &  & 0 & 0 &   &   &   &   & 0\\
  J0932$+$0343 &  3.9 &  2.0 & 0.5 & 1 &   & 0& 1 & 1 & 1 & 1 & 2 & 2 & 1 \\
  J0932$+$0109 & 22.9 & 21.9 & 1.0 & 1 & 1 &  & 1 & 1 & 0 & 0 & 3 &   & 1\\
  J0929$+$0401 &  4.1 &  7.4 & 1.8 & 1 &   &  & 1 & 0 &   &   & 1 &   & 1\\
  J0935$+$0429 &  6.7 &  3.2 & 0.5 & 1 &   &  & 1 & 1 &   & 1 & 2 & 1 & 1\\
\hline
  J0845$+$0339 &      &      &     & 1 &   &  &   &   &   &   &   &   & \\
  J0904$+$0355 &      &  2.6 &     & 1 & 1 &  & 1 & 1 & 1 & 1 & 3 & 2 & 1\\
  J0920$+$0042 &  0.7 &      &     & 1 &   &  &   &   & 1 & 1 &   & 2 &\\
\hline\end{tabular}
\end{table*}

\clearpage
\newpage


\section{Spectra and color-magnitude diagrams of eFEDS CWDB objects}

The following graphs show the SDSS-V spectra and the CMDs, if available, of all CWDB objects discussed in this paper. we firstly show the aCWDBs (and candidates) from Tab.~\ref{t:cvs}, then the CWDBs from Tab.~\ref{t:other}. The order of appearance is the same as in the named tables.

\begin{figure*}[]
\begin{minipage}[c]{0.45\hsize}
\resizebox{\hsize}{!}{\includegraphics{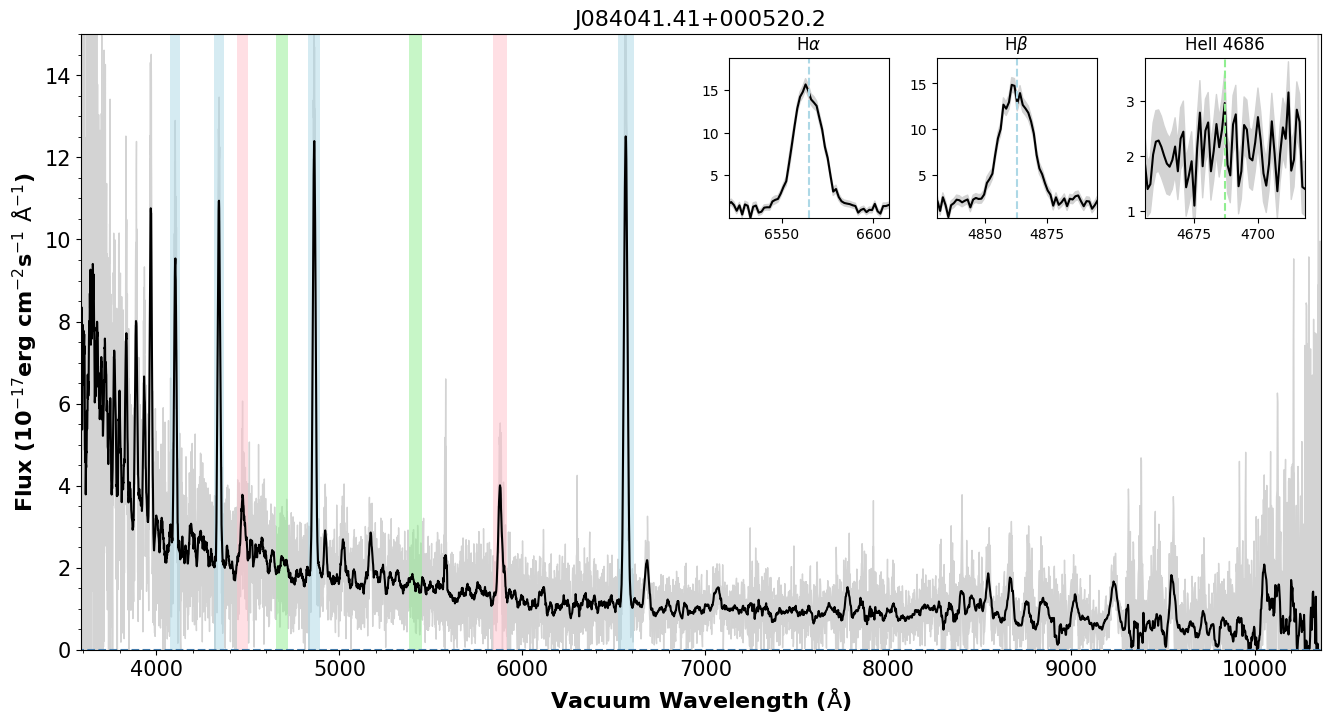}}
\end{minipage}
\hfill

\begin{minipage}[c]{0.45\hsize}
\resizebox{\hsize}{!}{\includegraphics{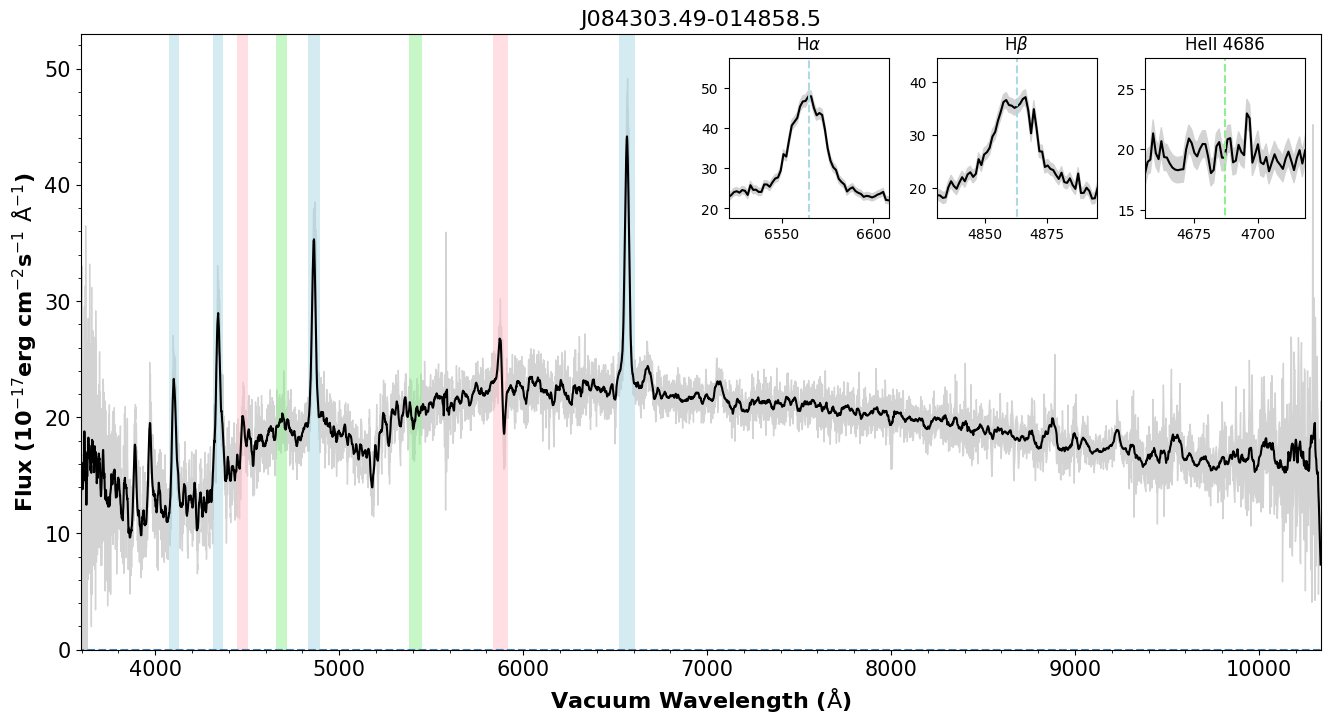}}
\end{minipage}
\hfill
\begin{minipage}[c]{0.321\hsize}
\resizebox{\hsize}{!}{\includegraphics{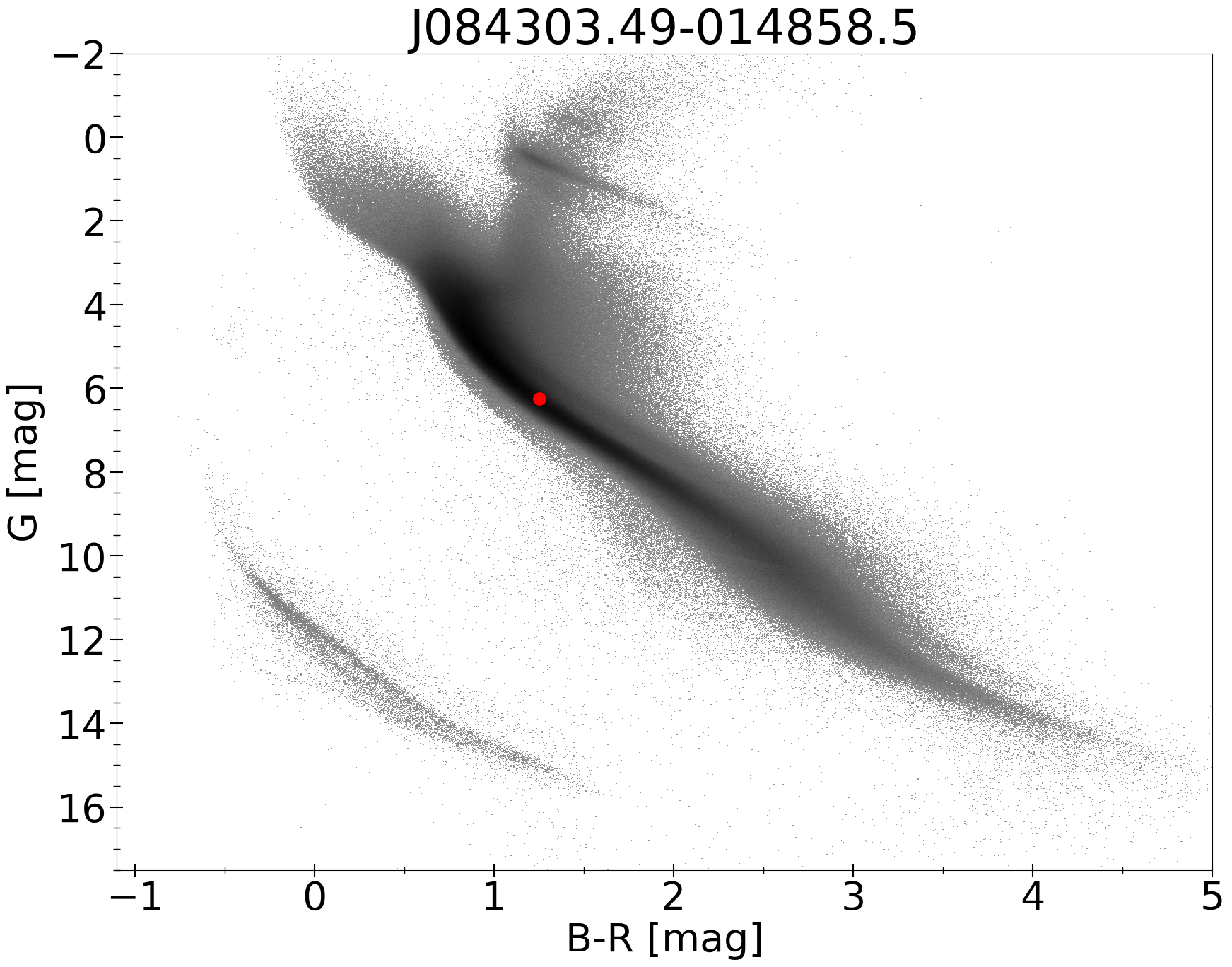}}
\end{minipage}



\begin{minipage}[c]{0.45\hsize}
\resizebox{\hsize}{!}{\includegraphics{specpngs/CV_self_spec-00000-59345-04545021122.png}}
\end{minipage}
\hfill
\begin{minipage}[c]{0.321\hsize}
\resizebox{\hsize}{!}{\includegraphics{hrds/hrd_131-0004_2-6554.png}}
\end{minipage}

\begin{minipage}[c]{0.45\hsize}
\resizebox{\hsize}{!}{\includegraphics{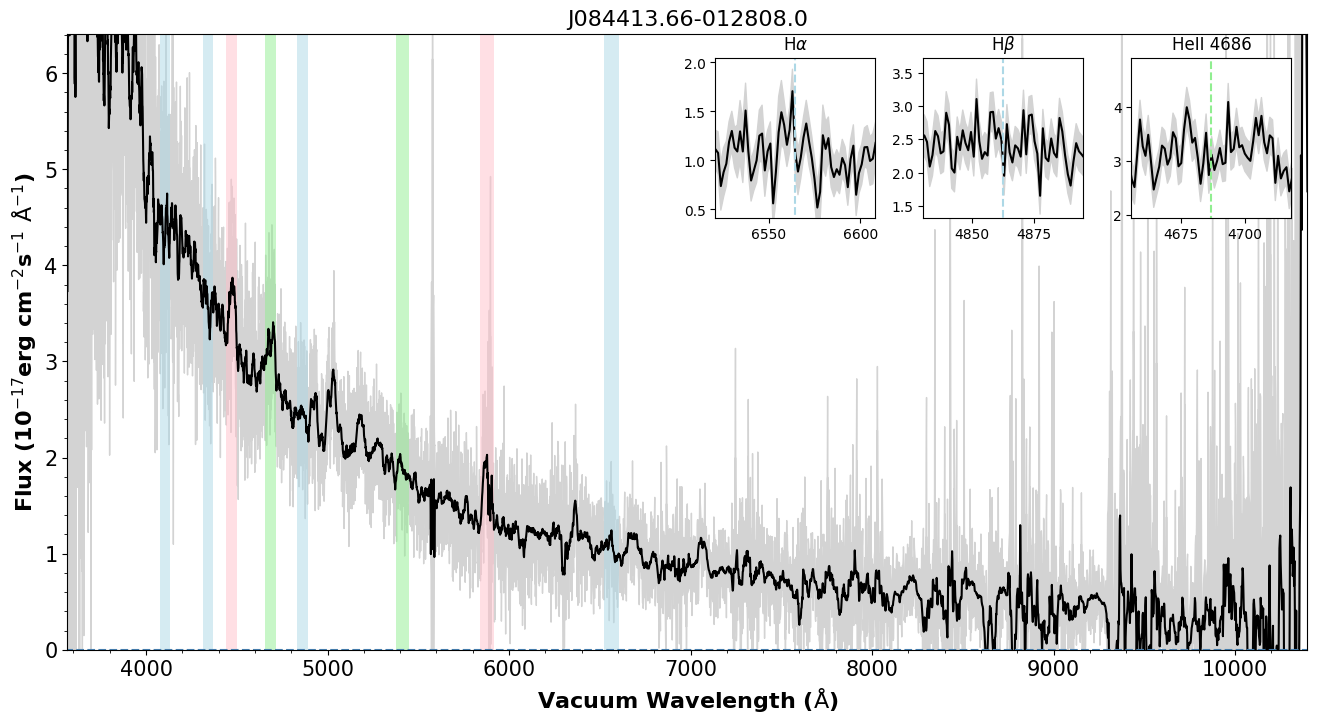}}
\end{minipage}
\hfill
\begin{minipage}[c]{0.321\hsize}
\resizebox{\hsize}{!}{\includegraphics{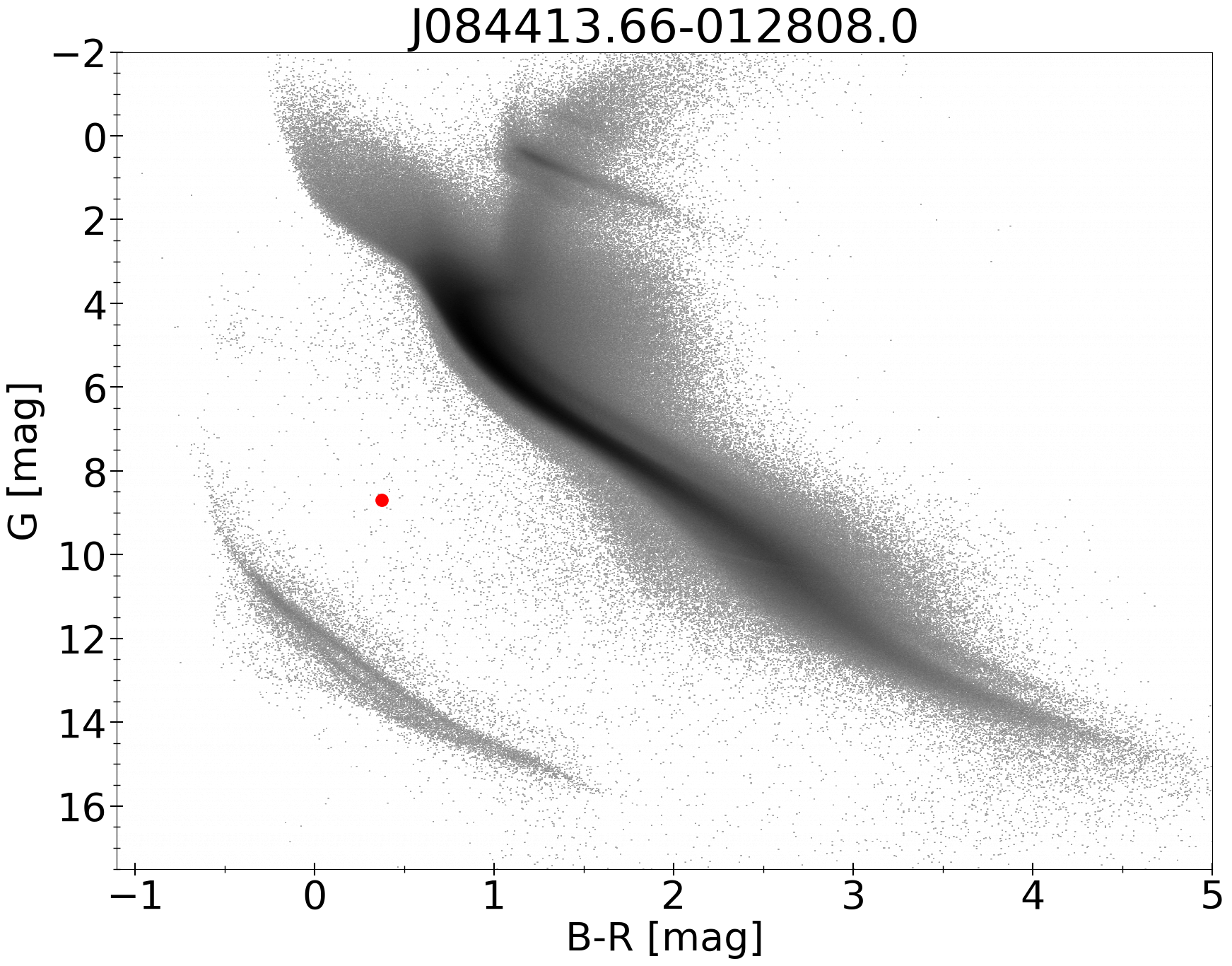}}
\end{minipage}

\begin{minipage}[c]{0.45\hsize}
\resizebox{\hsize}{!}{\includegraphics{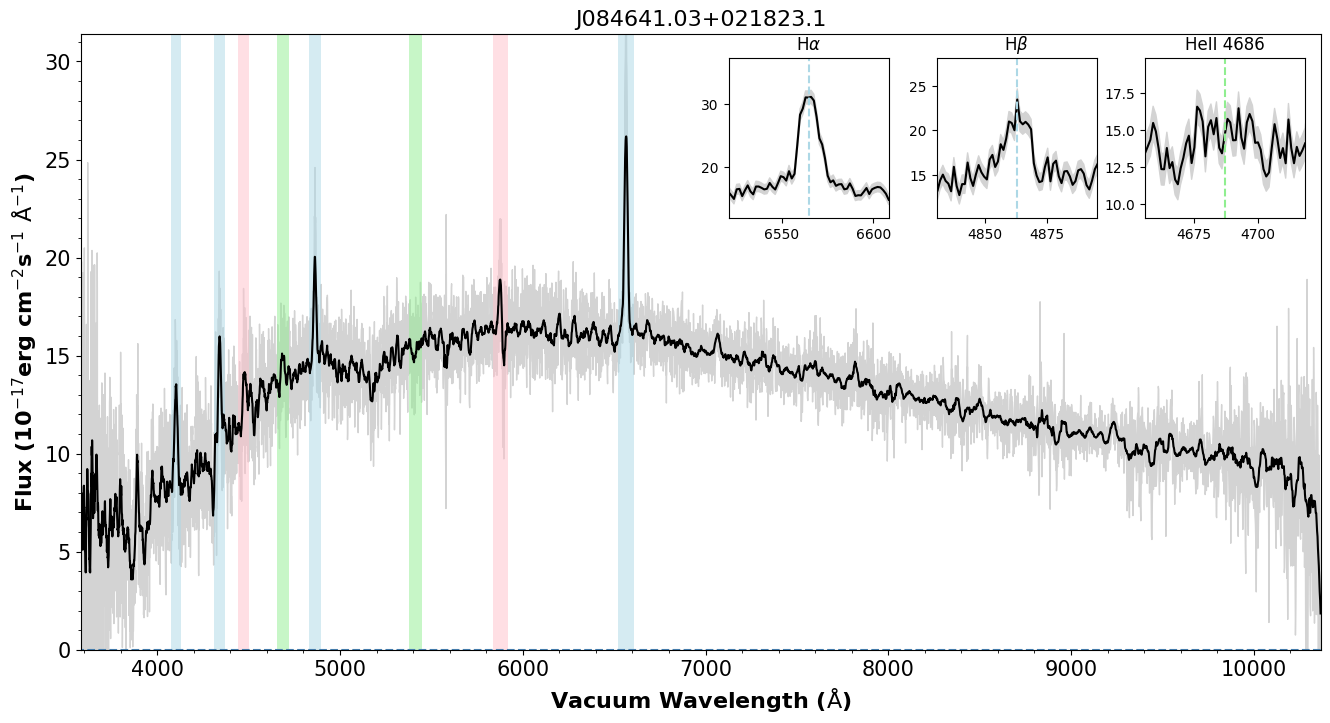}}
\end{minipage}
\hfill
\hfill
\begin{minipage}[c]{0.321\hsize}
\resizebox{\hsize}{!}{\includegraphics{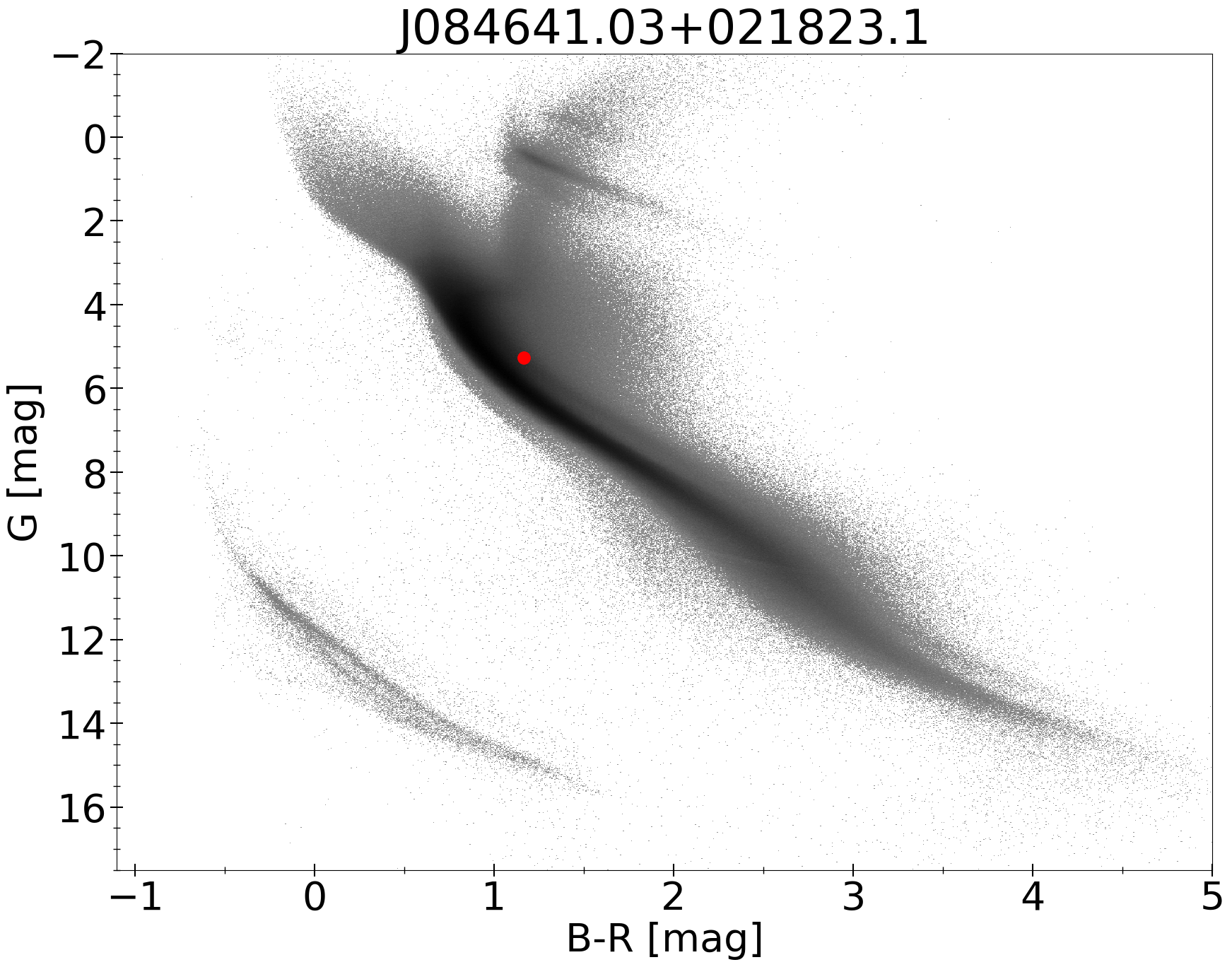}}
\end{minipage}

\end{figure*}



\begin{figure*}[]
\begin{minipage}[c]{0.45\hsize}
\resizebox{\hsize}{!}{\includegraphics{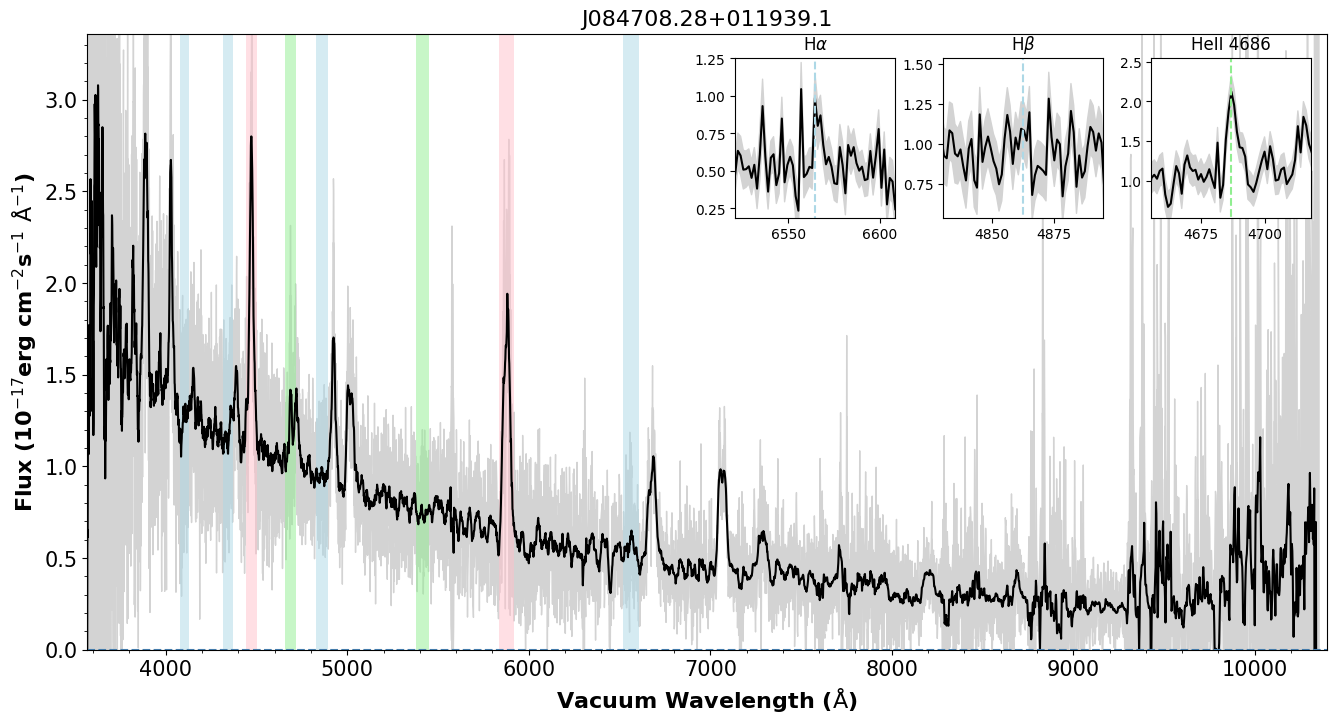}}
\end{minipage}
\hfill

\begin{minipage}[c]{0.45\hsize}
\resizebox{\hsize}{!}{\includegraphics{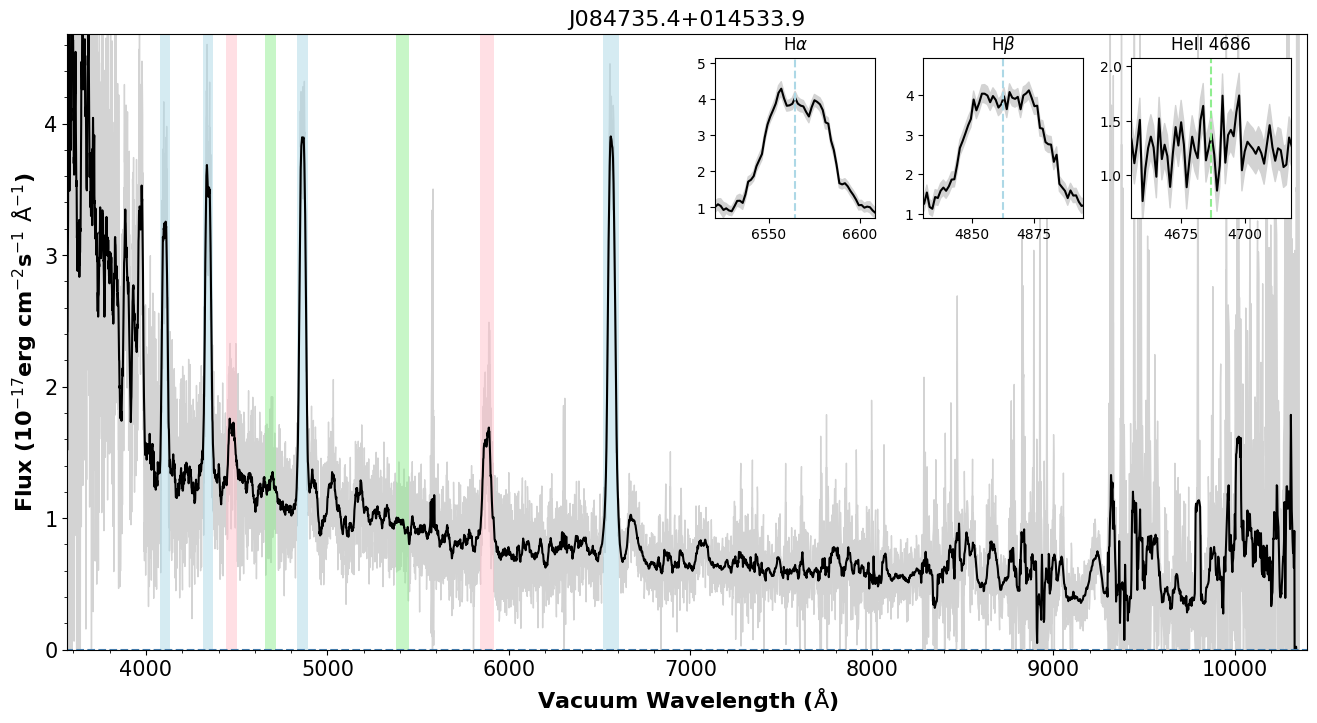}}
\end{minipage}
\hfill

\hfill
\begin{minipage}[c]{0.321\hsize}
\resizebox{\hsize}{!}{\includegraphics{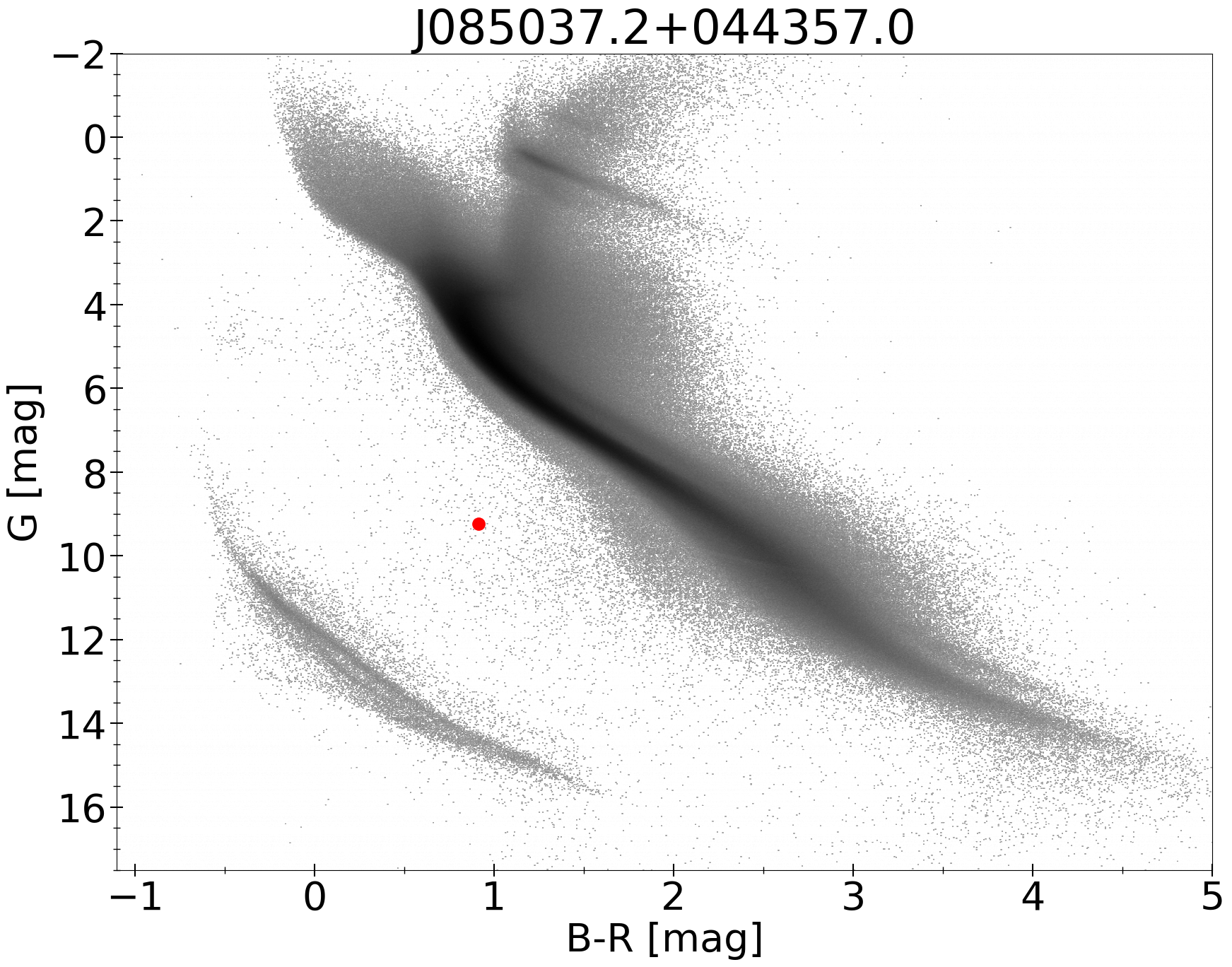}}
\end{minipage}

\begin{minipage}[c]{0.45\hsize}
\resizebox{\hsize}{!}{\includegraphics{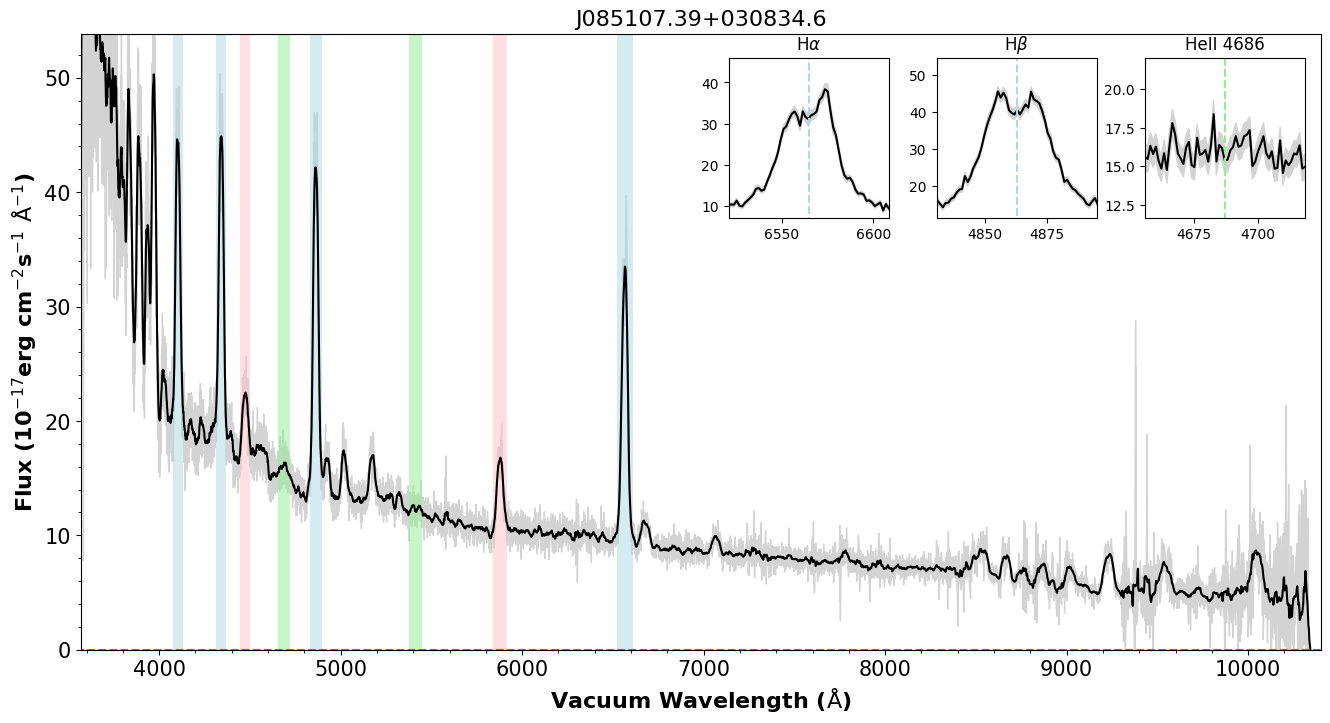}}
\end{minipage}
\hfill
\begin{minipage}[c]{0.321\hsize}
\resizebox{\hsize}{!}{\includegraphics{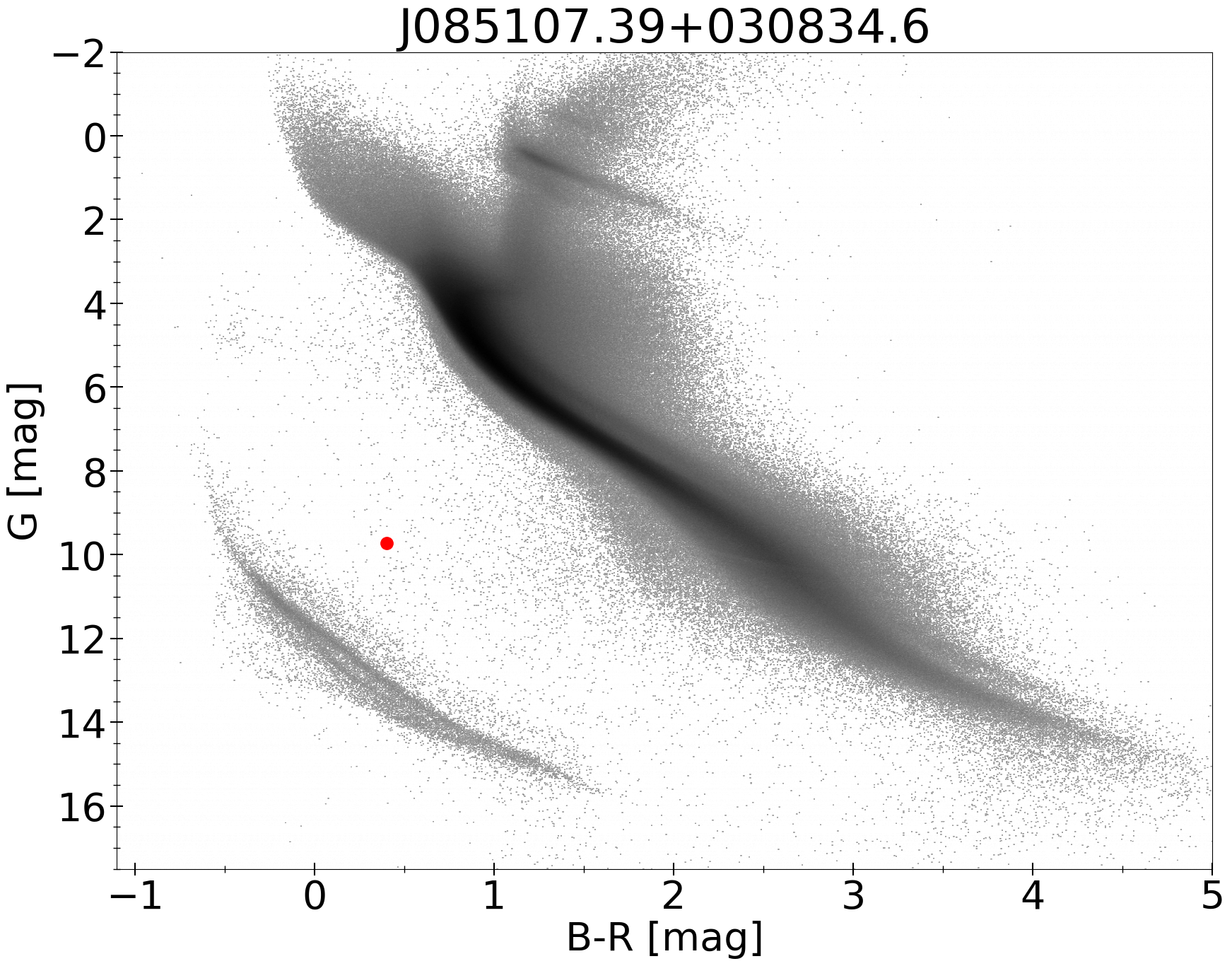}}
\end{minipage}

\begin{minipage}[c]{0.45\hsize}
\resizebox{\hsize}{!}{\includegraphics{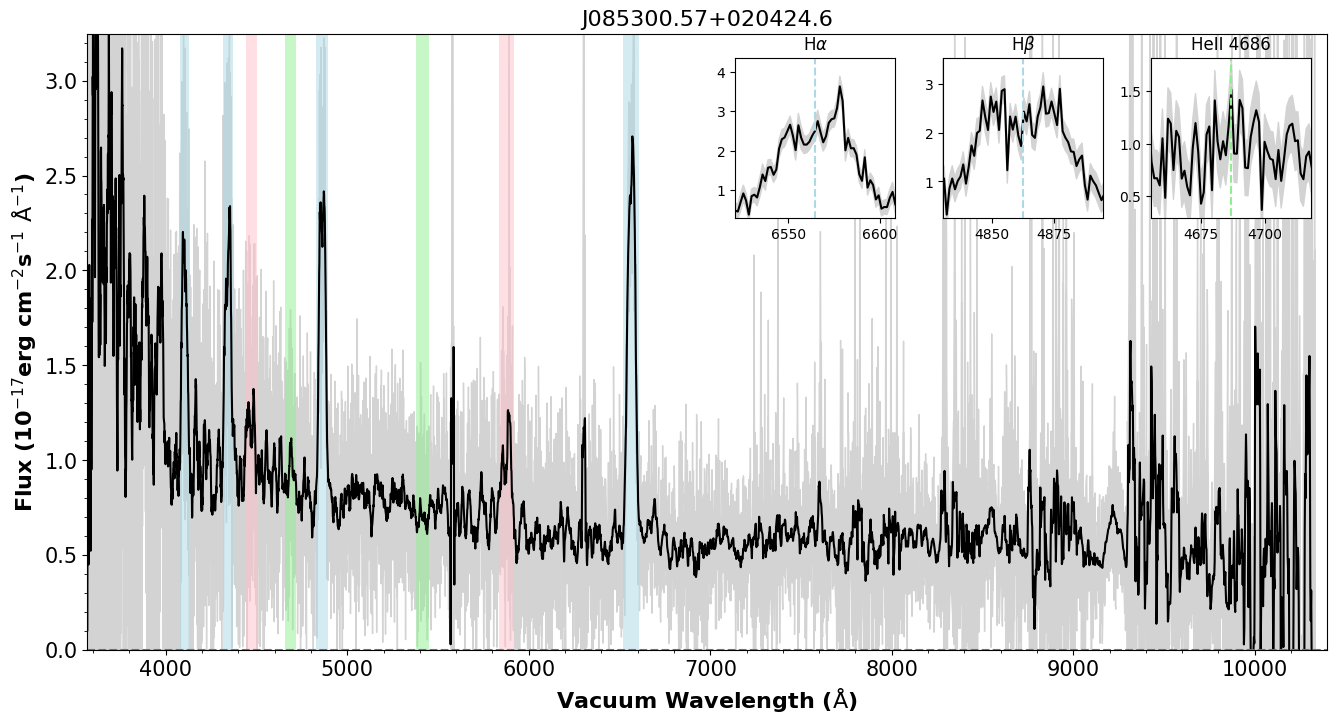}}
\end{minipage}
\hfill

\end{figure*}


\begin{figure*}[t]

\begin{minipage}[c]{0.45\hsize}
\resizebox{\hsize}{!}{\includegraphics{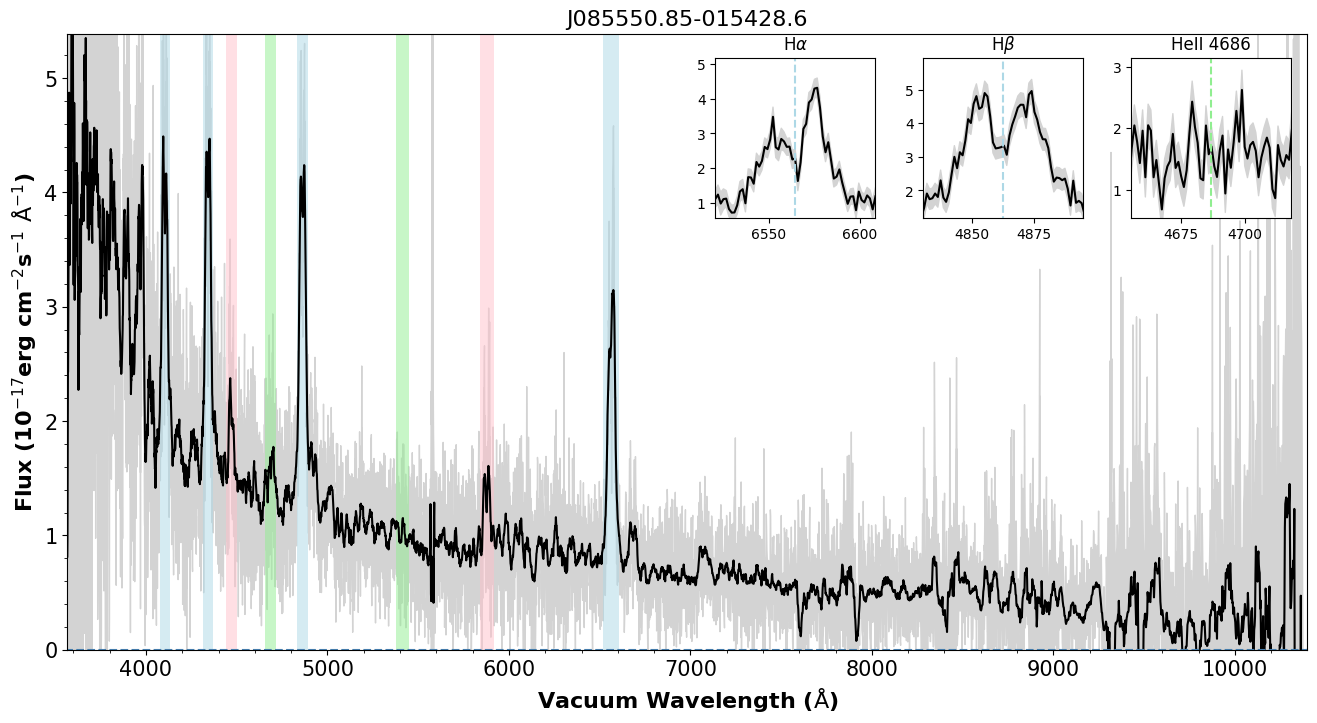}}
\end{minipage}
\hfill
\begin{minipage}[c]{0.321\hsize}
\resizebox{\hsize}{!}{\includegraphics{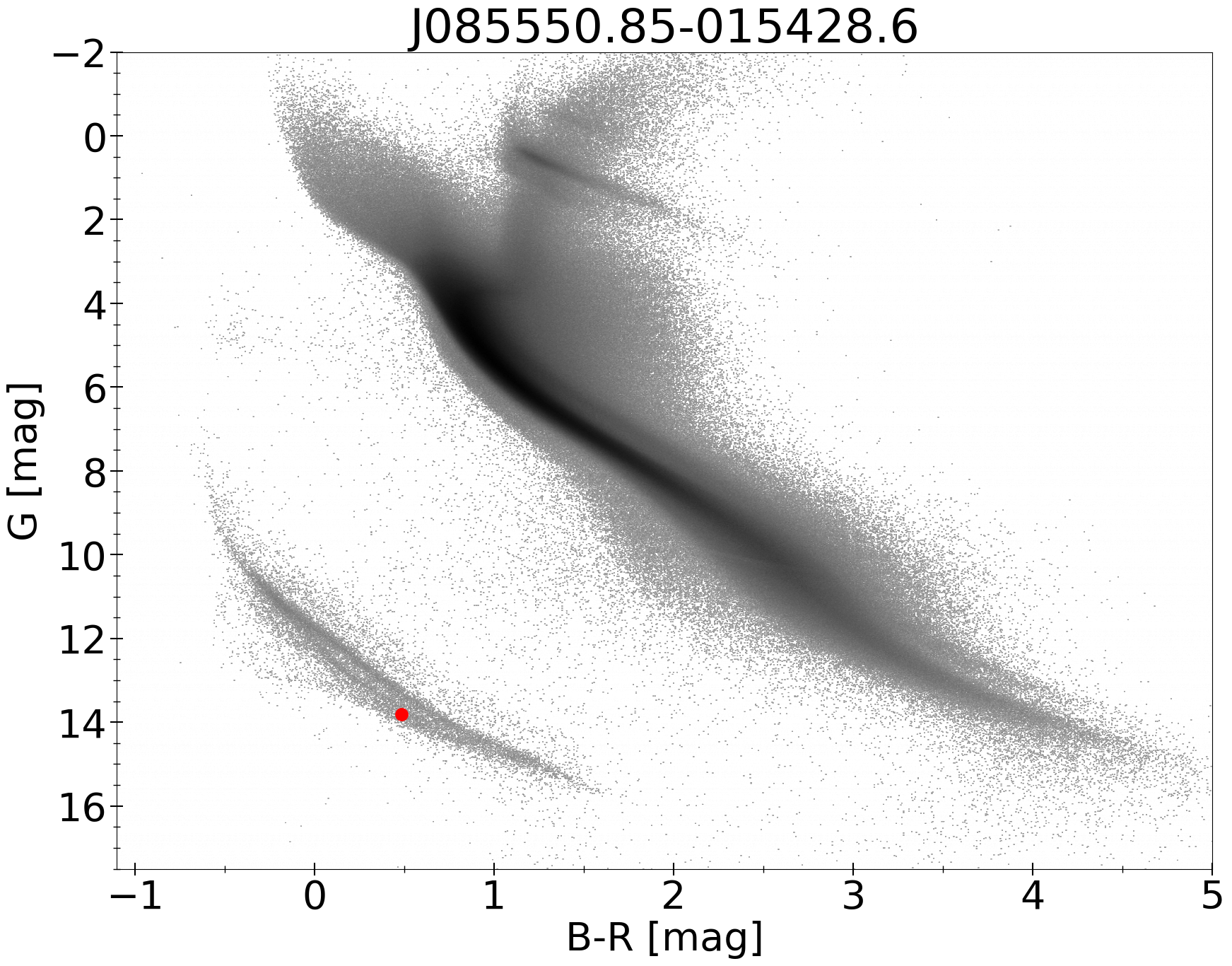}}
\end{minipage}

\begin{minipage}[c]{0.45\hsize}
\resizebox{\hsize}{!}{\includegraphics{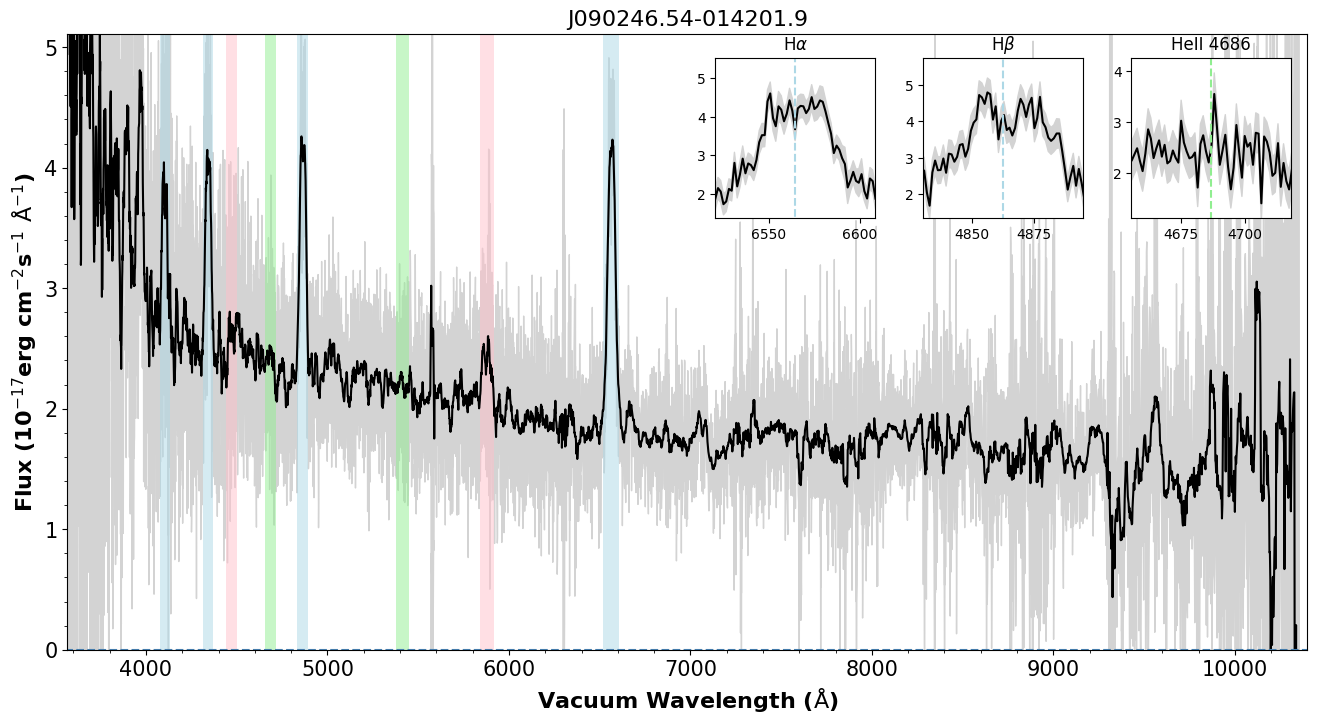}}
\end{minipage}
\hfill
\begin{minipage}[c]{0.321\hsize}
\resizebox{\hsize}{!}{\includegraphics{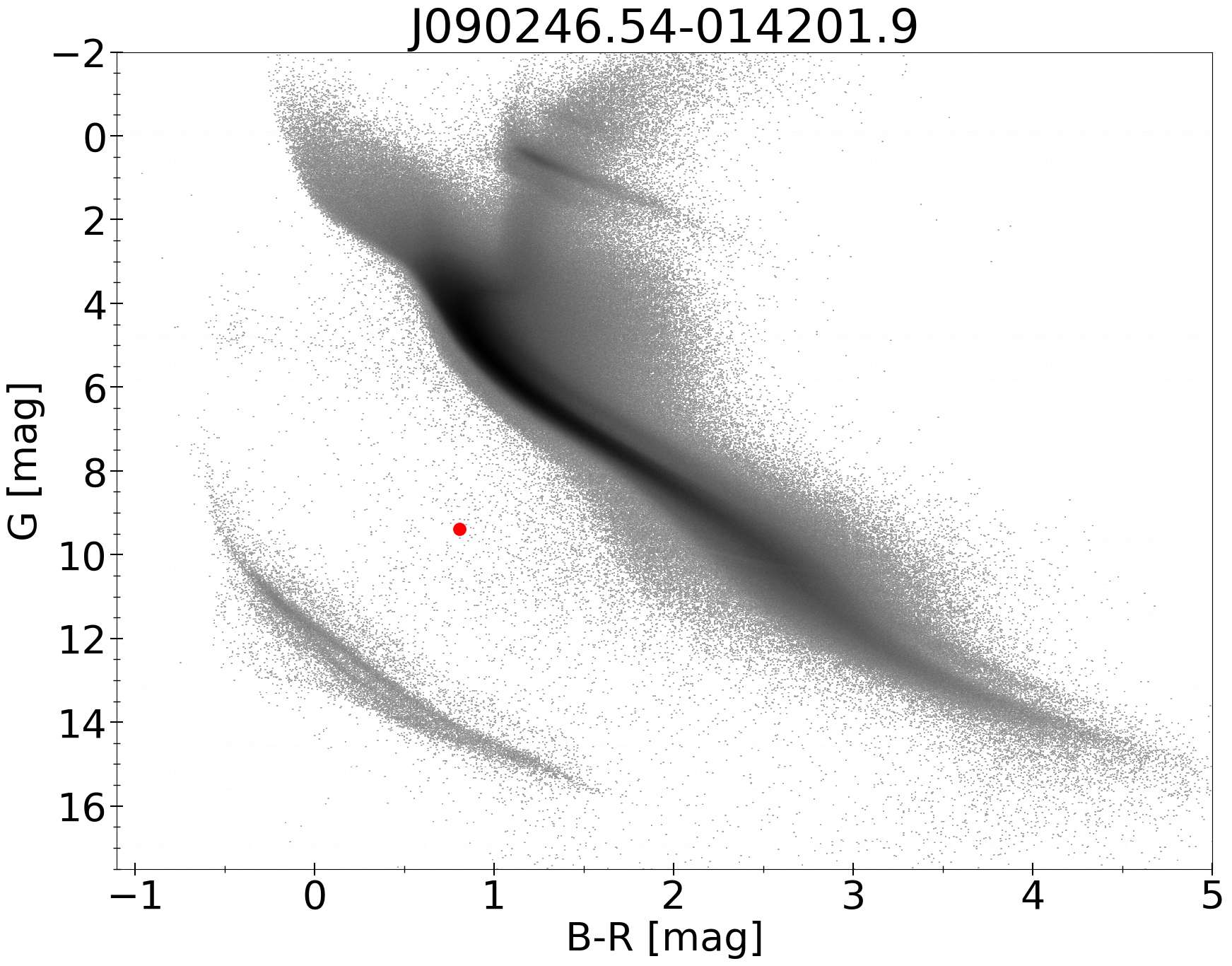}}
\end{minipage}

\begin{minipage}[c]{0.45\hsize}
\resizebox{\hsize}{!}{\includegraphics{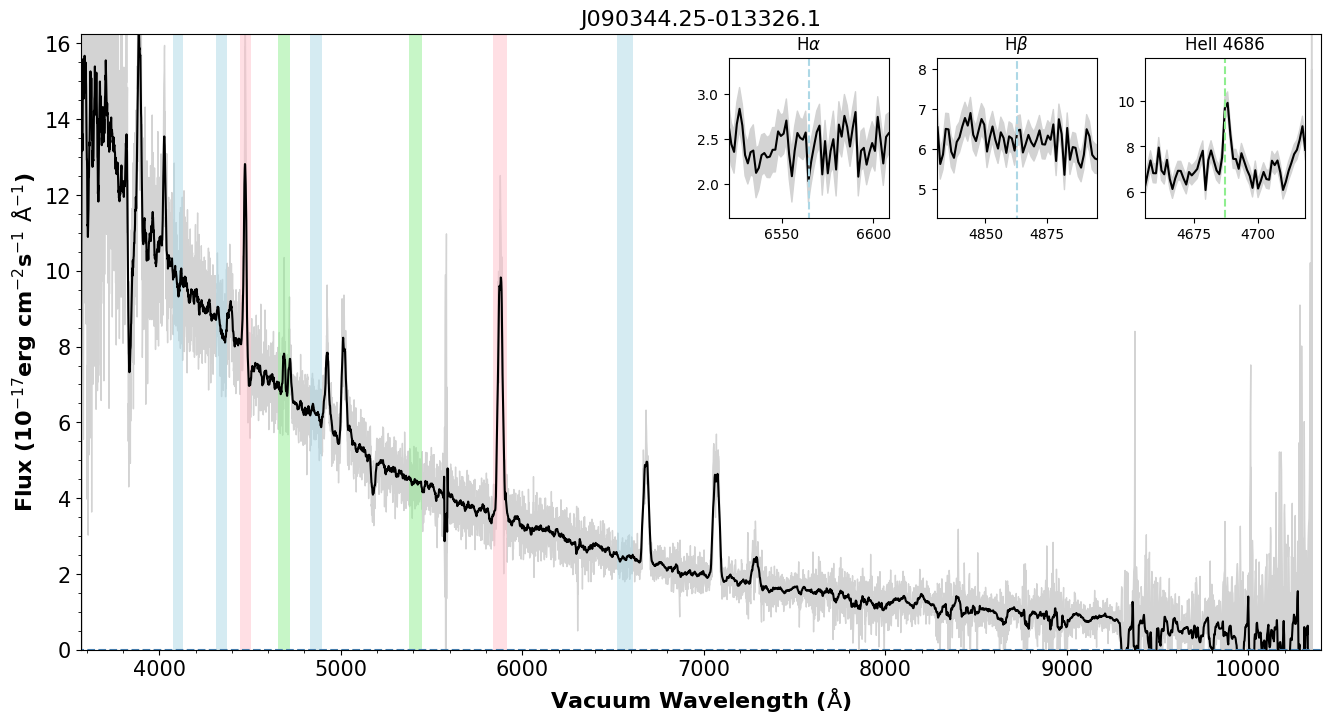}}
\end{minipage}
\hfill
\begin{minipage}[c]{0.321\hsize}
\resizebox{\hsize}{!}{\includegraphics{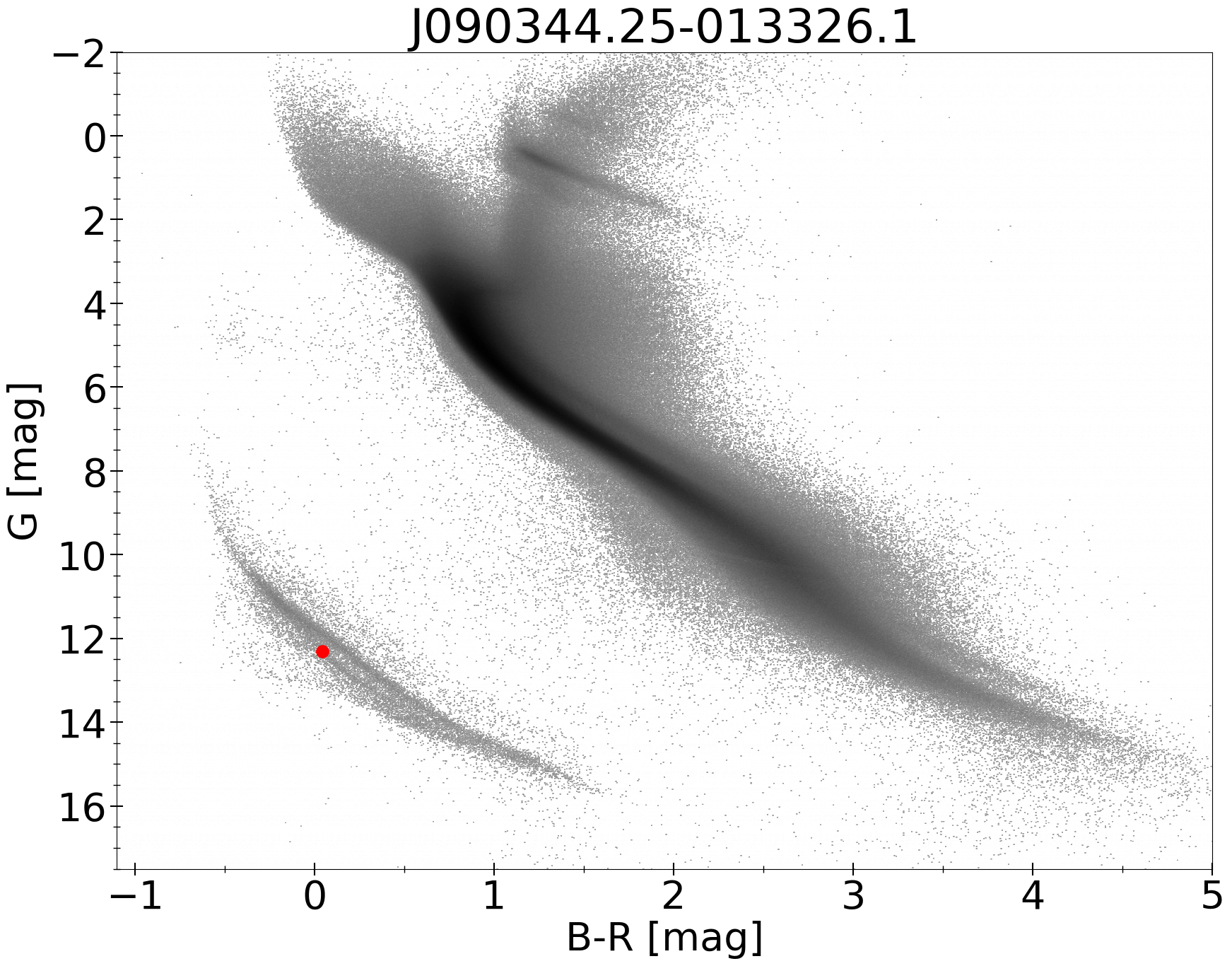}}
\end{minipage}

\begin{minipage}[c]{0.45\hsize}
\resizebox{\hsize}{!}{\includegraphics{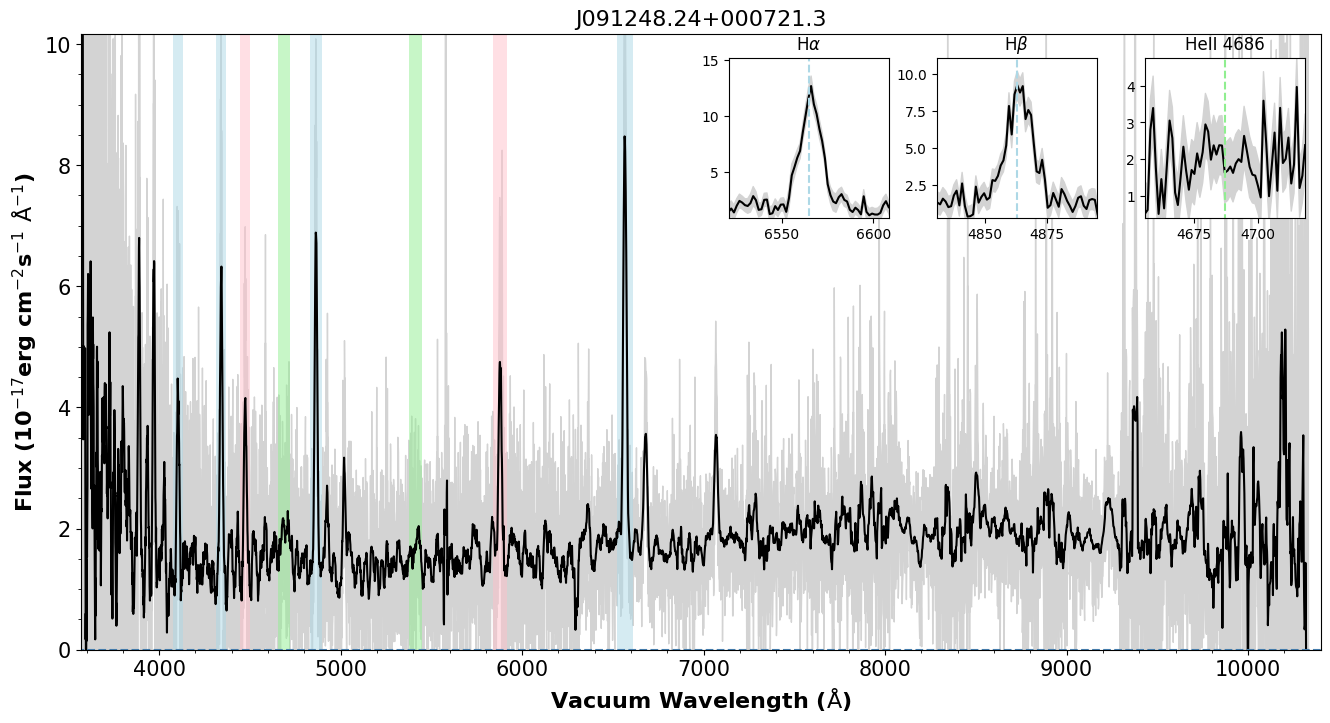}}
\end{minipage}
\hfill
\begin{minipage}[c]{0.321\hsize}
\resizebox{\hsize}{!}{\includegraphics{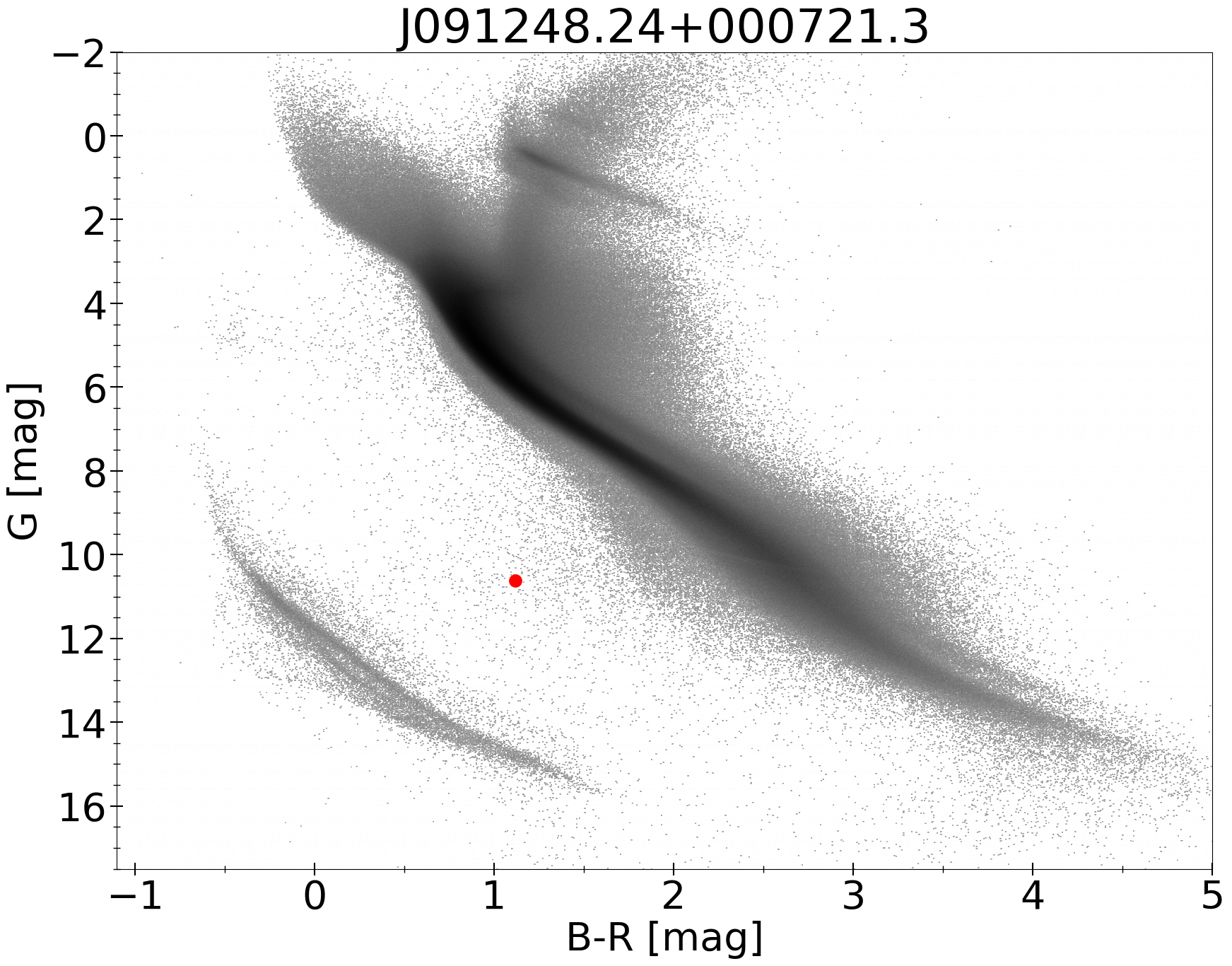}}
\end{minipage}

\begin{minipage}[c]{0.45\hsize}
\resizebox{\hsize}{!}{\includegraphics{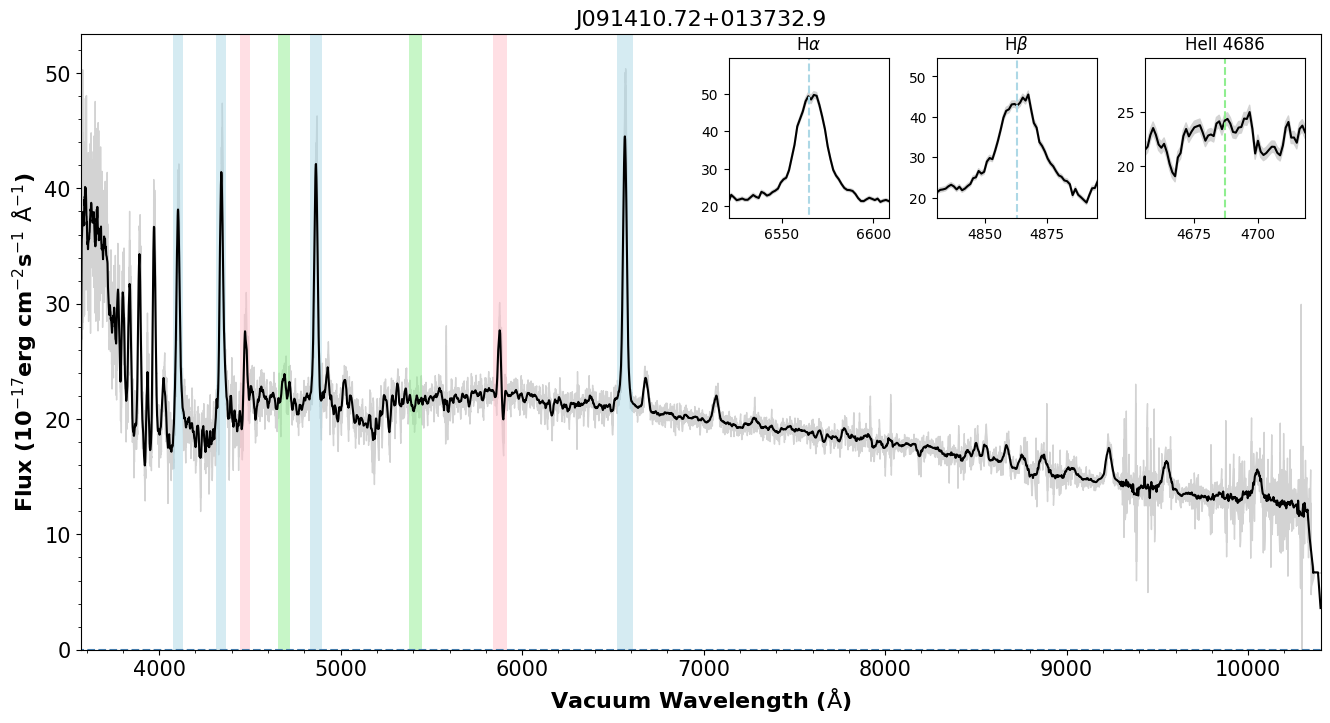}}
\end{minipage}
\hfill
\begin{minipage}[c]{0.321\hsize}
\resizebox{\hsize}{!}{\includegraphics{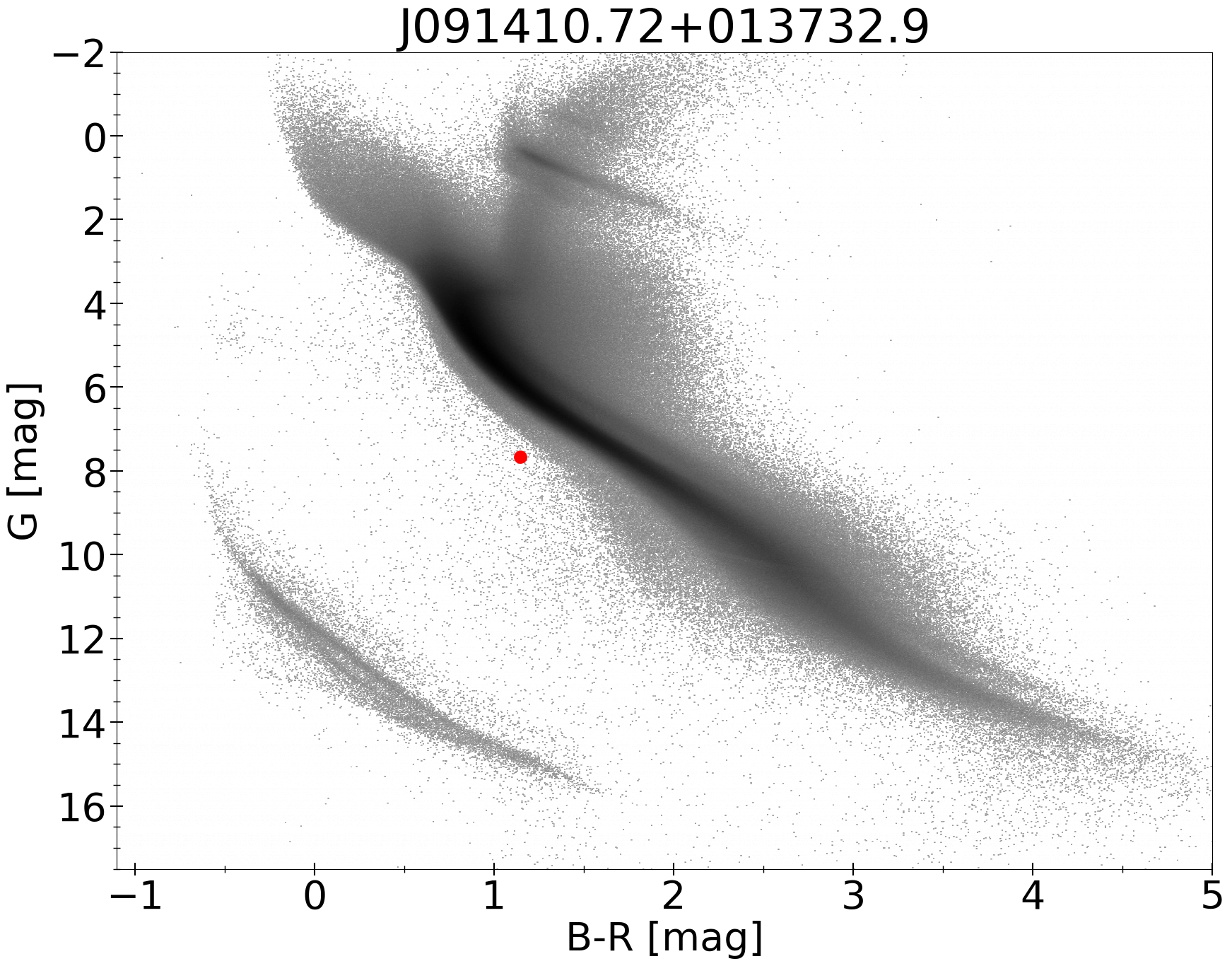}}
\end{minipage}

\end{figure*}

\begin{figure*}[t]

\begin{minipage}[c]{0.45\hsize}
\resizebox{\hsize}{!}{\includegraphics{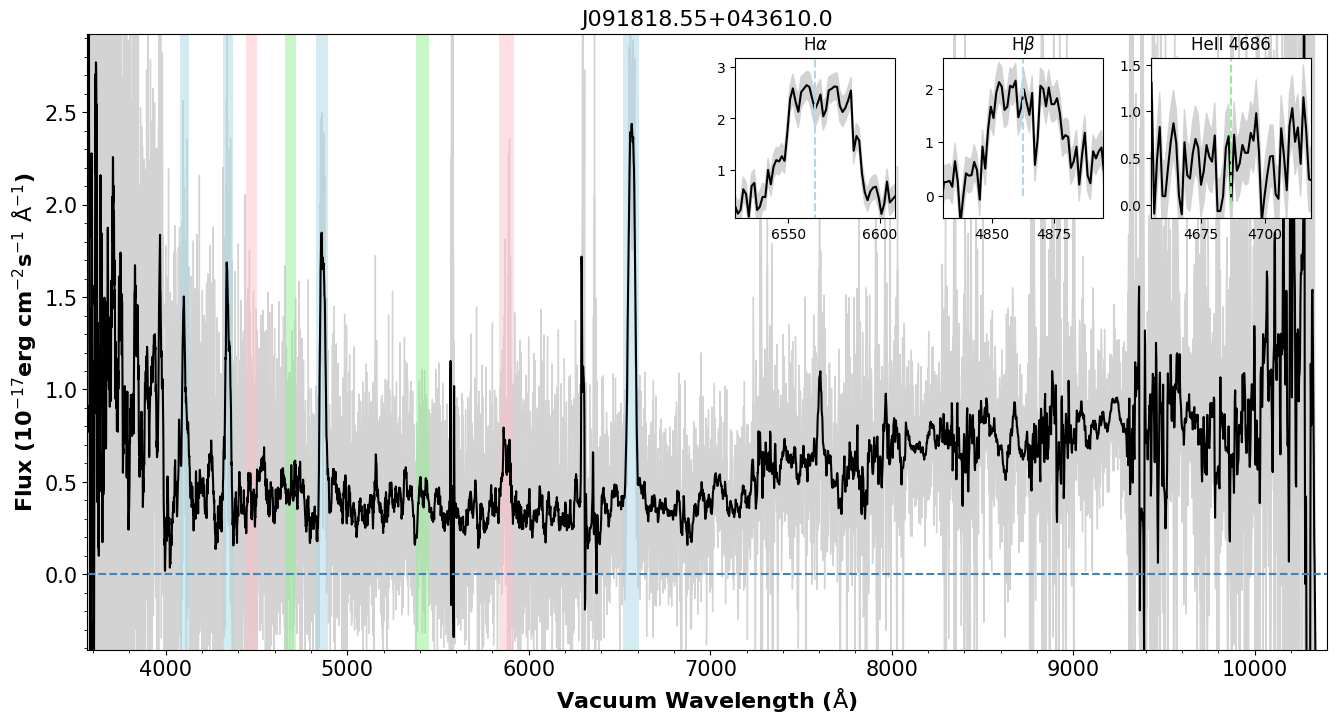}}
\end{minipage}
\hfill

\begin{minipage}[c]{0.45\hsize}
\resizebox{\hsize}{!}{\includegraphics{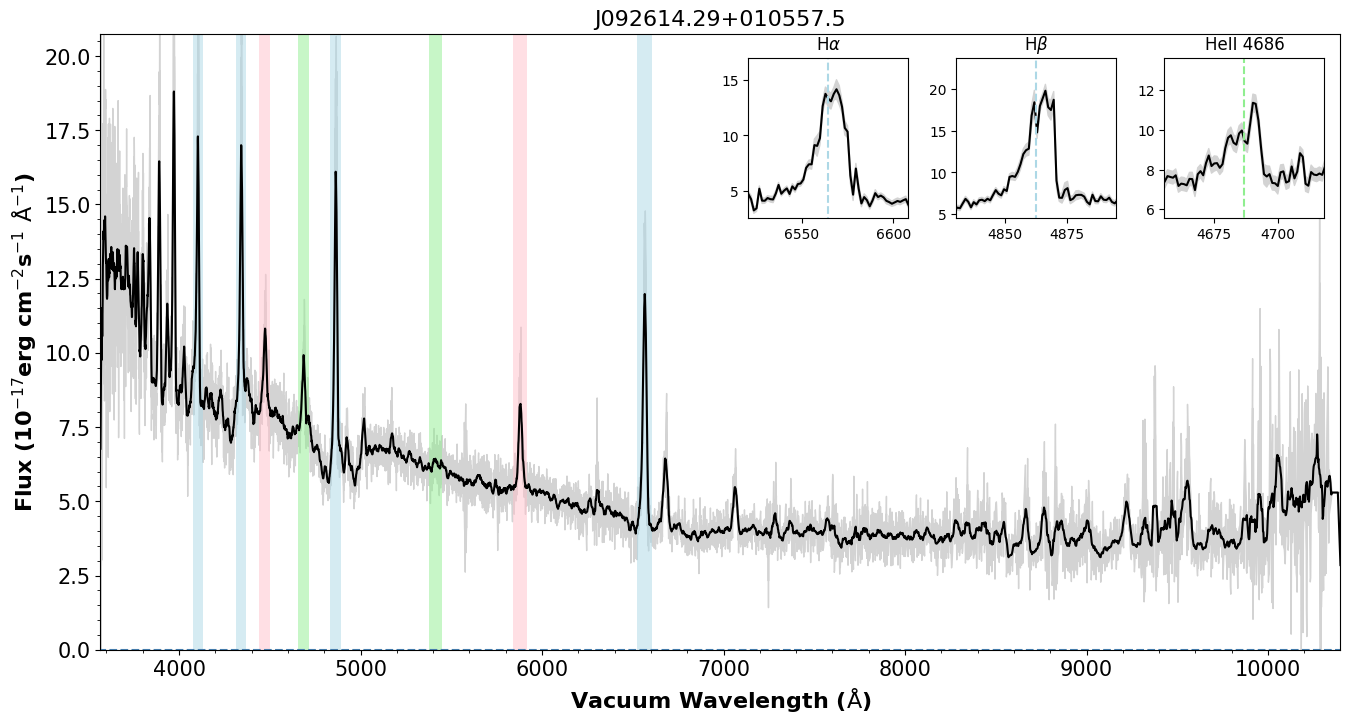}}
\end{minipage}
\hfill
\begin{minipage}[c]{0.321\hsize}
\resizebox{\hsize}{!}{\includegraphics{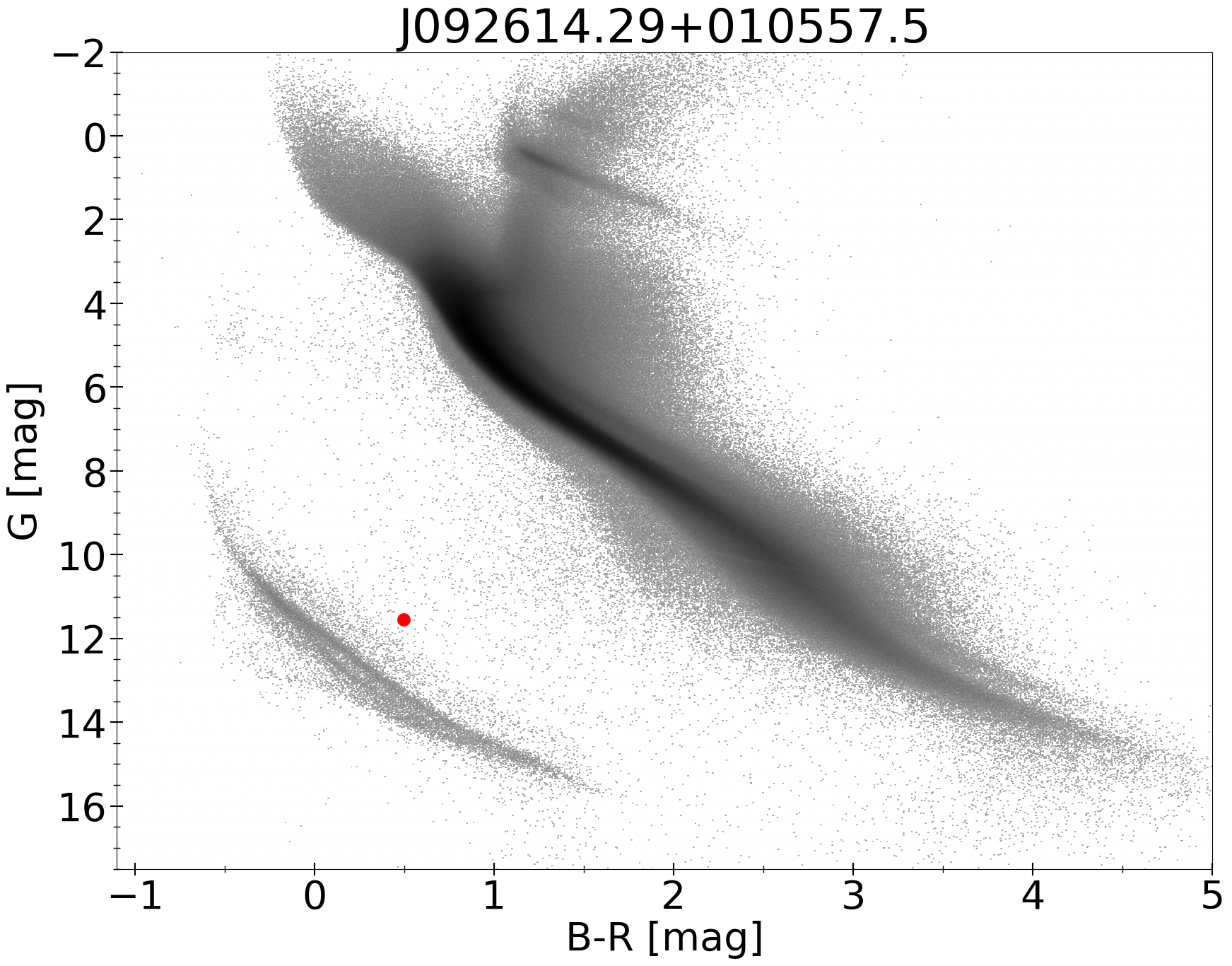}}
\end{minipage}

\begin{minipage}[c]{0.45\hsize}
\resizebox{\hsize}{!}{\includegraphics{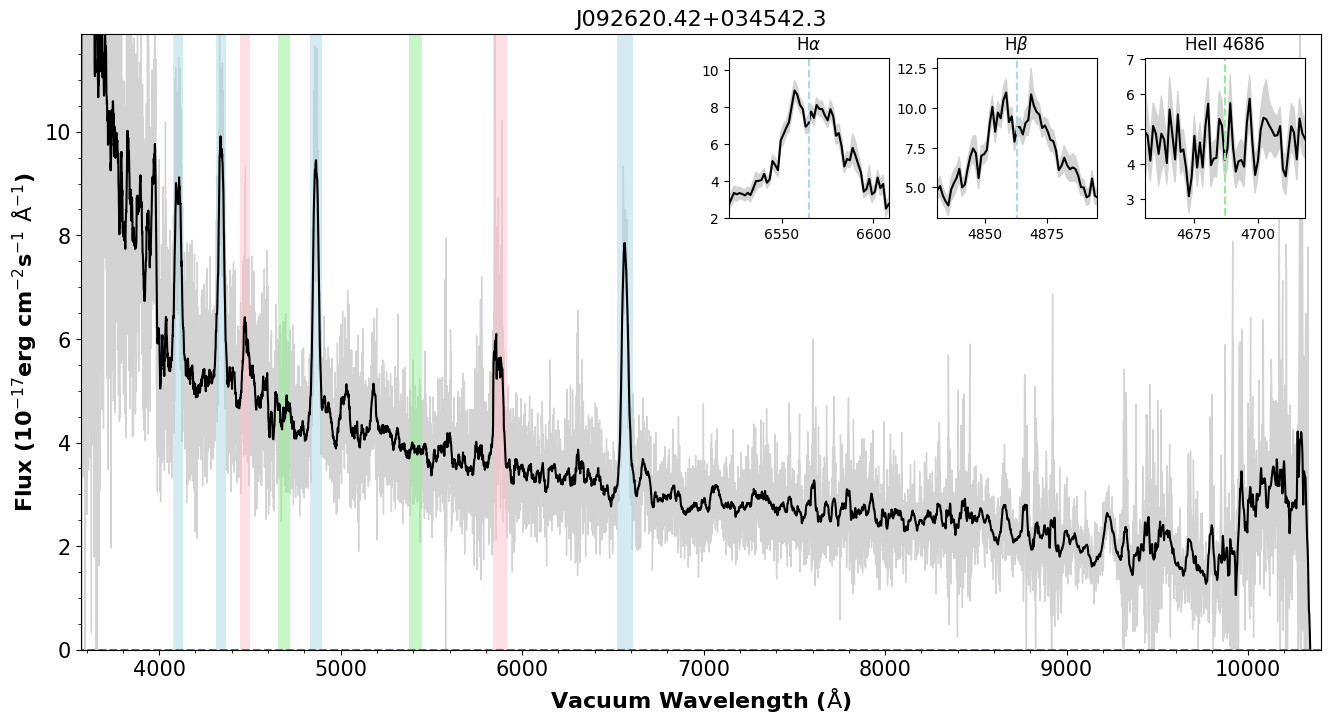}}
\end{minipage}
\hfill
\begin{minipage}[c]{0.321\hsize}
\resizebox{\hsize}{!}{\includegraphics{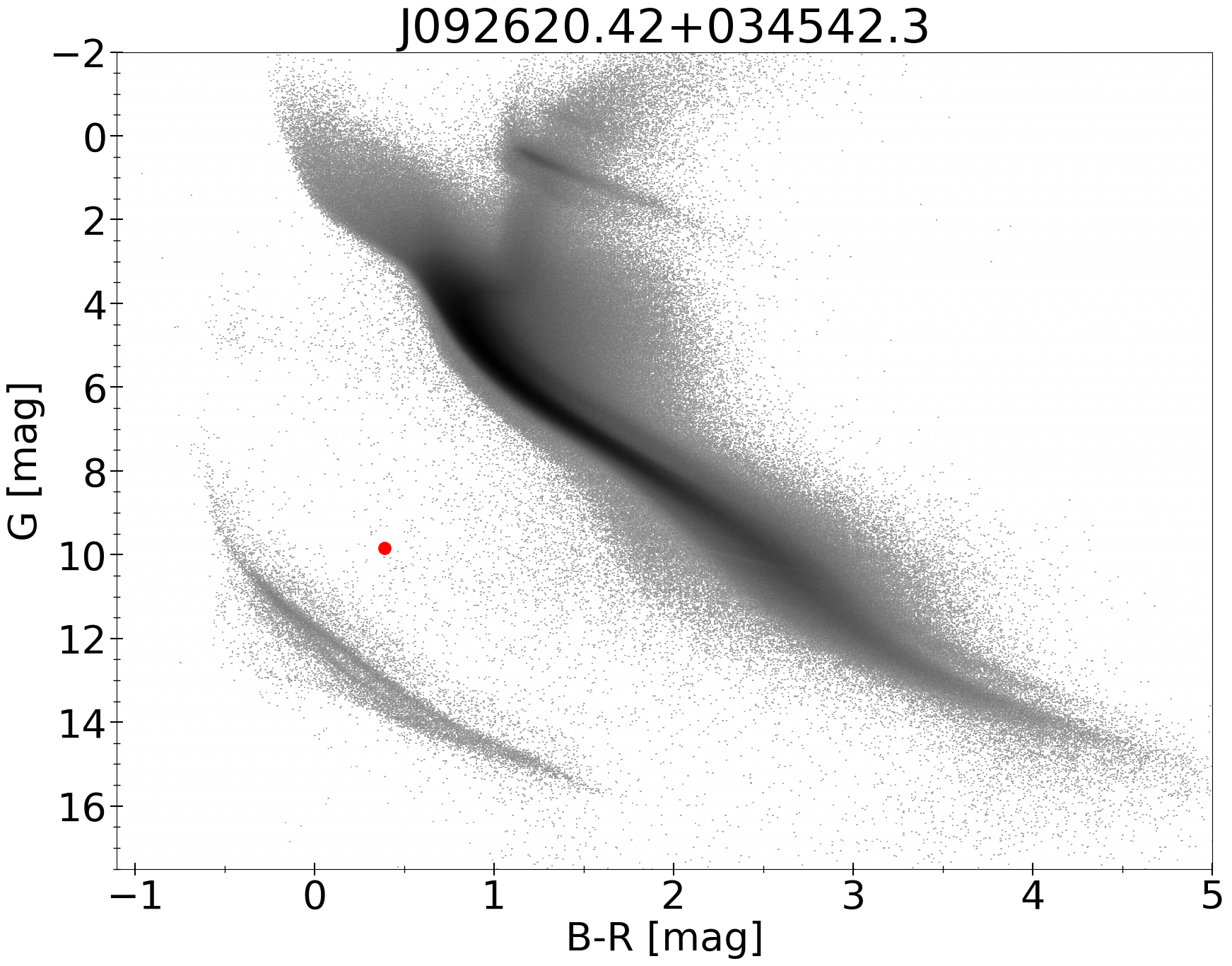}}
\end{minipage}

\begin{minipage}[c]{0.45\hsize}
\resizebox{\hsize}{!}{\includegraphics{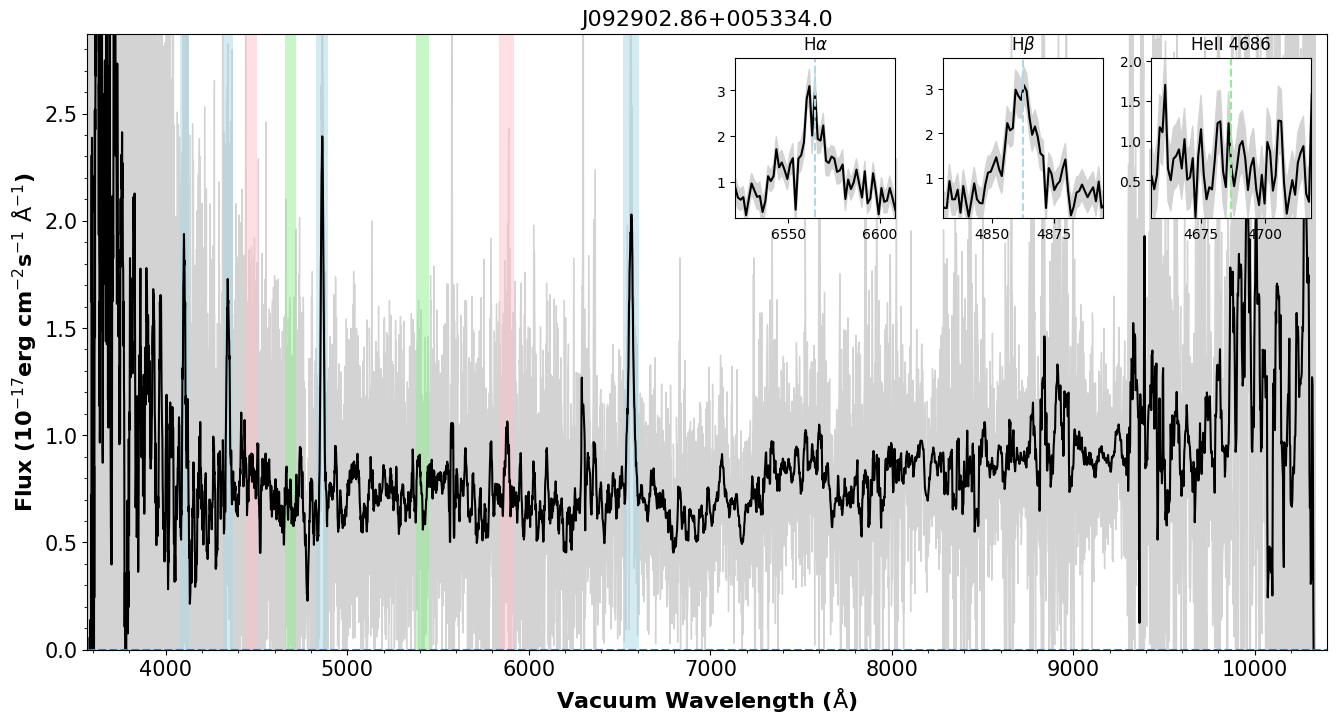}}
\end{minipage}
\hfill

\begin{minipage}[c]{0.45\hsize}
\resizebox{\hsize}{!}{\includegraphics{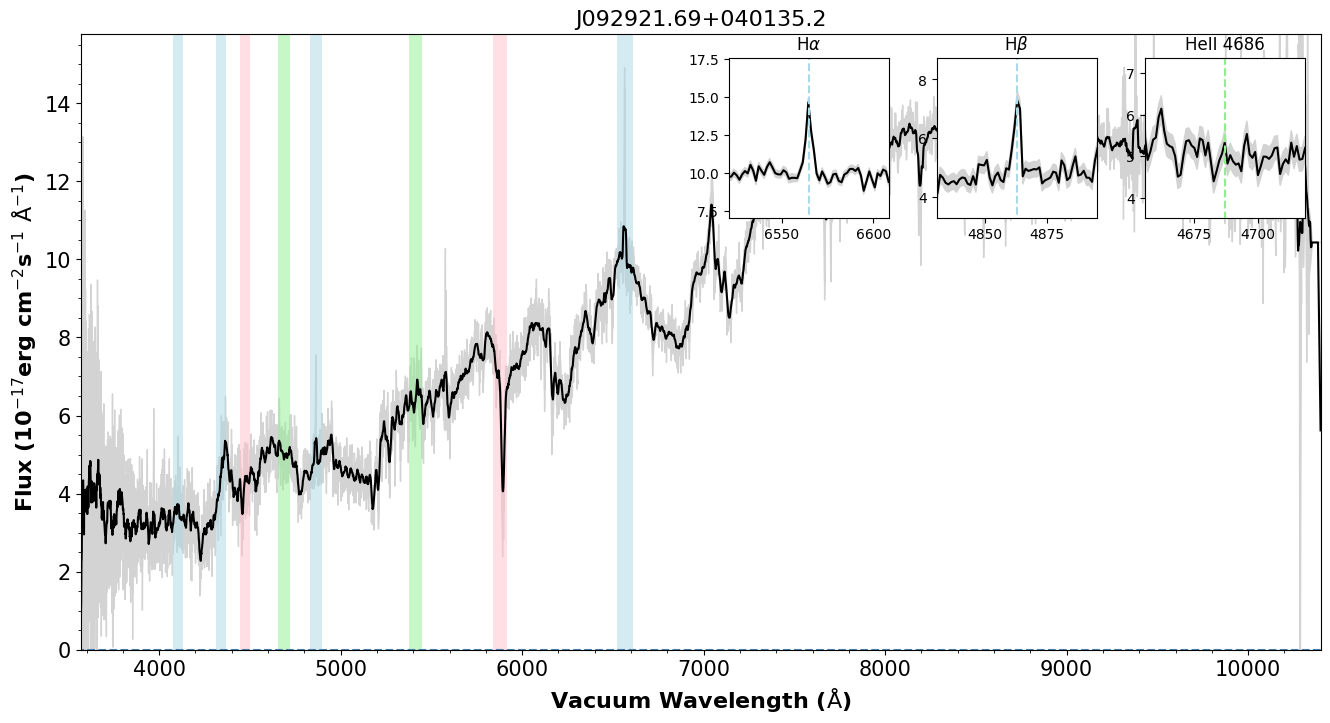}}
\end{minipage}
\hfill
\begin{minipage}[c]{0.321\hsize}
\resizebox{\hsize}{!}{\includegraphics{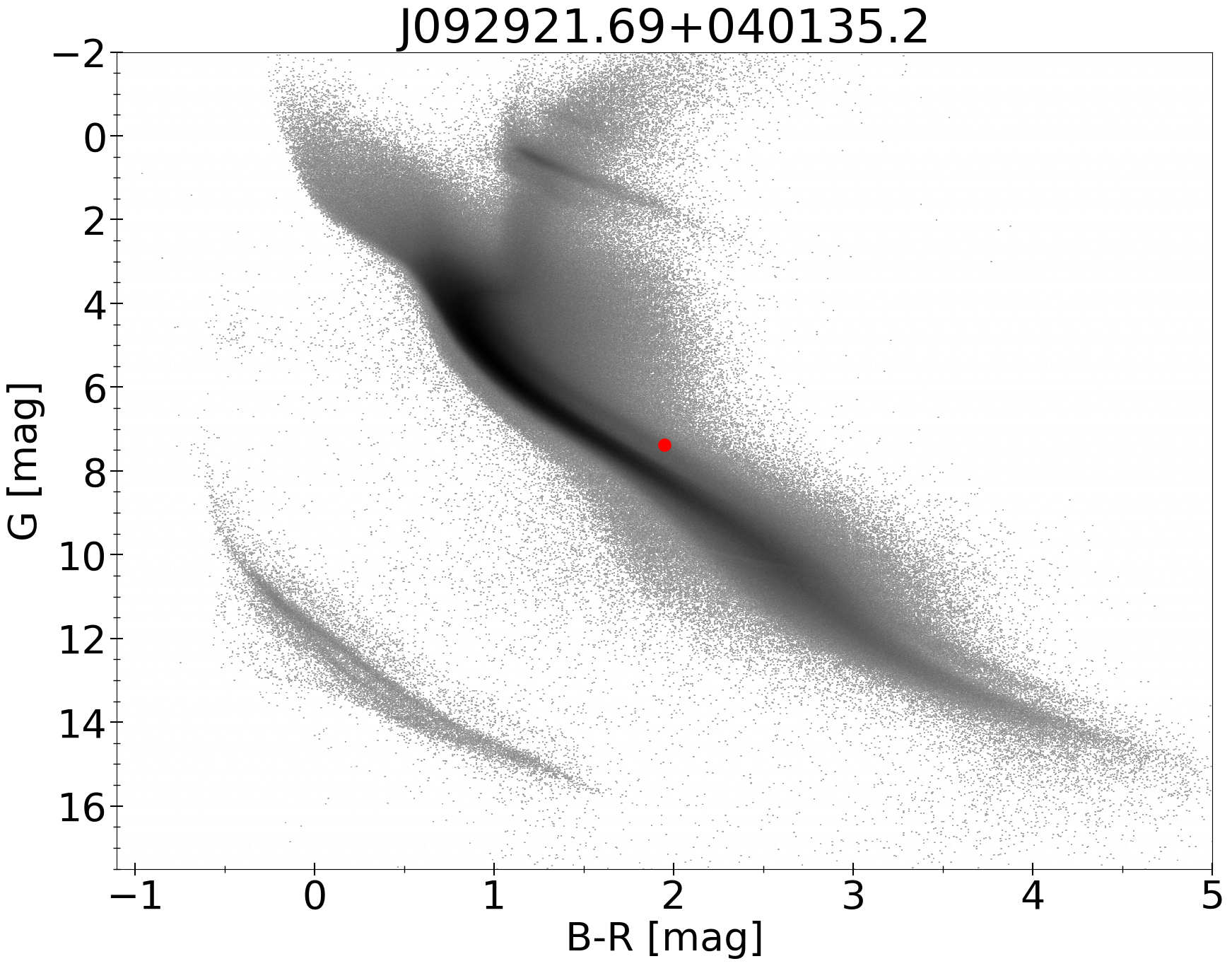}}
\end{minipage}

\end{figure*}

\begin{figure*}[t]

\begin{minipage}[c]{0.45\hsize}
\resizebox{\hsize}{!}{\includegraphics{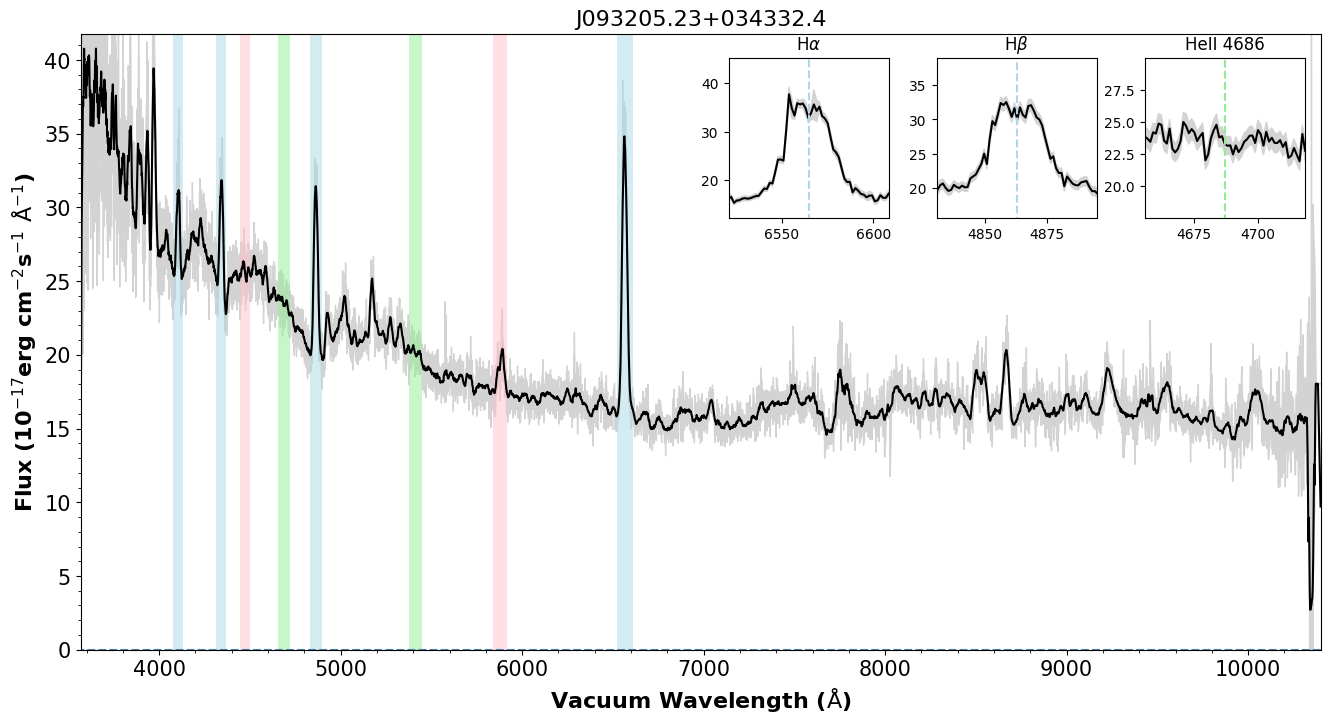}}
\end{minipage}
\hfill
\begin{minipage}[c]{0.321\hsize}
\resizebox{\hsize}{!}{\includegraphics{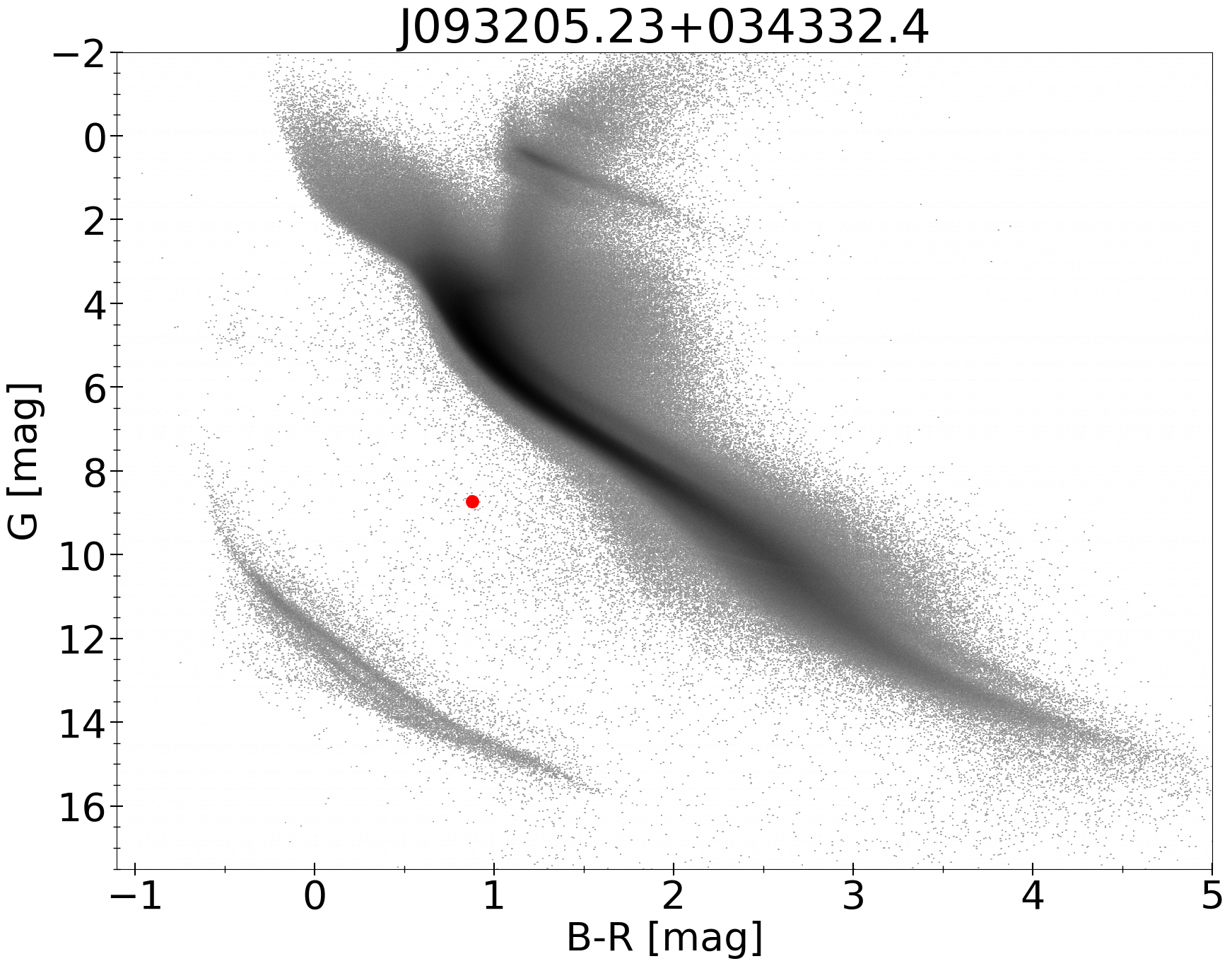}}
\end{minipage}

\begin{minipage}[c]{0.45\hsize}
\resizebox{\hsize}{!}{\includegraphics{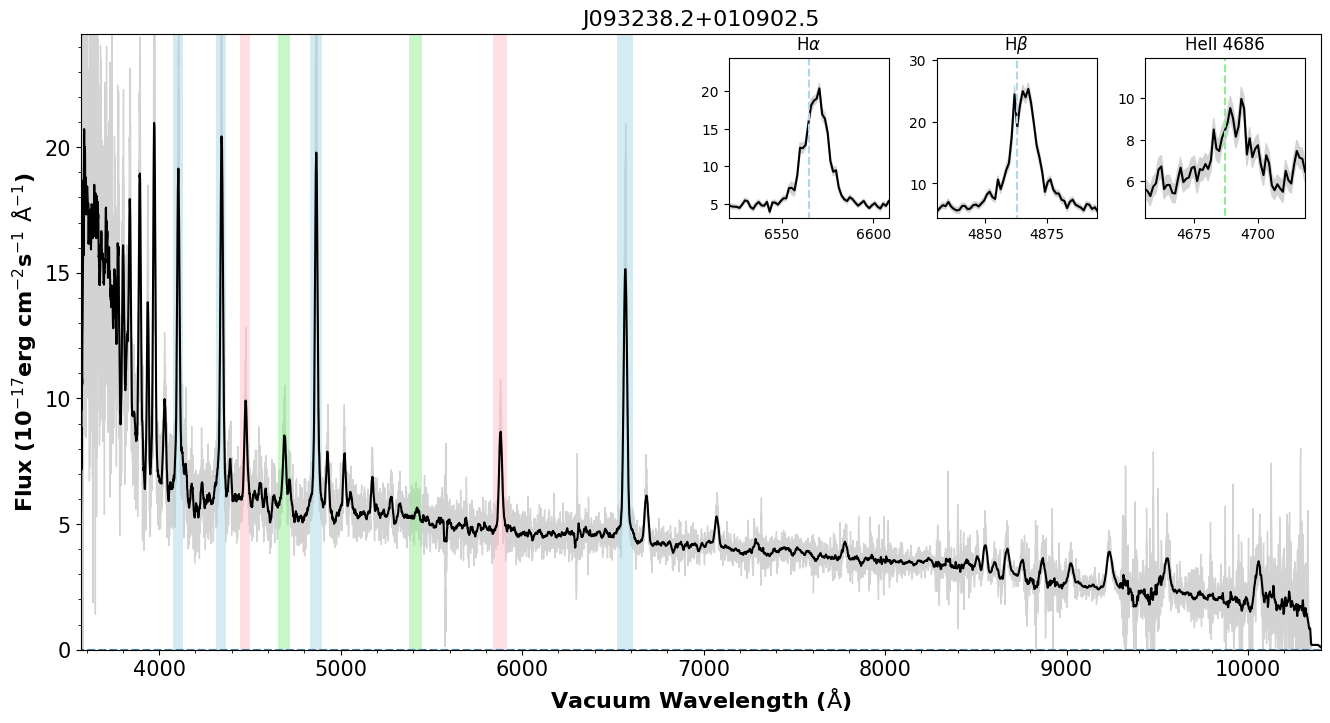}}
\end{minipage}
\hfill
\begin{minipage}[c]{0.321\hsize}
\resizebox{\hsize}{!}{\includegraphics{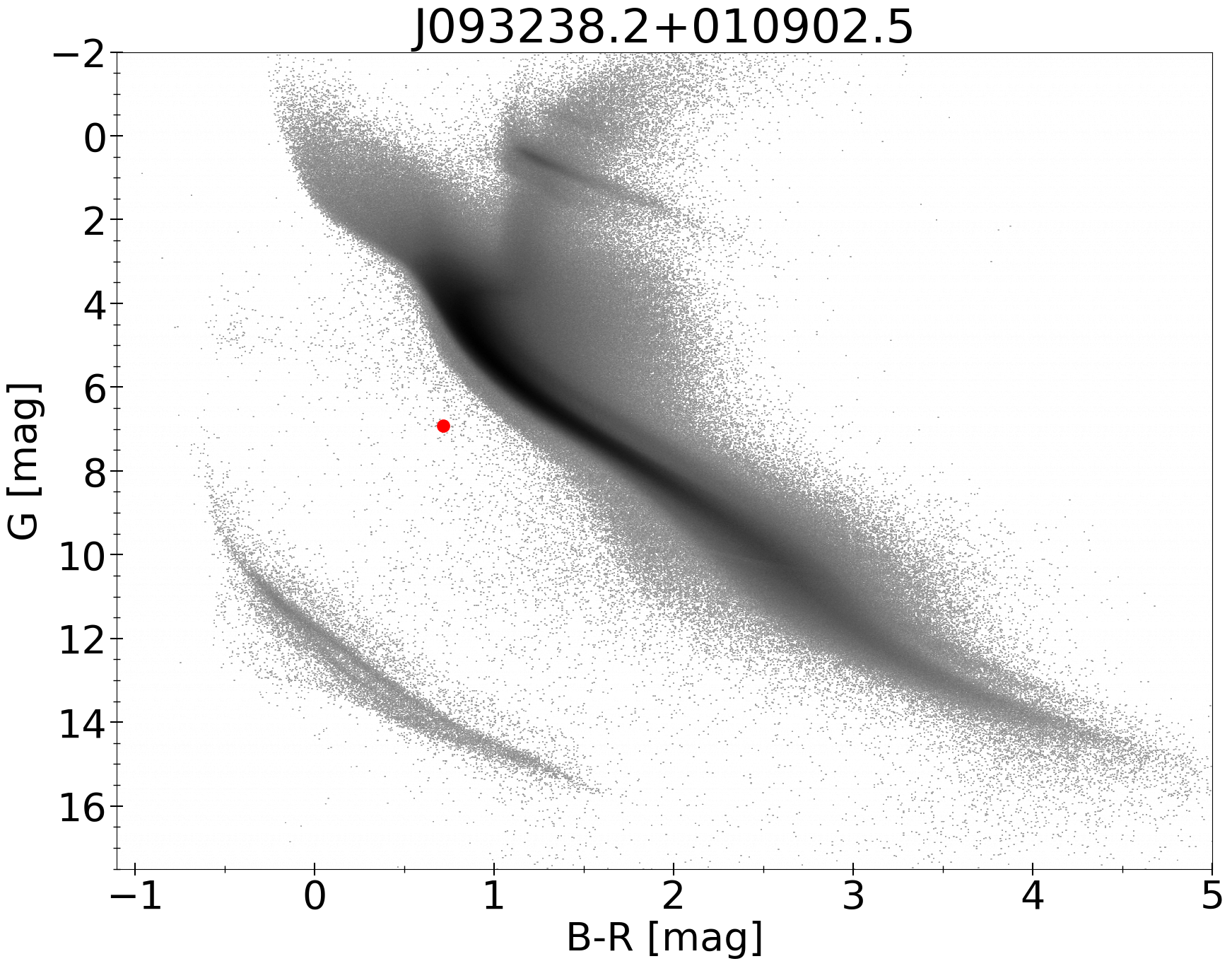}}
\end{minipage}

\begin{minipage}[c]{0.45\hsize}
\resizebox{\hsize}{!}{\includegraphics{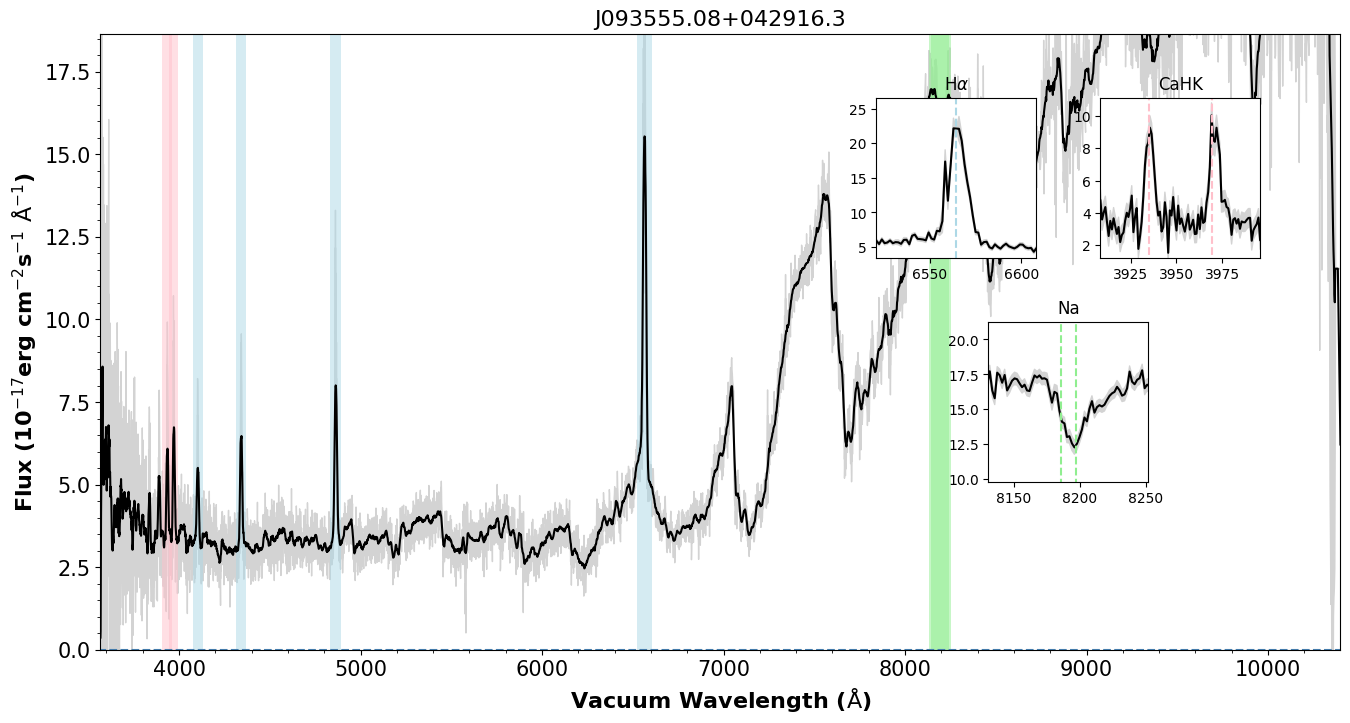}}
\end{minipage}
\hfill
\begin{minipage}[c]{0.321\hsize}
\resizebox{\hsize}{!}{\includegraphics{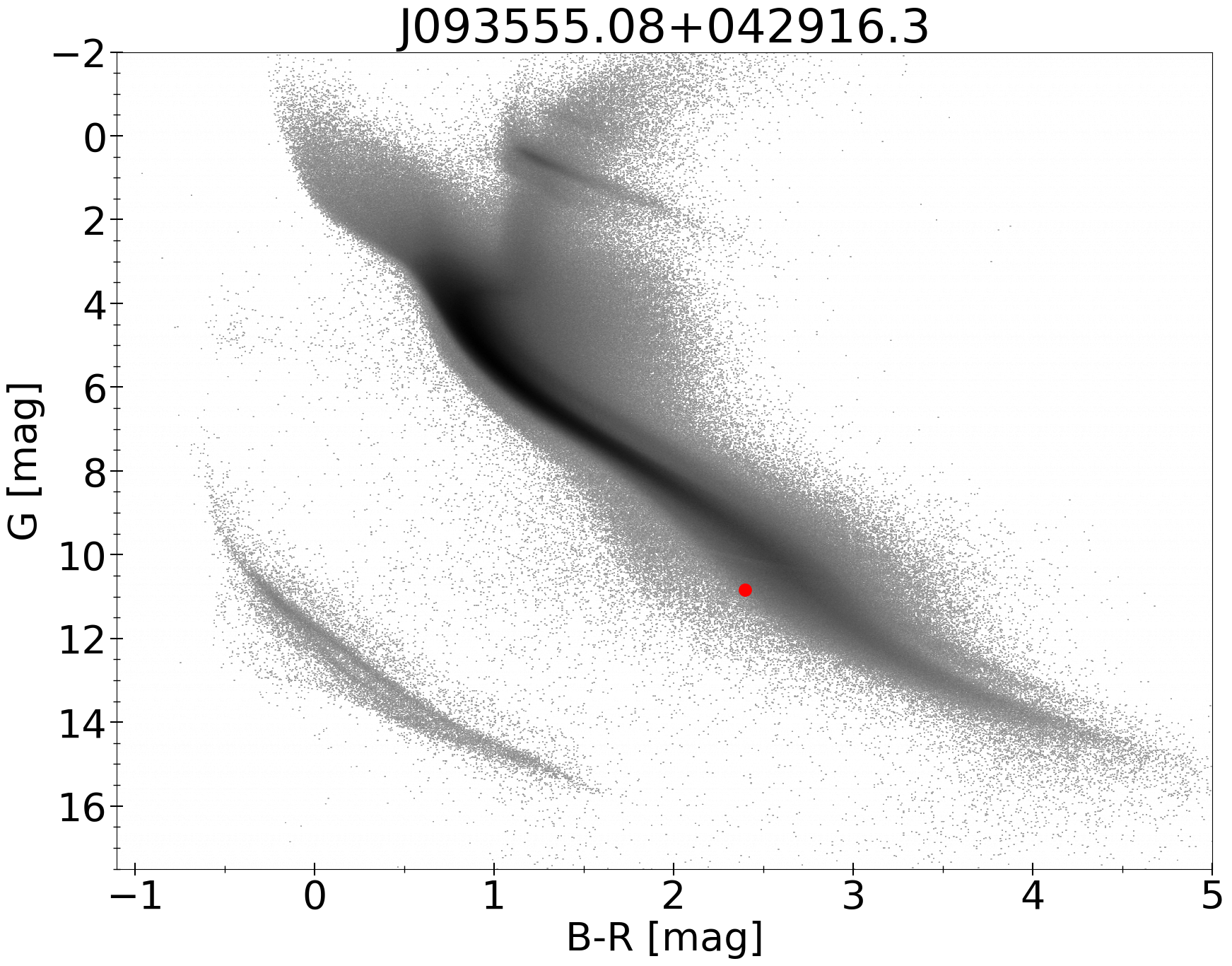}}
\end{minipage}

\caption{Spectra and CMDs of X-ray selected aCWDBs and aCWDB candidates (first 23 entries in Tab.~\ref{t:cvs}).} 
\end{figure*}

\begin{figure*}[t]

\begin{minipage}[c]{0.45\hsize}
\resizebox{\hsize}{!}{\includegraphics{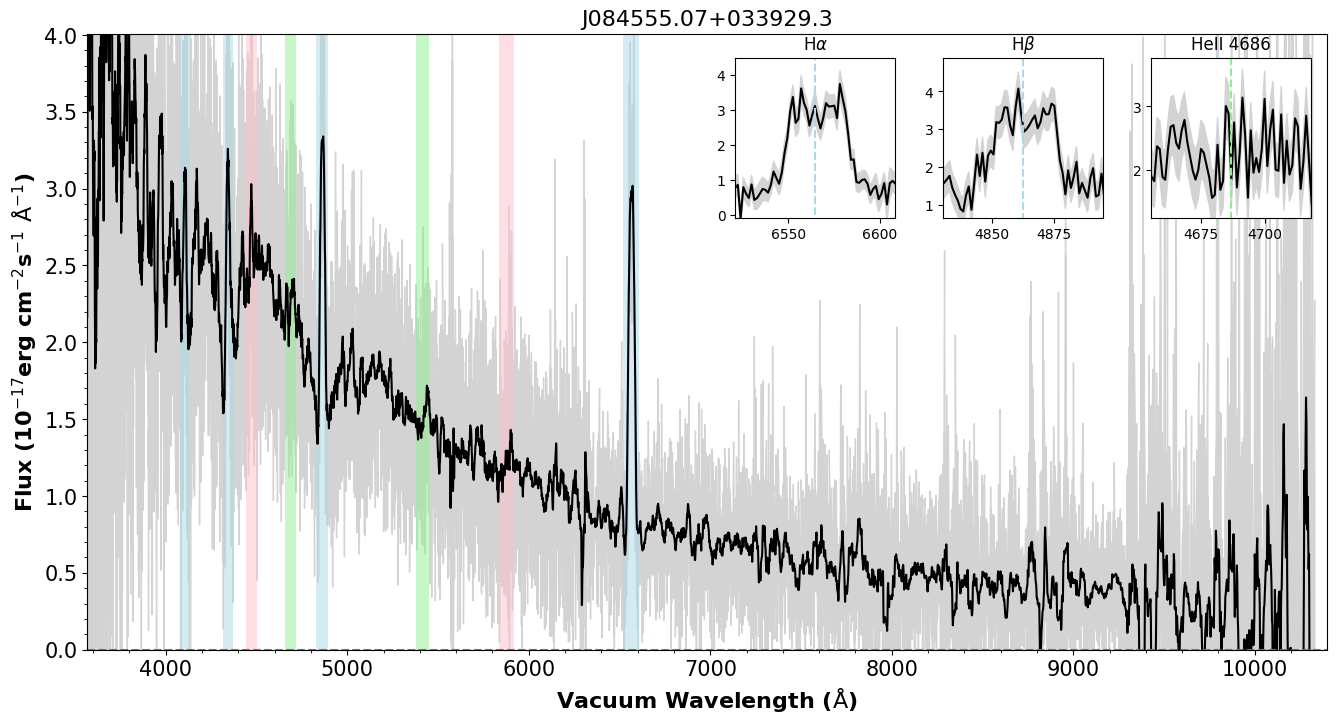}}
\end{minipage}
\hfill
\begin{minipage}[c]{0.321\hsize}
\resizebox{\hsize}{!}{\includegraphics{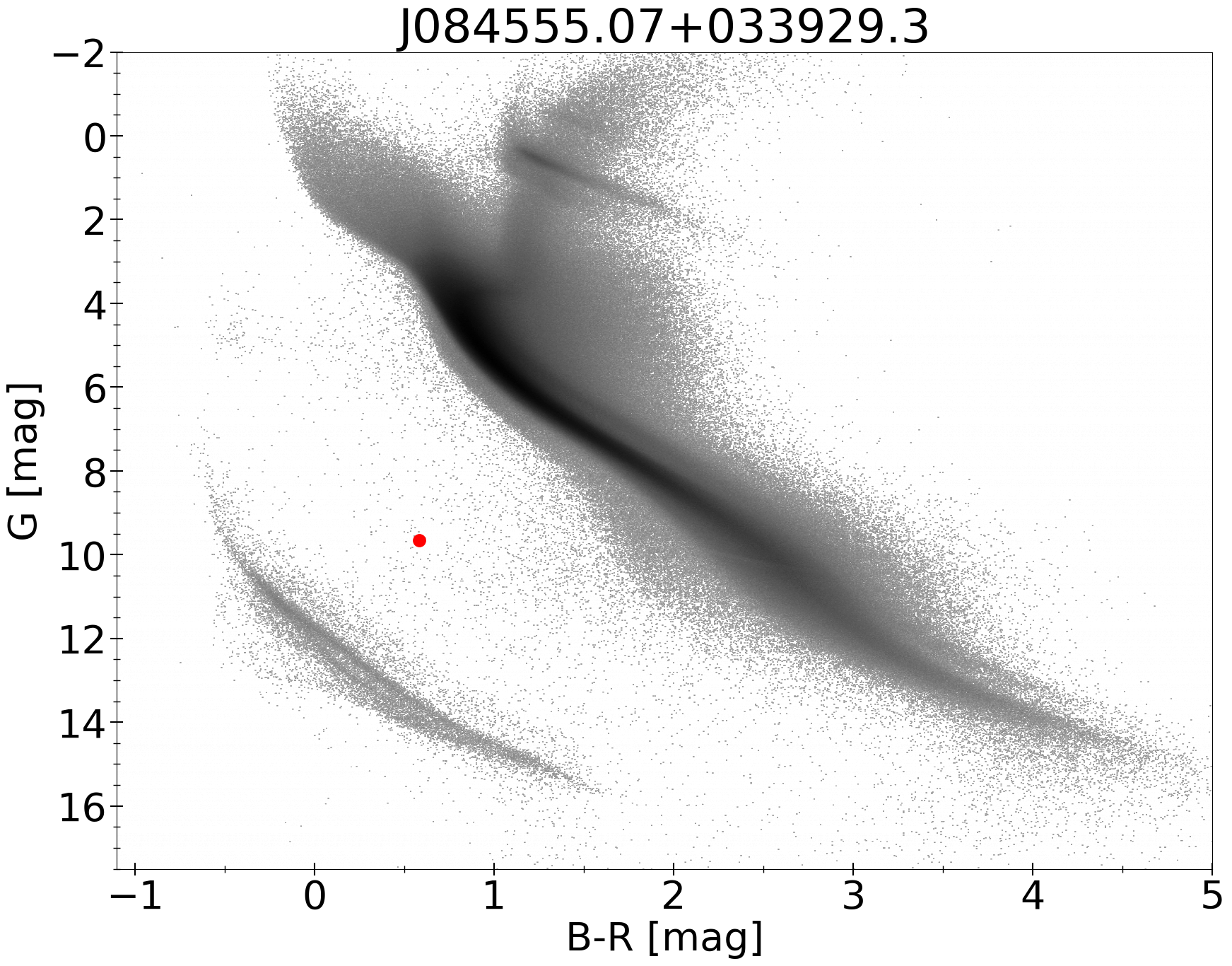}}
\end{minipage}

\begin{minipage}[c]{0.45\hsize}
\resizebox{\hsize}{!}{\includegraphics{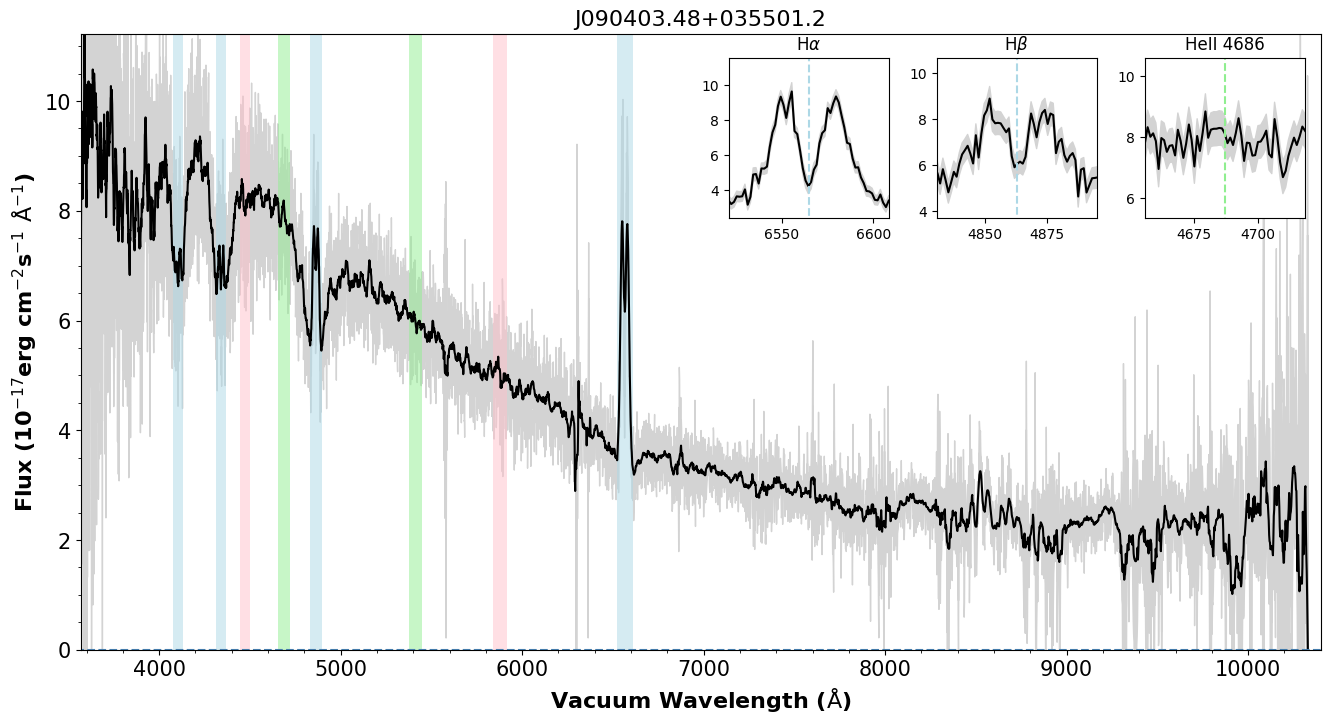}}
\end{minipage}
\hfill
\begin{minipage}[c]{0.321\hsize}
\resizebox{\hsize}{!}{\includegraphics{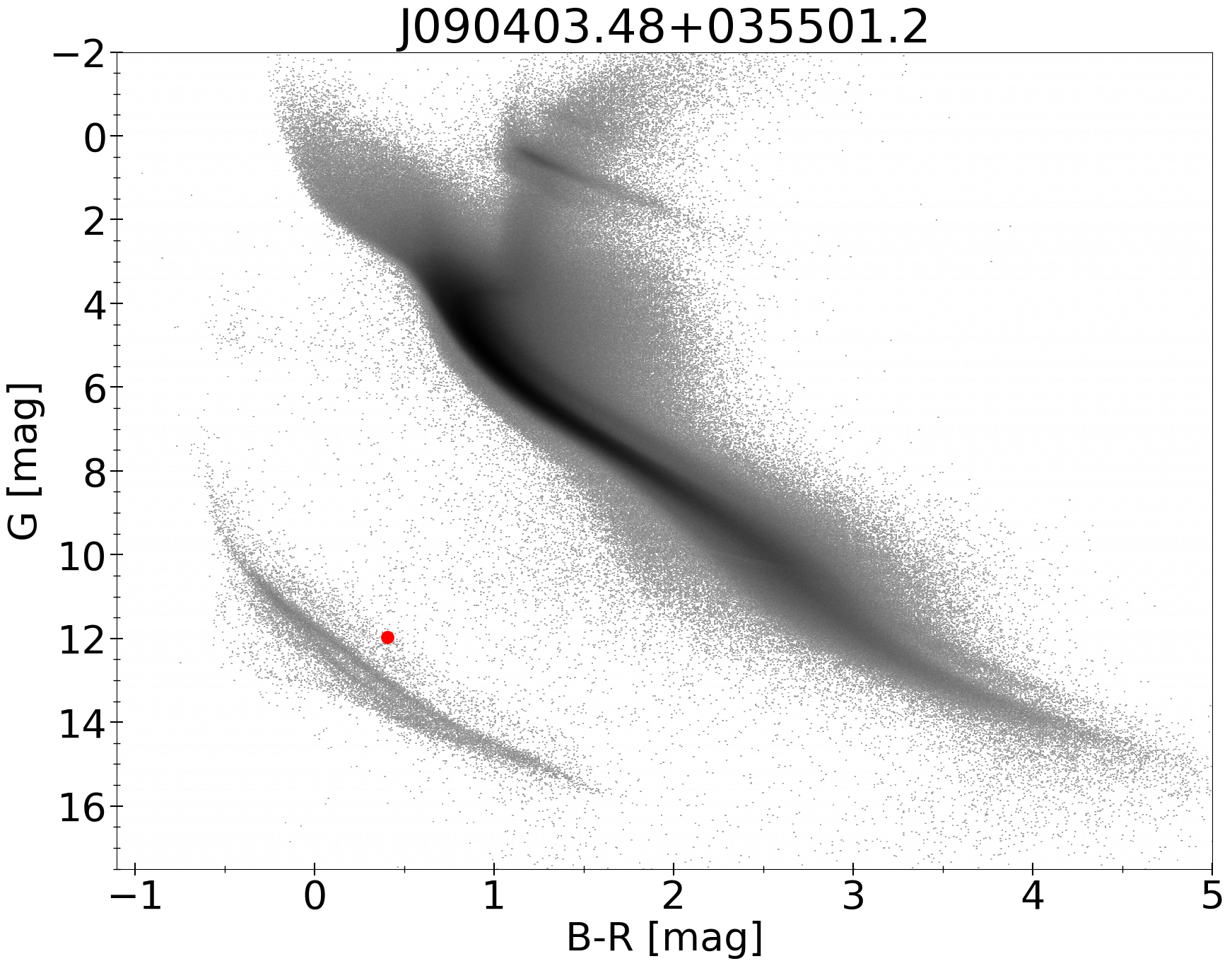}}
\end{minipage}

\begin{minipage}[c]{0.45\hsize}
\resizebox{\hsize}{!}{\includegraphics{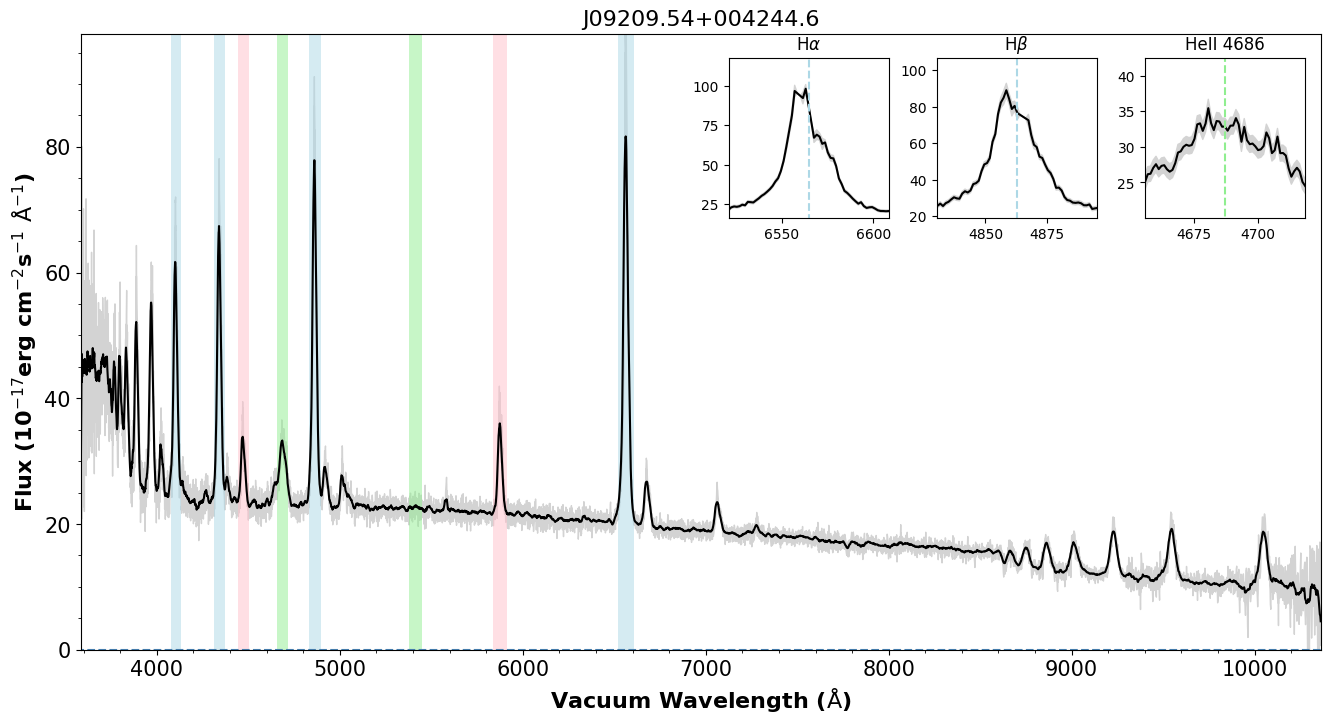}}
\end{minipage}
\hfill
\begin{minipage}[c]{0.321\hsize}
\resizebox{\hsize}{!}{\includegraphics{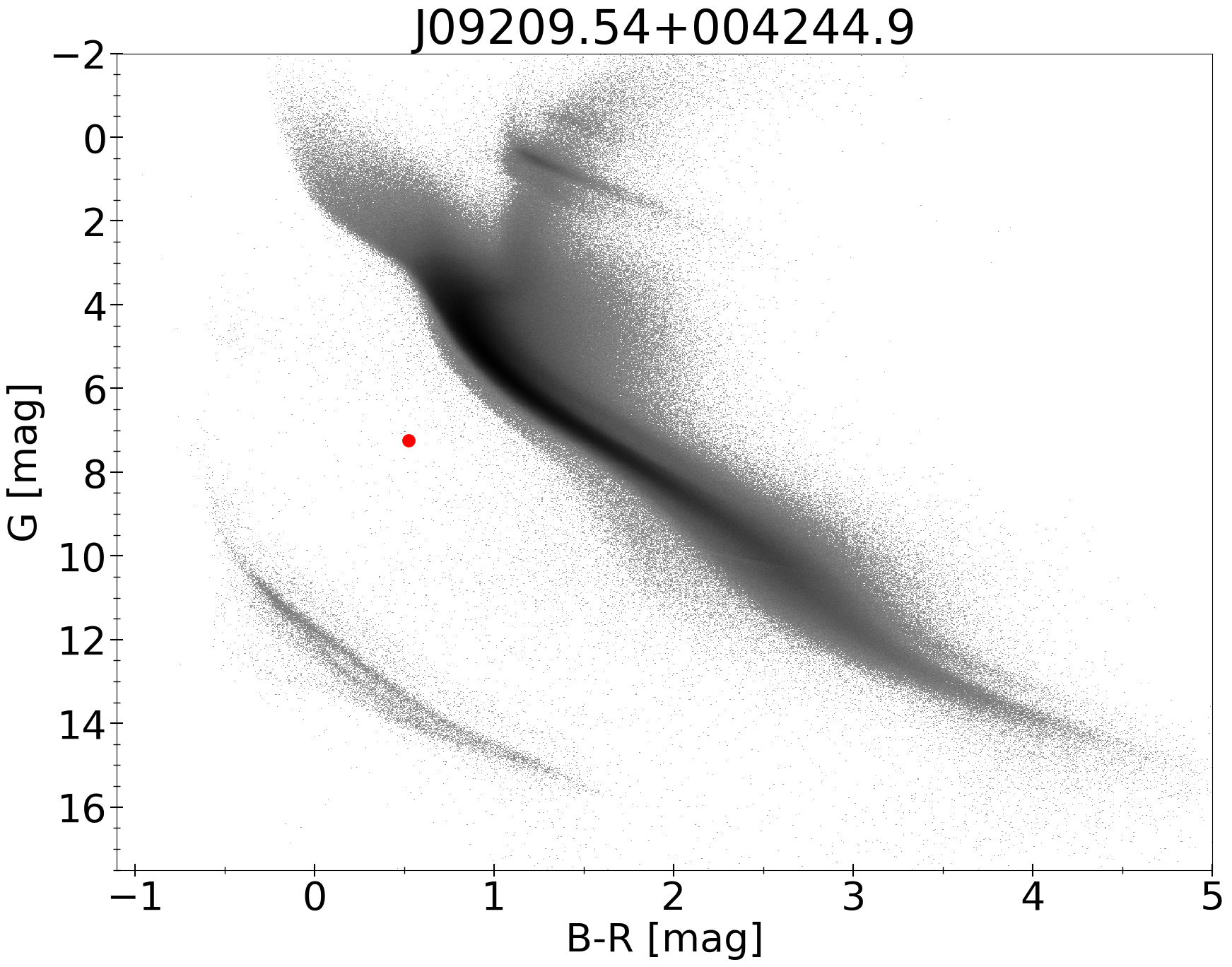}}
\end{minipage}

\caption{Spectra and CMDs of non X-ray selected aCWDBs (bottom section of Tab.~\ref{t:cvs})}
\end{figure*}


\begin{figure*}[t]

\begin{minipage}[c]{0.45\hsize}
\resizebox{\hsize}{!}{\includegraphics{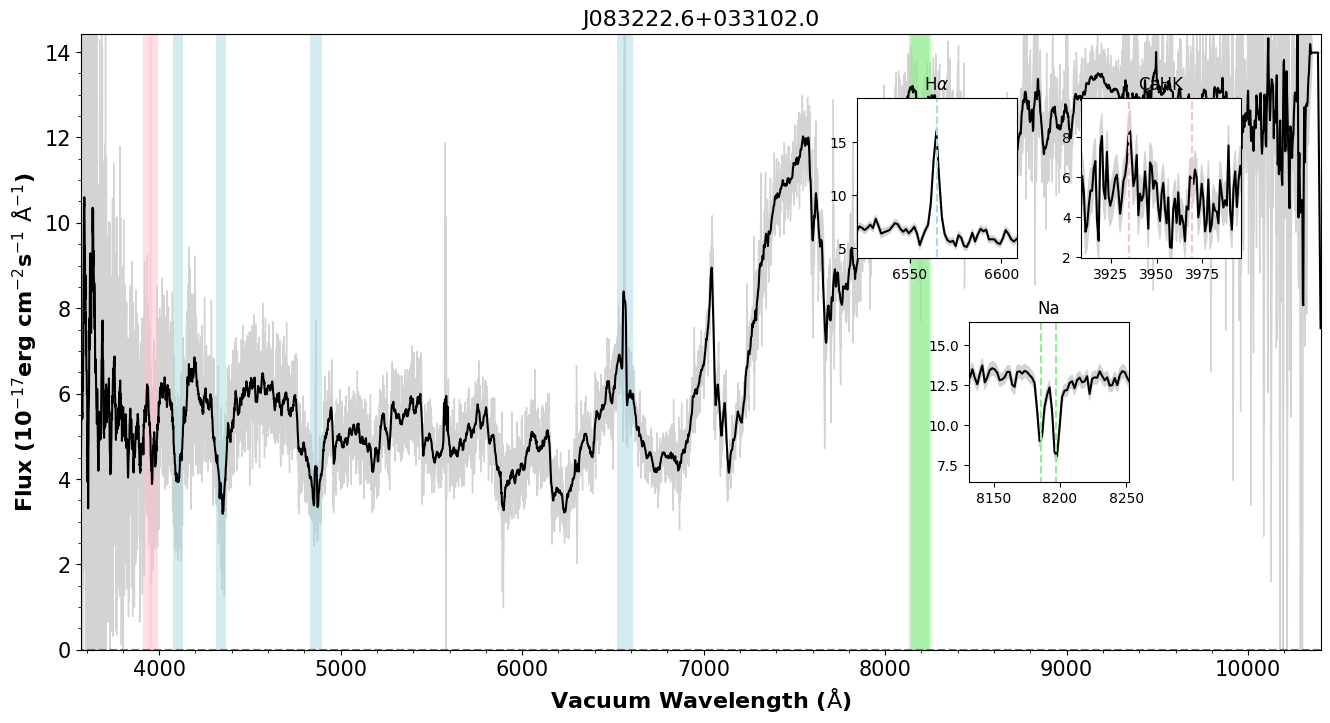}}
\end{minipage}
\hfill
\begin{minipage}[c]{0.321\hsize}
\resizebox{\hsize}{!}{\includegraphics{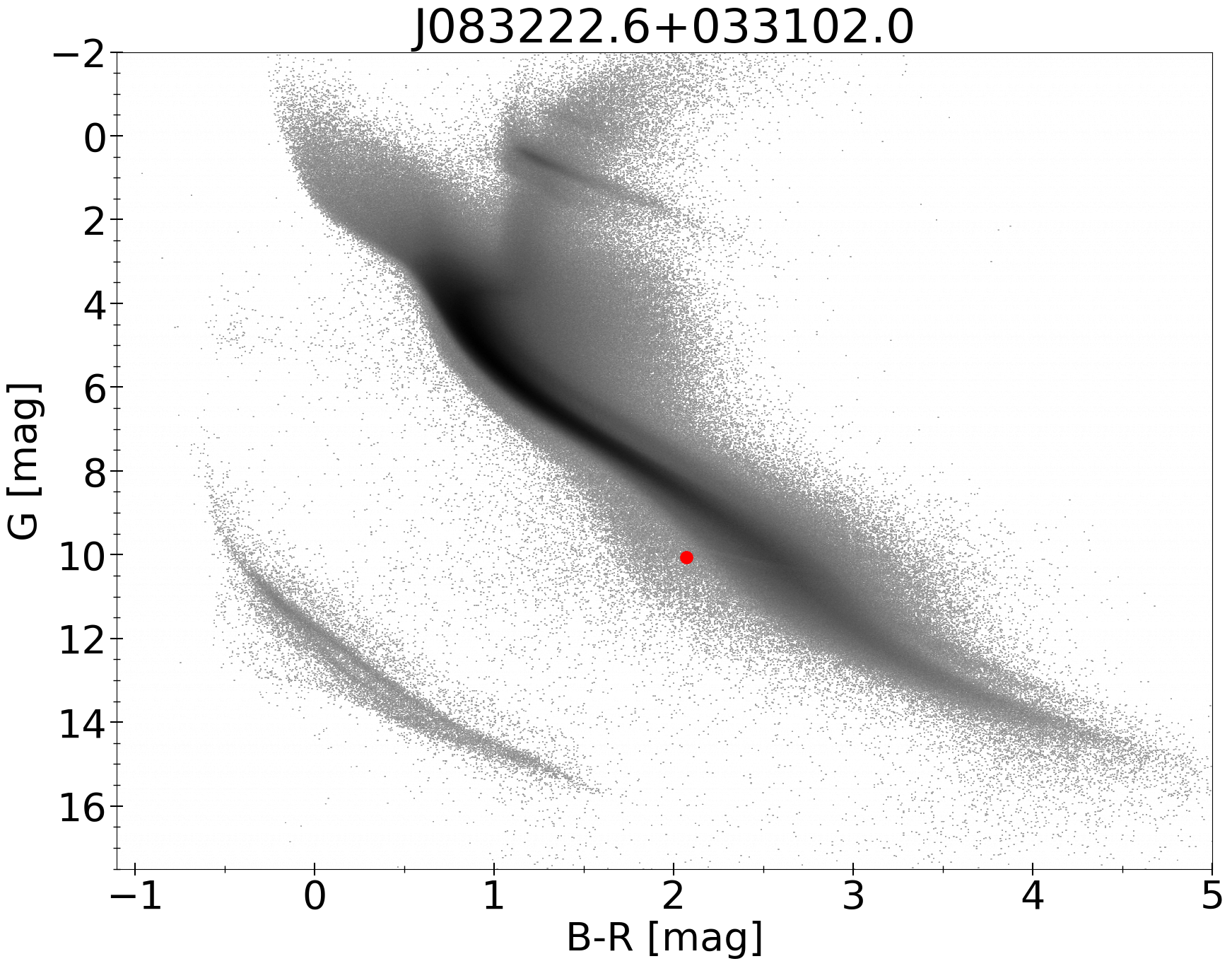}}
\end{minipage}

\begin{minipage}[c]{0.45\hsize}
\resizebox{\hsize}{!}{\includegraphics{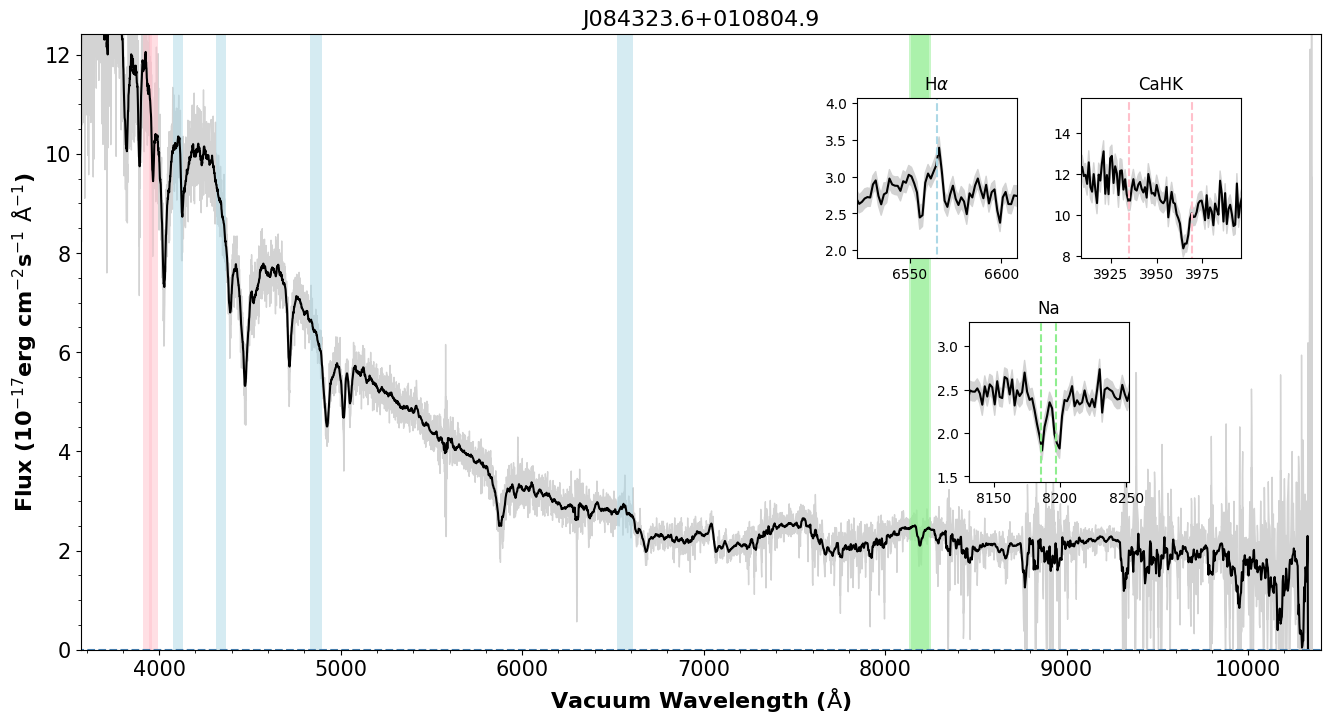}}
\end{minipage}
\hfill
\begin{minipage}[c]{0.321\hsize}
\resizebox{\hsize}{!}{\includegraphics{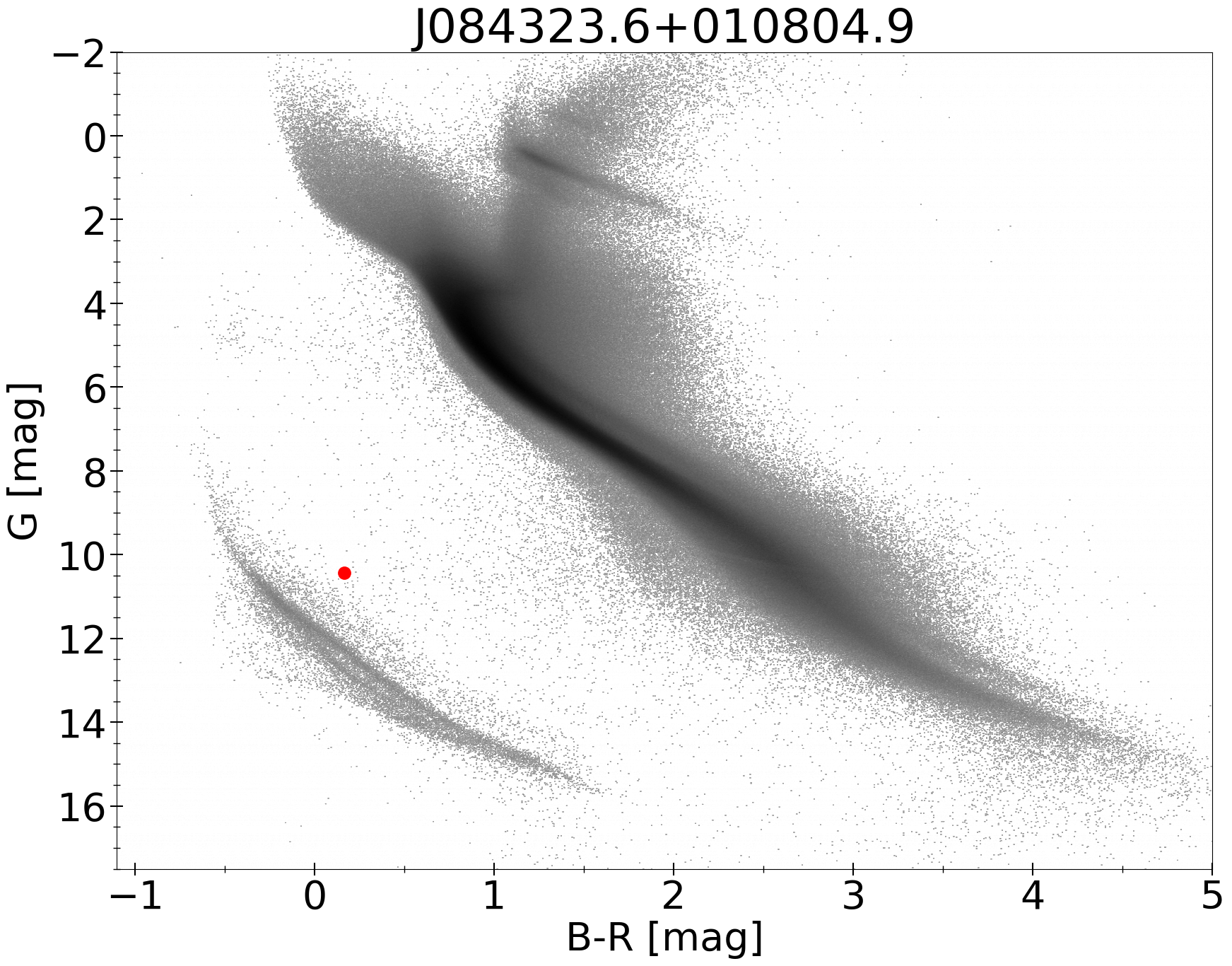}}
\end{minipage}

\begin{minipage}[c]{0.45\hsize}
\resizebox{\hsize}{!}{\includegraphics{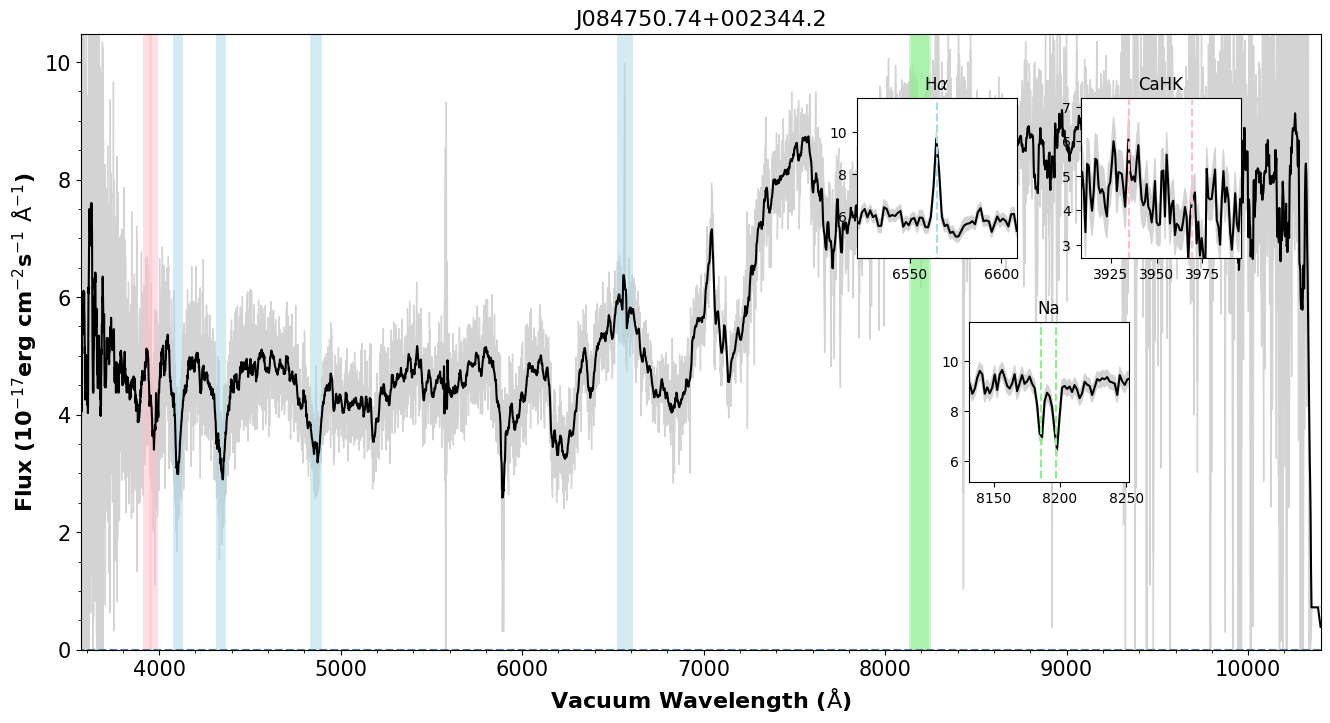}}
\end{minipage}
\hfill
\begin{minipage}[c]{0.321\hsize}
\resizebox{\hsize}{!}{\includegraphics{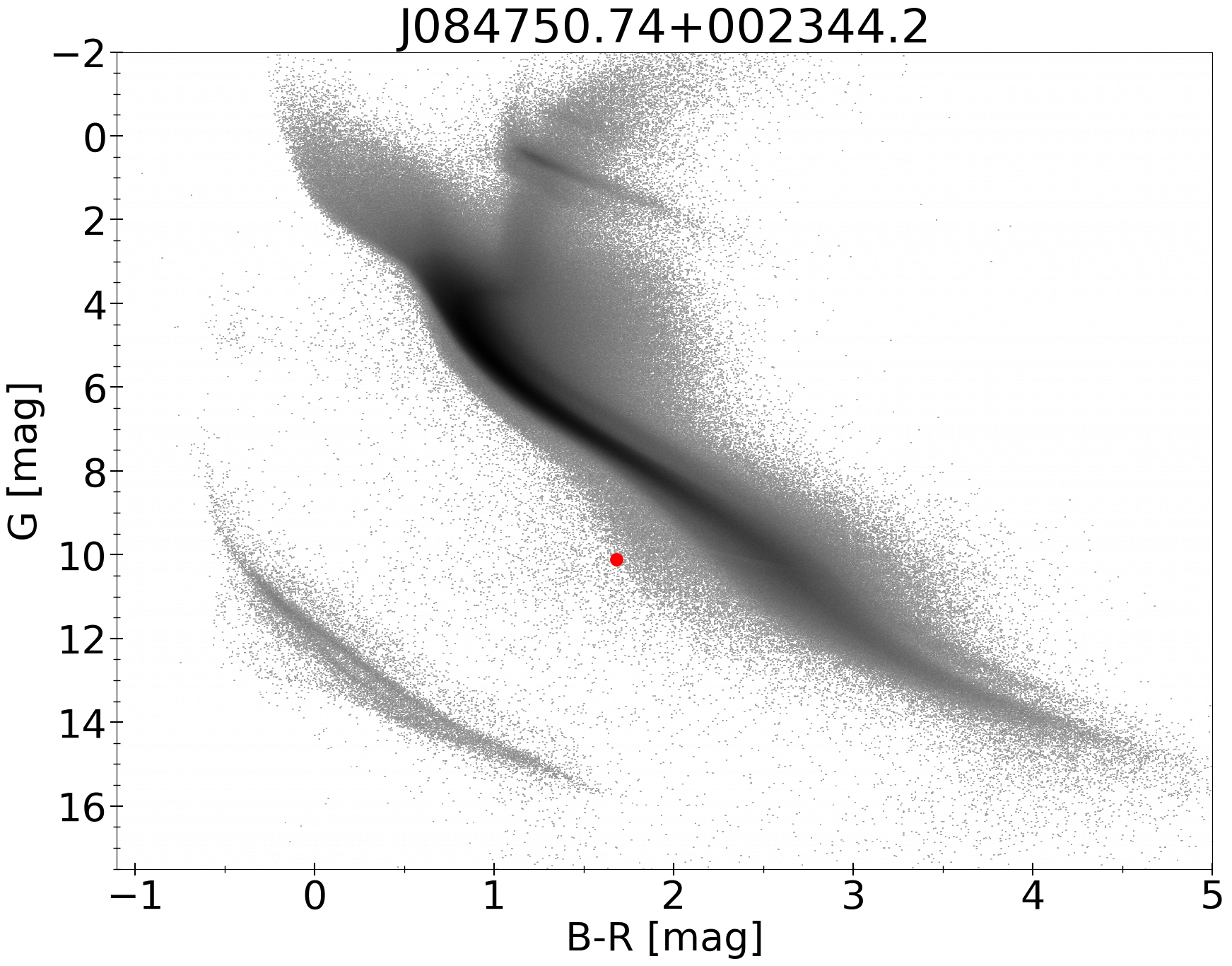}}
\end{minipage}

\begin{minipage}[c]{0.45\hsize}
\resizebox{\hsize}{!}{\includegraphics{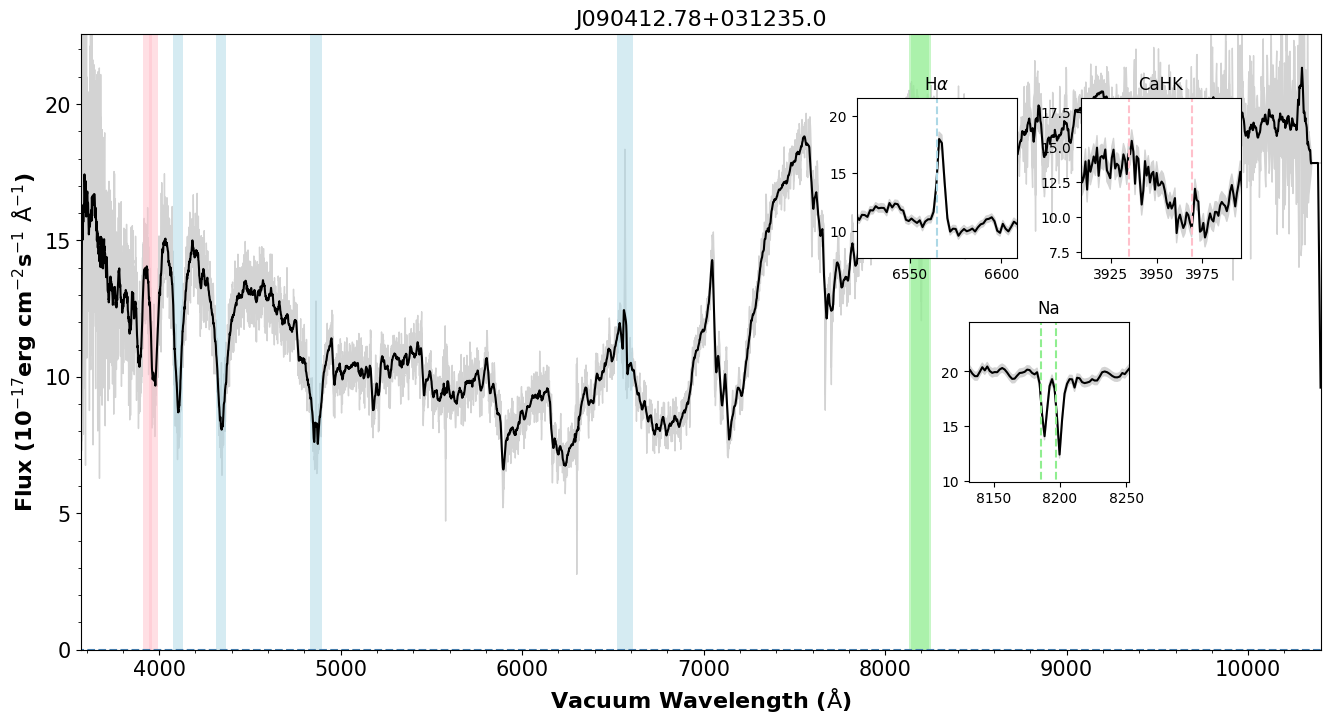}}
\end{minipage}
\hfill
\begin{minipage}[c]{0.321\hsize}
\resizebox{\hsize}{!}{\includegraphics{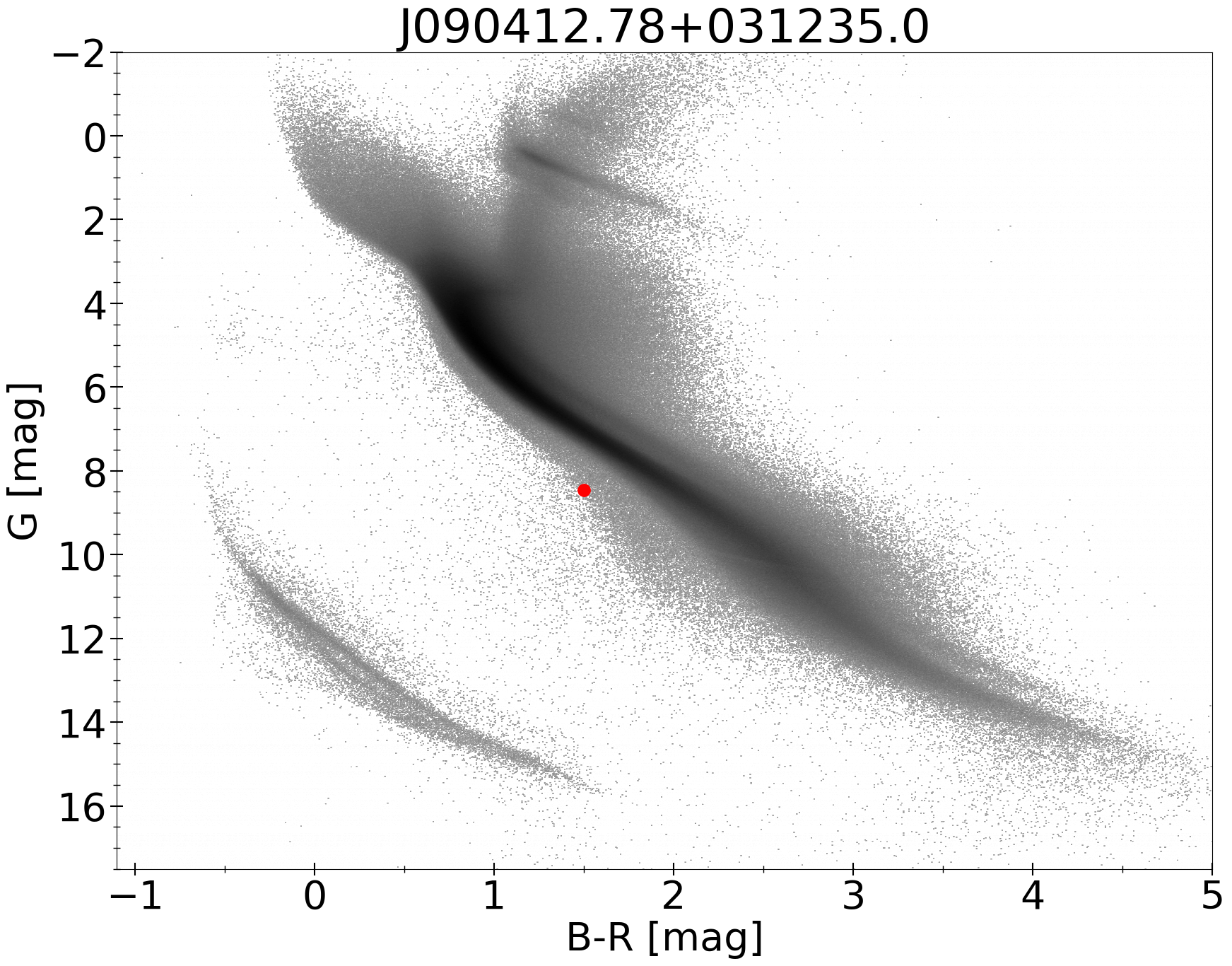}}
\end{minipage}

\begin{minipage}[c]{0.45\hsize}
\resizebox{\hsize}{!}{\includegraphics{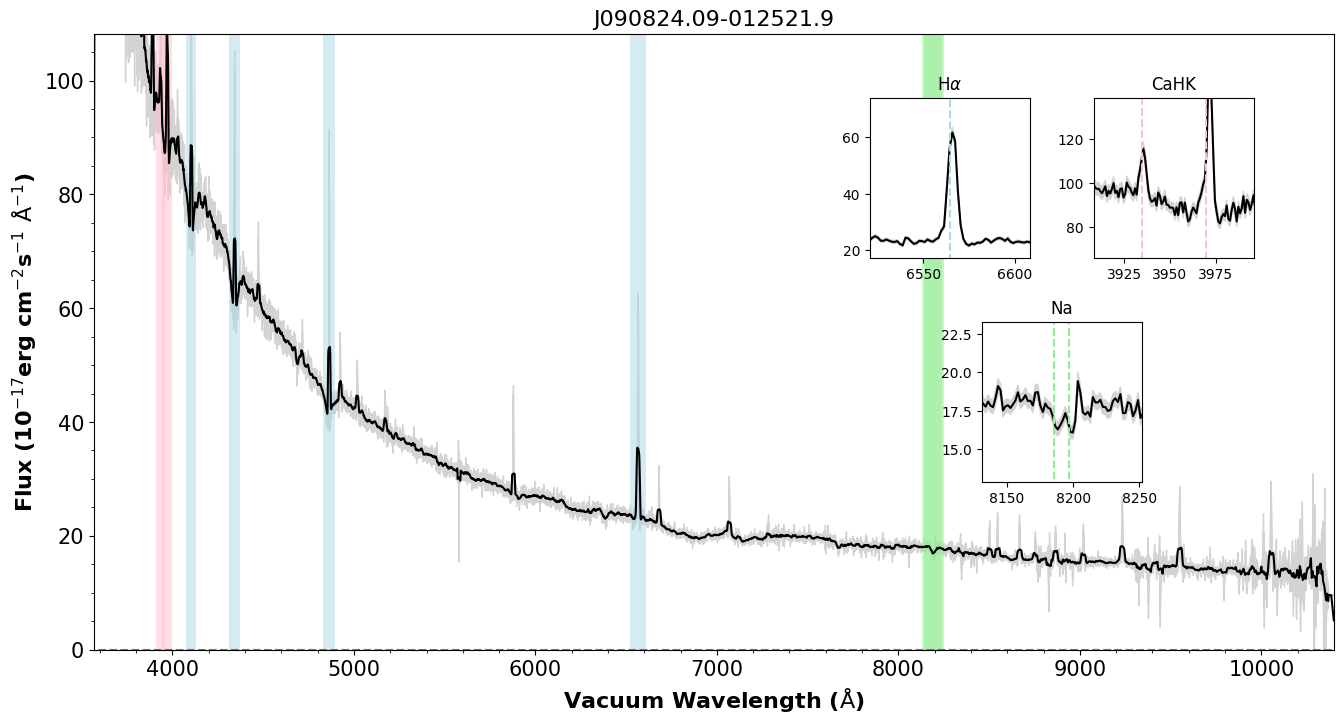}}
\end{minipage}
\hfill
\begin{minipage}[c]{0.321\hsize}
\resizebox{\hsize}{!}{\includegraphics{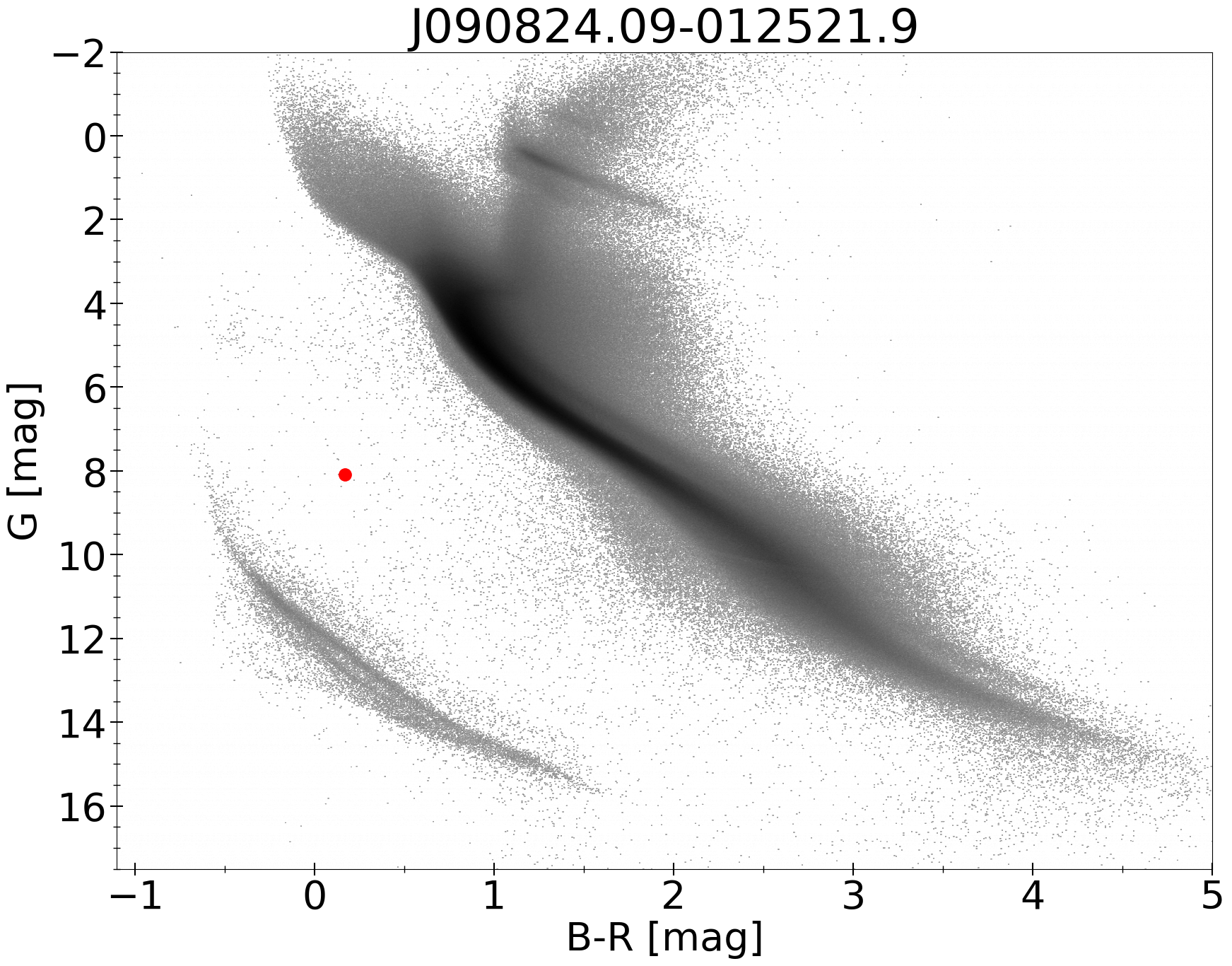}}
\end{minipage}

\end{figure*}

\begin{figure*}
\begin{minipage}[c]{0.45\hsize}
\resizebox{\hsize}{!}{\includegraphics{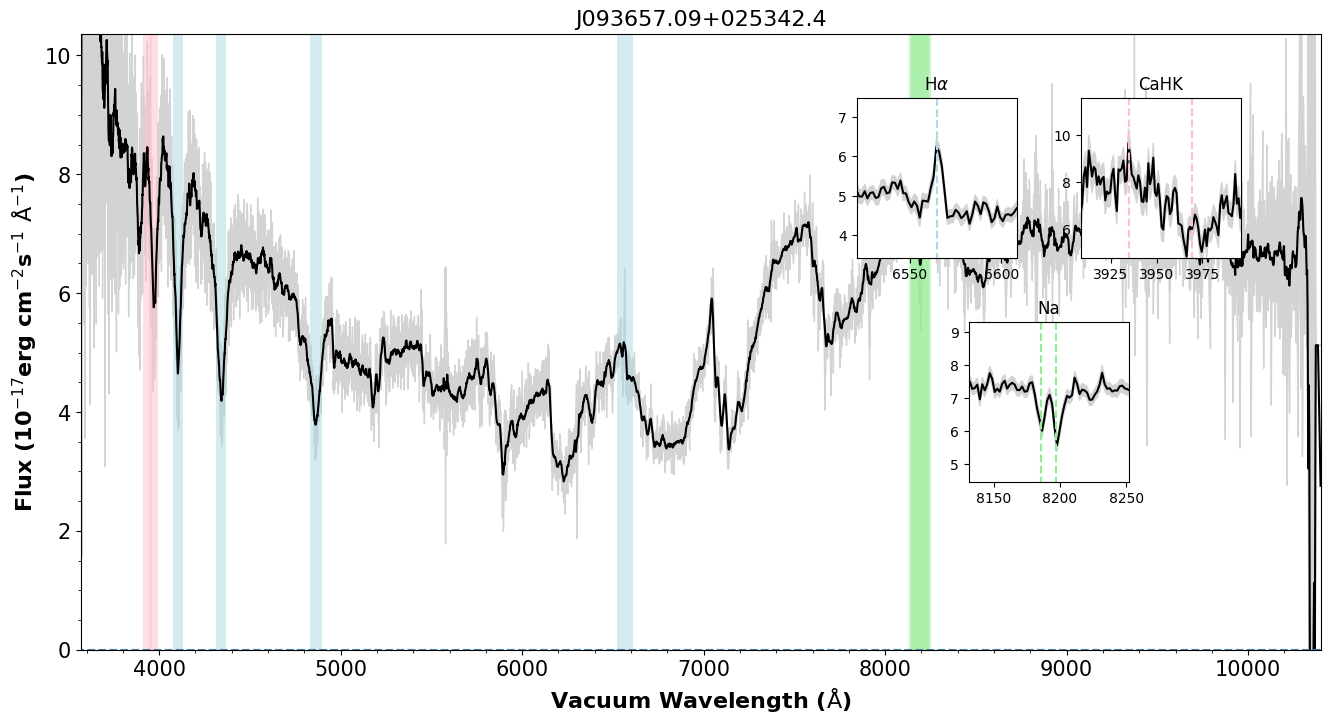}}
\end{minipage}
\hfill
\begin{minipage}[c]{0.321\hsize}
\resizebox{\hsize}{!}{\includegraphics{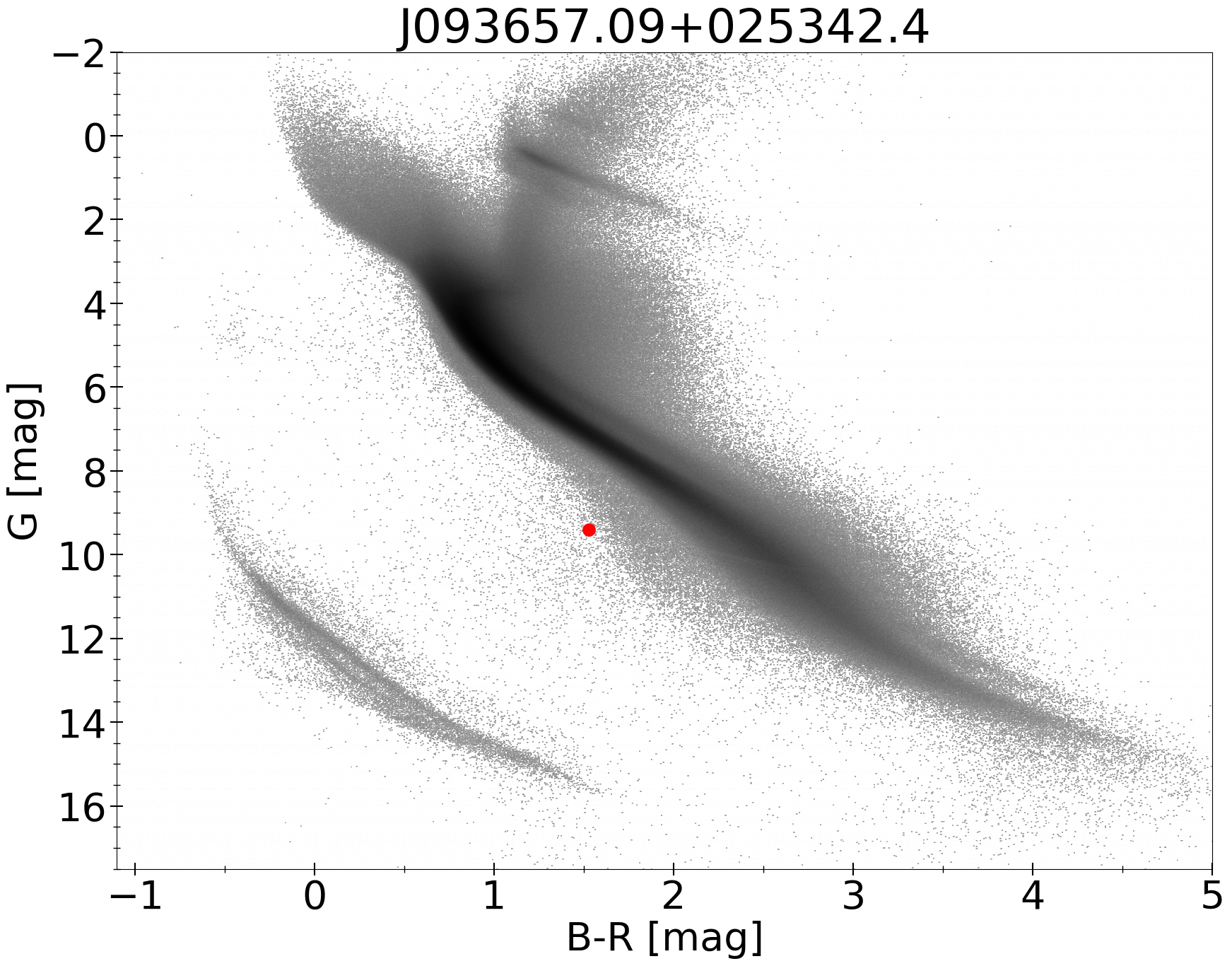}}
\end{minipage}

\caption{Spectra and CMDs of non-accreting CWDBs in the eFEDS field (Tab.~\ref{t:other})}

\end{figure*}

\end{appendix}

\end{document}